%% file: template.tex
\documentclass{article}

\usepackage{arxiv}
\usepackage[utf8]{inputenc} 
\usepackage[T1]{fontenc}    
\usepackage[hidelinks]{hyperref}
\usepackage{url}            
\usepackage{booktabs}       
\usepackage{amsfonts}       
\usepackage{nicefrac}       
\usepackage{microtype}      
\usepackage{graphicx}
\usepackage{natbib}
\usepackage{doi}

\usepackage{framed}
\usepackage{amsmath}
\usepackage[most]{tcolorbox}
\usepackage[dvipsnames]{xcolor}
\colorlet{shadecolor}{orange!15}

\usepackage{xcolor}
\usepackage{soul}
\sethlcolor{orange!15}

\usepackage{float}
\floatstyle{boxed} 
\restylefloat{figure}
\usepackage{mathrsfs}

\usepackage{multirow}
\usepackage{multicol}
\usepackage{longtable}
\usepackage{subcaption}
\usepackage{array}
\usepackage{makecell}
\usepackage{rotating}

\DeclareMathOperator{\arctantwo}{arctan_2}
\usepackage{relsize}

\title{Biochemical Network Motifs Under Periodic Forcing: A Selective Catalogue of Transfer Functions and Frequency Response Properties}

\author{ {Nguyen H.N. Tran} \\
	Department of Mechanical Engineering and Mechanics\\
	Drexel University\\
	Philadelphia, PA 19103 \\
	\texttt{nt625@drexel.edu} \\
	\And
	{Federico Frascoli} \\
	Department of Mathematics \\
	Swinburne University of Technology \\
	Melbourne, Australia \\
	\texttt{ffrascoli@swin.edu.au} \\
	\AND
        {Andrew H.A. Clayton} \\
	Department of Physics and Astronomy \\
	Swinburne University of Technology \\
	Melbourne, Australia \\
	\texttt{aclayton@swin.edu.au} \\
}

\date{}

\renewcommand{\shorttitle}{Biochemical Network Motifs Under Periodic Forcing}

\begin{document}
\maketitle

\begin{abstract}
    Biological cells house the internal molecular machinery responsible for their physiological functions. These machinery components -- the unique molecular species -- undergo ordered interactions in time. \textbf{Biochemical networks} provide a representation of these ordered interactions -- akin to \textit{blueprints} of the underlying molecular processes. Embedded within these blueprints are frequently recurring architectures. These recurring architectures are known as \textbf{biochemical network motifs}, and serve as a \textit{basis} for constructing larger networks. They are a natural \textit{graphical alphabet} of biochemical networks. Understanding the function of each \textit{letter} in this alphabet (motif) in an attempt to gain insight into how their combinations create the \textit{words} (biochemical networks) that drive cellular functions, has been a longstanding pursuit of systems biology.
    One specific objective within this pursuit is understanding how individual motifs respond to pulsatile and oscillatory signals. This is especially relevant because biochemical networks are often activated by signals that, in nature, occur in the form of pulses and oscillations. A powerful analytical tool for studying such dynamics is the transfer function -- a compact frequency-domain description of input–output dynamics.
    In this work, we derive transfer functions for a set of commonly studied network motifs and characterise their responses to pulsatile and oscillatory inputs. The novelty of this review does not lie in the introduction of new mathematical theorems or biological discoveries, but in bridging well-established frequency domain formalisms from control theory with the analysis of biochemical networks under periodic forcing. In doing so, our contributions are threefold:
    \begin{enumerate}
    \item \textbf{A systematic derivation and compilation of transfer functions for common network motifs} --- consolidating results scattered across the literature and establishing a consistent formalism for motif-level transfer functions.
    \item \textbf{Contextualisation of these transfer functions within biological models} --- extending abstract transfer functions to concrete biological settings so that the results are readily applicable without extensive mathematical labour.
    \item \textbf{Resolution of ambiguity between biological and control-theoretic treatments of feedback} --- clarifying how feedback loops should be understood within the transfer function formalism and reconciling differences between biology literature and control-oriented literature. This is done by formalising the notion of an \textit{intrinsic transfer function}.
    \end{enumerate}
\end{abstract}


\tableofcontents

\include{1_Introduction}
\include{2_Background}
\include{4_Discussion}
\include{4a_SimpleRegulation}
\include{4b_Cascades}
\include{4c_Feedforward}
\include{4d_Autoregulation}
\include{4e_Feedback}
\include{5_Appendix}
\include{6_AppendixB}
\include{7_AppendixC}

\bibliographystyle{plain}
\bibliography{references}  






\end{document}

%% file: 1_Introduction.tex
\section{Introduction and literature}
\label{sec:Introduction}
The study of external forcing on biochemical networks is a well-established method for probing their internal mechanisms and manipulating their behaviour. Much like in physics—where systems are perturbed to reveal underlying structure --- external stimuli in biology can uncover the principles governing regulatory dynamics. While step or constant inputs are common practice, the application of periodic forcing is more recent with the advent of tools like optogenetics and microfluidics. Its theoretical discussions in the biological literature are also relatively underexplored.

Coupled with this is the study of large biochemical networks which remains challenging for its complexity. Reductionist approaches offer one path through this complexity. One such approach is the use of network motifs --- sub-networks that significantly recur within larger systems. While these motifs have been thoroughly studied in the time domain, they have received less attention in the frequency domain --- for systems governed by pulsatile, rhythmic, or oscillatory signals, as is often the case in biological signaling. By examining network motifs through the lens of frequency response, we uncover a complementary layer of insight --- one that is often obscured in purely time-domain approaches.

Towards the frequency domain description, we need a frequency-response function. The transfer function from control theory becomes a readily available tool for this. Originally developed in electrical engineering, transfer functions describe how a system responds to inputs of different frequencies without requiring knowledge of its internal structure. This ``black box'' approach made its way into biology as early as the 20th century, notably in the study of visual systems and later in bacterial chemotaxis \cite{shimizu_modular_2010}.

A few scattered studies on frequency response and transfer functions of specific network architectures have already appeared in prior literature, scattered throughout theoretical and experimental literature, following the growing popularity of reductionist systems biology in the early 2000s and 2010s \cite{geva-zatorsky_fourier_2010,hersen_signal_2008,mitchell_oscillatory_2015,shin_linear_2010,toettcher_using_2013}. The goal here is to extend this perspective by providing a compendium of the most common motif architectures, along with general derivations of their transfer functions. These derivations are presented in a way that is agnostic to specific biochemical contexts or timescales, allowing readers to analyse any given motif and immediately form a qualitative, intuition-driven prediction of its frequency response characteristics. For example, simply by observing the number of nodes in a signaling cascade, one can anticipate how rapidly signals will attenuate. Or, by recognising multiple branches converging on a single node, one can infer the potential existence of an optimal forcing frequency. This framework aims to empower the reader with a kind of visual intuition.

A key motivation of this work is to concretely define and formalise what it means to be a feedback loop. There is much conflict and ambiguity in the biological literature with its control-oriented treatments, and this often leads to confusion for the newly initiated. We discuss this further in the mini-introduction of Section \ref{Sec:autoreg} on autoregulations --- a motif so canonical and simple, yet at the heart of much ambiguity. 

\textbf{Commentary: A Reflection on What is Fundamental}

The reader, familiar with biological models, may raise objections.

As with my many debates with my biologist friends:

“There is no such thing as an arbitrary network motif from A to B in biology. The interaction from gene A to gene B is fundamentally different from that between proteins. How can you propose such a general idea of a biochemical network?”

The mathematical modeller, on the other hand, might argue:

“This is all arbitrary! We have freedom in how we model. How can you make any general claims at all?”

But herein lies the pursuit of fundamentals.

There is only one truth we can rely on: measurement.

Given a scale in (state) space and time, we ask: What can we observe? What can we quantify? With a known set of species and a resolution at which their interactions become distinguishable, we inherit an ordering. That ordering --- the sequence and direction of influence in time (and therefore in timescales and frequencies as we will see later) --- is not a modelling choice. It comes from the physical structure of the system and our act of measurement. From the perspective of bottom-up theory, it is equivalent to assuming a network architecture.

We allow ourselves one modest assumption: that we are observing the system at a scale where the rate of change of each species can be described by a first-order derivative in time --- regardless of the underlying molecular mechanism. All we ask is to live in a regime where such a first-order differential description holds.

And then the question becomes:
What is fundamental to this structure of ordered, first-order interactions in time?

It is this: the form of the equations local to equilibrium.

When we linearise the system --- regardless of the form of its original description --- we arrive at a linear time-invariant (LTI) representation. The numerical values may change. The LTI structure, however, does not. This is what we will exploit in studying complex biological systems.

Without such abstraction, we risk drowning in molecular detail. My adviser used to ask me: what does an A0-sized diagram of the ERK pathway really tell us? At some point, too much information becomes noise --- a point of diminishing returns.

This is where the value of reductionist systems biology becomes apparent. While the field may have moved away from pure reductionism following its surge in the 2000s and 2010s, the insight into structure-function relationships continue to offer powerful tools for reasoning about complex biological processes.

%% file: 2_Background.tex
\section{Background}

Here, we highlight core ideas that will be used throughout this review. As these concepts are well established and widely covered in classical texts, we refer the reader to standard control theory references as well as Section 2 of a prior pedagogical paper \cite{tran_transfer_2023} for a more biochemically oriented presentation. Our goal in this section is not to re-derive these results, but rather to emphasise the key intuition behind each concept and relate them back to our context of biochemical networks.

In summary, biochemical reactions are physical processes that evolve in continuous time. As such, they can be described by continuous time-invariant ordinary differential equations (ctiODEs). At equilibrium, the dynamics of these biochemical reactions converge into simpler dynamics described by linear regimes of these ctiODEs. The structure of ctiODEs in these linear regimes are known as linear ctiODEs (lctiODEs). lctiODEs have an equivalent representation in complex frequency space -- a space built by a continuum of oscillations whose envelopes are allowed to exponentially grow and decay over time or ``generalised oscillations'' if you will. This equivalent representation is known as the transfer function and serves as a frequency response function of the system. Such a function allows us to predict outputs due to periodically forced inputs, once a system has settled into equilibrium.

\subsection{First order ctiODE system --- the language of biochemical networks}
A first order ctiODE system is a set of equations relating the rate of change of each state variable to the input and state variables present in the system. That is, given a set of input variables $\{u_1,u_2,\ldots,u_n\}$ and set of state variables $\{x_1,x_2,\ldots,x_n\}$, the corresponding first order ctiODE system takes the form:

\begin{equation}
    \begin{aligned}
    \frac{dx_1}{dt} &= F_1(u_1,u_2,...,u_m,x_1,x_2,...,x_n) \\
    \frac{dx_2}{dt} &= F_2(u_1,u_2,...,u_m,x_1,x_2,...,x_n) \\
    & \hspace{6em} \vdots \\
    \frac{dx_n}{dt} &= F_n(u_1,u_2,...,u_m,x_1,x_2,...,x_n) \text{ .}
\end{aligned}
\label{ctiODE}
\end{equation}

As one progressively zooms into a system of biochemical reactions in both \textit{space} and \textit{time}, a resolution is eventually reached where the \textit{spatial} presence of each species at each moment in \textit{time} becomes observable. At this resolution, we are able to track the population of each species over time. By doing so, we gain access to a fundamental physical quantity: the \textit{speed} at which each population changes over time. So by sufficiently zooming into our physical system, three core observables become unlocked: population, time, and the speed at which populations change. The existence of these three quantities implies that there must be an underlying set of governing rules relating them. Because of its general structure, it is not unreasonable to propose that these rules take the mathematical form of Equation \eqref{ctiODE} where each $u_i$ and each $x_j$ represent a unique biochemical species. In this way, Equation \eqref{ctiODE} is then the \textbf{definition of a biochemical network} written in the language of dynamical systems.

\subsection{First order lctiODE system --- the language of biochemical networks at equilibrium}
Suppose our biochemical network, defined by Equation \eqref{ctiODE}, settles into an equilibrium state denoted by $\{\overline{x_1}, \overline{x_2}, \dots, \overline{x_n}\} = \{\overline{\textbf{x}_j}\}$ due to a set of inputs $\{\overline{u_1}, \overline{u_2}, \dots, \overline{u_m}\} = \{\overline{\textbf{u}_i}\}$. The dynamics of the biochemical network local to this equilibrium is given by:

\begin{equation}
    \begin{aligned}
        \frac{d(\Delta x_1)}{dt} &= \frac{\partial F_1}{\partial u_1}\bigg|_{(\overline{\mathbf{u}_i},\overline{\mathbf{x}_j})} \Delta u_1 + \dots + \frac{\partial F_1}{\partial u_m}\bigg|_{(\overline{\mathbf{u}_i},\overline{\mathbf{x}_j})} \Delta u_m + \frac{\partial F_1}{\partial x_1}\bigg|_{(\overline{\mathbf{u}_i},\overline{\mathbf{x}_j})} \Delta x_1 + \dots + \frac{\partial F_1}{\partial x_n}\bigg|_{(\overline{\mathbf{u}_i},\overline{\mathbf{x}_j})} \Delta x_n \\
        \frac{d(\Delta x_2)}{dt} &= \frac{\partial F_2}{\partial u_1}\bigg|_{(\overline{\mathbf{u}_i},\overline{\mathbf{x}_j})} \Delta u_1 + \dots + \frac{\partial F_2}{\partial u_m}\bigg|_{(\overline{\mathbf{u}_i},\overline{\mathbf{x}_j})} \Delta u_m + \frac{\partial F_2}{\partial x_1}\bigg|_{(\overline{\mathbf{u}_i},\overline{\mathbf{x}_j})} \Delta x_1 + \dots + \frac{\partial F_2}{\partial x_n}\bigg|_{(\overline{\mathbf{u}_i},\overline{\mathbf{x}_j})} \Delta x_n \\
        & \hspace{17em} \vdots \\
        \frac{d(\Delta x_n)}{dt} &= \frac{\partial F_n}{\partial u_1}\bigg|_{(\overline{\mathbf{u}_i},\overline{\mathbf{x}_j})} \Delta u_1 + \dots + \frac{\partial F_n}{\partial u_m}\bigg|_{(\overline{\mathbf{u}_i},\overline{\mathbf{x}_j})} \Delta u_m + \frac{\partial F_n}{\partial x_1}\bigg|_{(\overline{\mathbf{u}_i},\overline{\mathbf{x}_j})} \Delta x_1 + \dots + \frac{\partial F_n}{\partial x_n}\bigg|_{(\overline{\mathbf{u}_i},\overline{\mathbf{x}_j})} \Delta x_n
    \end{aligned}
    \label{lctiODE}
\end{equation}

where $\Delta u_i = (u_i - \overline{u_i})$ and $\Delta x_j = (x_j - \overline{x_j})$. Equation \eqref{lctiODE} is what we have previously called lctiODE and is the \textbf{definition of a biochemical network \textit{at equilibrium}} written in the language of dynamical systems. Before proceeding further, it is fruitful to take a step back and realise that we can do away with the distinction between inputs $u$ and states $x$ entirely. They are both just biochemical species --- one that we can control directly and the other indirectly. What we have previously called ``input'' is simply a species with no incoming rates, and what we have called  ``output'' is one that both accepts and gives rates. We can collectively denote all species as $\zeta_k$. The entire network can then be described as a set of coupled differential equations for each $\zeta_k$, and so

\begin{equation*}
    \boldsymbol{\zeta} = \left\{\zeta_1, \ldots, \zeta_m, \zeta_{m+1}, \ldots, \zeta_{m+n}\right\}=\left\{ u_1, \ldots, u_m, x_1, \ldots, x_n\right\} \text{ .}
\end{equation*}

The form of our lctiODEs defining our biochemical network at equilibrium can now be written more compactly as:

\begin{equation}
    \begin{aligned}
        \frac{d(\Delta \zeta_1)}{dt} &= 
        \frac{\partial F_1}{\partial \zeta_1}\bigg|_{(\overline{\boldsymbol{\zeta}})} \Delta \zeta_1 
        + \dots 
        + \frac{\partial F_1}{\partial \zeta_m}\bigg|_{(\overline{\boldsymbol{\zeta}})} \Delta \zeta_m 
        + \frac{\partial F_1}{\partial \zeta_{m+1}}\bigg|_{(\overline{\boldsymbol{\zeta}})} \Delta \zeta_{m+1} 
        + \dots 
        + \frac{\partial F_1}{\partial \zeta_{m+n}}\bigg|_{(\overline{\boldsymbol{\zeta}})} \Delta \zeta_{m+n} \\
        \frac{d(\Delta \zeta_2)}{dt} &= 
        \frac{\partial F_2}{\partial \zeta_1}\bigg|_{(\overline{\boldsymbol{\zeta}})} \Delta \zeta_1 
        + \dots 
        + \frac{\partial F_2}{\partial \zeta_m}\bigg|_{(\overline{\boldsymbol{\zeta}})} \Delta \zeta_m 
        + \frac{\partial F_2}{\partial \zeta_{m+1}}\bigg|_{(\overline{\boldsymbol{\zeta}})} \Delta \zeta_{m+1} 
        + \dots 
        + \frac{\partial F_2}{\partial \zeta_{m+n}}\bigg|_{(\overline{\boldsymbol{\zeta}})} \Delta \zeta_{m+n} \\
        & \hspace{17em} \vdots \\
        \frac{d(\Delta \zeta_n)}{dt} &= 
        \frac{\partial F_n}{\partial \zeta_1}\bigg|_{(\overline{\boldsymbol{\zeta}})} \Delta \zeta_1 
        + \dots 
        + \frac{\partial F_n}{\partial \zeta_m}\bigg|_{(\overline{\boldsymbol{\zeta}})} \Delta \zeta_m 
        + \frac{\partial F_n}{\partial \zeta_{m+1}}\bigg|_{(\overline{\boldsymbol{\zeta}})} \Delta \zeta_{m+1} 
        + \dots 
        + \frac{\partial F_n}{\partial \zeta_{m+n}}\bigg|_{(\overline{\boldsymbol{\zeta}})} \Delta \zeta_{m+n}
    \end{aligned}
    \label{lctiODE_zeta}
\end{equation}

Each $\Delta \zeta_k$ term represents the deviation of $\zeta_k$, at a given point in time, from its equilibrium value $\overline{\zeta_k}$. A positive $\Delta \zeta_k$ means that the system contains $|\Delta \zeta_k|$ units of $\zeta_k$ in surplus above equilibrium $\overline{\zeta_k}$. Conversely, a negative $\Delta \zeta_k$ means that the system contains $|\Delta \zeta_k|$ units of $\zeta_k$ in deficit below $\overline{\zeta_k}$. Each coefficient $\partial F_i/\partial \zeta_k|_{(\overline{\boldsymbol{\zeta}})}$ effectively a reaction rate constant --- capturing the inverse timescale that its deviation variable $\Delta \zeta_k$ contributes to the dynamics of $\Delta \zeta_i$. The product of this coefficient and its deviation variable encapsulates the rate at which $\Delta \zeta_k$ drives changes in $\Delta \zeta_i$ --- that is, the local reaction rate of $\zeta_i$ due to $\zeta_k$ near equilibrium $\overline{\boldsymbol{\zeta}}$. Together, each line in system \eqref{lctiODE_zeta} states that the fluctuation speed of each biochemical state is simply a linear combination of the fluctuations in all other biochemical species in the system.

\begin{figure}[ht]
    \centering
    \includegraphics[width=0.8\linewidth]{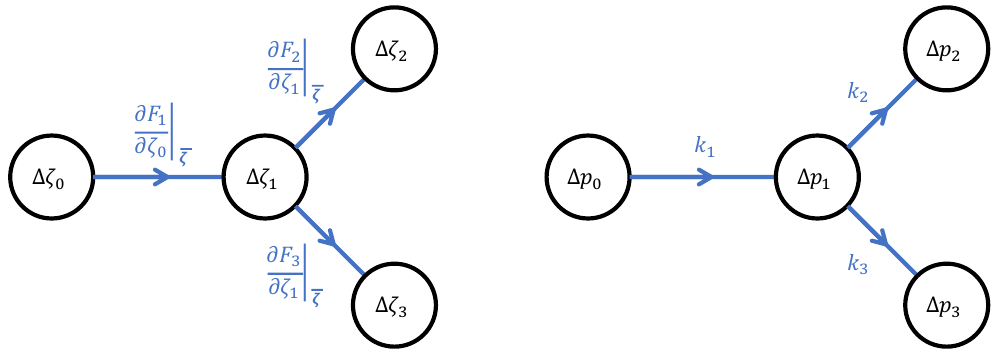}
    \begin{subequations} \label{lctiODEexamp}
    \begin{align}
    \frac{dP_\text{0}}{dt}   &= -k_1 P_\text{0} \\
    \frac{dP_1}{dt} &= +k_1 P_\text{0} - k_2 P_1 - k_3 P_1 \\
    \frac{dP_2}{dt} &= +k_2 P_1 \\
    \frac{dP_3}{dt} &= +k_3 P_1
    \end{align}
    \end{subequations}
    \caption{(Left) An illustrative example of an lctiODE system with one input $\zeta_{0}$ and three state variables $\zeta_1,\zeta_2,\zeta_3$, at equilibrium $(\overline{\zeta_0},\overline{\zeta_1},\overline{\zeta_2},\overline{\zeta_\text{3}})$. (Right) A biochemical network illustrating the Equation \eqref{lctiODEexamp}'s mass action process, involving one input protein $P_0$ and three output proteins $P_1$ to $P_3$, at equilibrium.}
    \label{fig:lctiODE}
\end{figure}

In general, ctiODE systems do not guarantee the existence of an equilibrium for arbitrary inputs. That is, we cannot assume a biochemical network will always settle into equilibrium. However, when the input is constant, an equilibrium will exist. Our focus in this work is on periodic forcing. Periodic forcing signals have a constant time-average $\{\overline{\textbf{u}_i}\}$. This average input induces a corresponding steady-state output $\{\overline{\textbf{x}_j}\}$. Thus, for the purposes of studying periodic forcing, we can safely rely on Equation \eqref{lctiODE}. Figure \ref{fig:lctiODE} provides an illustrative example of a network representation of an lctiODE, alongside a simple biochemical network at equilibrium that shares the same graphical structure.

\subsection{Laplace transform — a lens into the spectral content of biochemical signals}
We are studying periodic forcing, and the natural language of periodic signals is the frequency domain. To analyse our system in this domain, we must convert its time-domain description into a representation in frequency space. The tool for this is the Laplace transform $\mathscr{L}$. Given a function $f(t)$, its Laplace transform is defined as:

\begin{equation}
    \mathscr{L} \{f(t)\}(s) = \int_{0}^{\infty} f(t) e^{-st} dt = F(s) \text{ .}
    \label{Ltransform}
\end{equation}

where $s$ is a complex number. The transform decomposes biochemical signals into a continuum of exponentially modulated sinusoids, of exponential modulation rate $\Re(s)$ and sinusoidal frequency $\Im(s)$ -- revealing a generalised frequency profile that is often obscured in the time-domain formulation. 

The complex number $s$ is often called the complex frequency. Like how angular frequency $\omega$ is just a number and does not have an inherent physical meaning without being embedded into the sinusoidal function $\sin{(\omega t)}$, the complex frequency $s$ is just an ordered pair of numbers $(\Re(s),\Im(s))$ and possesses no inherent physical meaning until it is embedded into the exponential function $e^{-st}$. It is only within this exponential context that the components of $s$ gain physical interpretation: the real part $\Re(s)$ being the envelope timescale (i.e. exponential decay/growth rate) and the imaginary part $\Im(s)$ being the carrier timescale (i.e. oscillatory frequency) of an exponentially modulated oscillation --- or, in the context at hand, a biochemical signal whose amplitude evolves exponentially in time while oscillating sinusoidally.

\subsection{Transfer function --- the action of biochemical networks}
The transfer function is the ratio of the Laplace transformed output to the Laplace transformed input. That is, given an input $\Delta\zeta_{\text{in}}(t) \in \{\Delta u_i,\Delta x_j\}$ and an output $\Delta\zeta_{\text{out}}(t) \in \{\Delta u_i,\Delta x_j\}$, the transfer function from input to output $T(s)$ is:

\begin{equation}
    T(s) = \frac{ \mathscr{L} \left\{ \Delta\zeta_{\text{out}}(t) \right\} (s) }{ \mathscr{L} \left\{ \Delta\zeta_{\text{in}}(t) \right\} (s)} = \frac{\Delta Z_{\text{out}}(s)}{\Delta Z_{\text{in}}(s)} \text{  .}
    \label{TF}
\end{equation}

Equation \eqref{TF} quantifies how a biochemical network transforms one exponentially modulated oscillation into another of the same exponential modulation rate and oscillation frequency. The transfer function is analogous to the ``action'' that a biochemical network takes on an input to produce an output. 

\subsection{The meaning of a Laplace transform on a biochemical network at equilibrium --- doing math with pictures} \label{sec:mathwithpictures}
Recall that biochemical networks at equilibrium are conveyed by lctiODEs. lctiODEs have a very specific structure and, as a consequence, their associated transfer functions are always ratios of complex polynomials. That is,

\begin{equation}
    T(s) = \frac{\alpha_n s^n+\alpha_{n-1} s^{n-1}+\cdots+\alpha_1 s+\alpha_0}{\beta_m s^m+\beta_{n-1} s^{n-1}+\cdots+\beta_1 s+\beta_0} =  K \cdot \frac{s^n+a_{n-1} s^{n-1}\cdots+a_1s+a_0}{s^m+b_{n-1} s^{n-1}+\cdots+b_1s+b_0}\text{ .}
    \label{TF2}
\end{equation}

From Equation \eqref{TF}, we see that the transfer function defines the relationship between transformed input perturbations and transformed output perturbations: 

\begin{equation*}
    \Delta Z_{\text{out}}(s) = T(s) \Delta Z_{\text{in}}(s) \text{ .}
\end{equation*}

\begin{figure}[ht]
    \centering
    \includegraphics[width=0.8\linewidth]{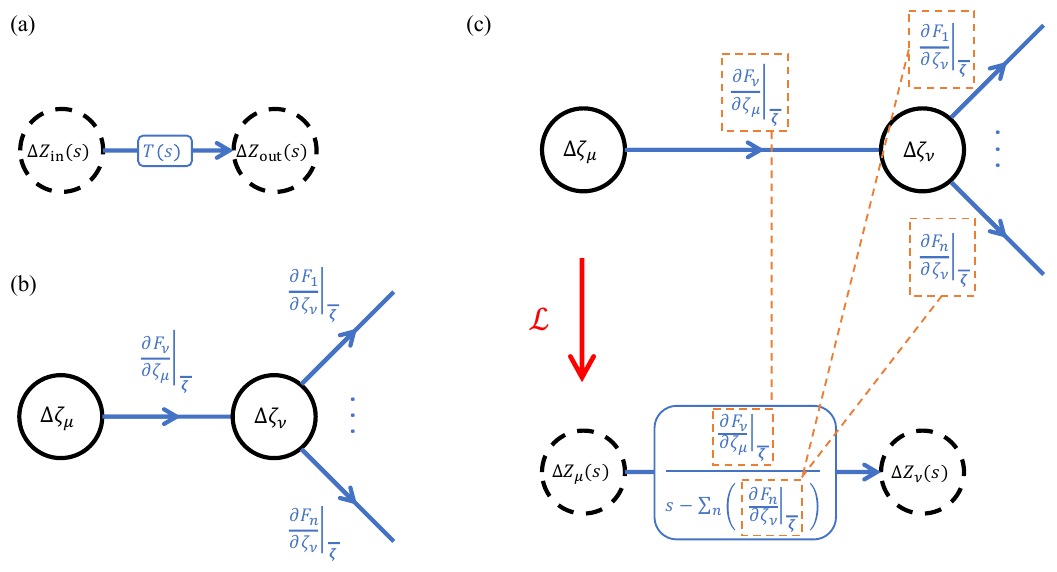}
    \caption{\textbf{(a)} The transfer function acts as a directed edge between transformed input- and transformed output-perturbations. \textbf{(b)} Graphical representation of a direct interaction between adjacent biochemical species. \textbf{(c)} Applying the Laplace transform to the biochemical network in part (b) transforms each node and aggregates the reaction rates from its surrounding edges into a single directed edge leading into it.}
    \label{fig:Laplace_zeta_Z}
\end{figure}

If we were to represent this relationship as a graph, the transfer function T(s) would appear as a directed edge from the node $Z_{\text{in}}(s)$ to $Z_{\text{out}}(s)$, as illustrated in Figure \ref{fig:Laplace_zeta_Z}(a). 

In the original time-domain lctiODE system (our biochemical network at equilibrium), the influence of any input node (species perturbation) $\Delta \zeta_\mu$ on an adjacent output node (species perturbation) $\Delta \zeta_\nu$ depends on the rate coming from $\Delta \zeta_\mu$ (production rate) given by

\begin{equation*}
    \left.\dfrac{\partial F_{\nu}}{\partial \zeta_\mu}\right|_{(\overline{\mathbf{x}_i},\overline{\mathbf{u}_j})} \text{ ,}
\end{equation*}

and outgoing rates from $\Delta \zeta_\nu$ (destruction rates from mechanisms such as degradation and downstream conversion) captured by

\begin{equation*}
    \left.\dfrac{\partial F_{1}}{\partial \zeta_\nu}\right|_{\overline{\boldsymbol{\zeta}}} \quad , \quad \dots  \quad, \quad \left.\dfrac{\partial F_{n}}{\partial \zeta_\nu}\right|_{\overline{\boldsymbol{\zeta}}}  \text{ .}
\end{equation*}

This is visualised as a directed edges, shown in Figure \ref{fig:Laplace_zeta_Z}(b). 

Given that the input node in Figure \ref{fig:Laplace_zeta_Z}(a) is simply the Laplace-transform of the input node in Figure \ref{fig:Laplace_zeta_Z}(b) --- and likewise for the output nodes --- finding a relationship between the transfer function $T(s)$ in Figure \ref{fig:Laplace_zeta_Z}(a) and the original biochemical rate terms in \ref{fig:Laplace_zeta_Z}(b) would allow us to directly relate the two network representations via the Laplace transform. More explicitly, it offers a concrete way to interpret what it means to “\textbf{Laplace transform a biochemical network}”.

To derive this relationship, we start from the system of equations defined in Equation \eqref{lctiODE}. Each line in the system relates an output node (appearing on the left-hand side of the equation) to all of its adjacent inputs and outputs (appearing on the right-hand side through their respective rates). For any given line, the transfer function between an input node $\Delta \zeta_\mu$ and an output node $\Delta \zeta_\nu$, defined by $F_\nu(\zeta_\mu) = d\zeta_\nu/dt$, is given by:

\begin{equation}
    T(s)
    =
    \dfrac{
    \left.\dfrac{\partial F_{\nu}}{\partial \zeta_\mu}\right|_{\overline{\boldsymbol{\zeta}}}
    }{
    s - \mathlarger{\mathlarger{\sum\limits_{n}}}
    \left.\dfrac{\partial F_n}{\partial \zeta_\nu}\right|_{\overline{\boldsymbol{\zeta}}}
    } 
    =
    \dfrac
    {
    \shortstack{production\\rate constant}
    }{
    s - \mathlarger{\mathlarger{\sum}}\, \bigg( \raisebox{-1ex}[0pt][0pt]{\shortstack{destruction\\rate constants}} \bigg)
    }\text{  .}
    \label{eq:TF_rate}
\end{equation}

The reader is advised to take caution here since destruction rates are always negative. Therefore, it is useful to mentally note the following in the context of biochemical networks:

\begin{equation}
    T(s)
    =
    \dfrac{
    \left.\dfrac{\partial F_{\nu}}{\partial \zeta_\mu}\right|_{{\overline{\boldsymbol{\zeta}}}}
    }{
    s + \mathlarger{\mathlarger{\sum\limits_{n}}}
    \Bigg| \left( \left. \dfrac{\partial F_n}{\partial \zeta_\nu}\right|_{{\overline{\boldsymbol{\zeta}}}} \right) \Bigg|
    } 
    =
    \dfrac
    {
    \shortstack{production\\rate constant}
    }{
    s + \mathlarger{\mathlarger{\sum}}\, \Bigg| \raisebox{-1ex}[0pt][0pt]{\shortstack{destruction\\rate constants}} \Bigg|
    }\text{  .}
    \label{eq:TF_rate_note}
\end{equation}

We have now explicitly expressed the transfer function $T(s)$ in Figure \ref{fig:Laplace_zeta_Z}(a) in terms of the original biochemical rate terms from Figure \ref{fig:Laplace_zeta_Z}(b), thereby establishing a direct mapping between the two graphical representations. We can now answer what it means to Laplace transform a biochemical network. \textbf{Laplace transforming a biochemical network produces a new network in which node positions remain fixed, but the local reaction rate constants are consolidated into single directed edges representing transfer functions. It transforms networks of biochemical concentrations $\{u_i(t),x_j(t)\}=\{\zeta_k(t)\}$ connected by rates into networks of \textbf{perturbations} in biochemical concentrations $\{\Delta U_i(s),\Delta X_j(s)\}=\{\Delta Z_k(s)\}$ connected by these perturbation descriptors known as transfer functions.} This transformation is illustrated in Figure \ref{fig:Laplace_zeta_Z}(c) for the case of a two node biochemical network.

\begin{figure}[ht]
    \centering
    \includegraphics[width=0.7\linewidth]{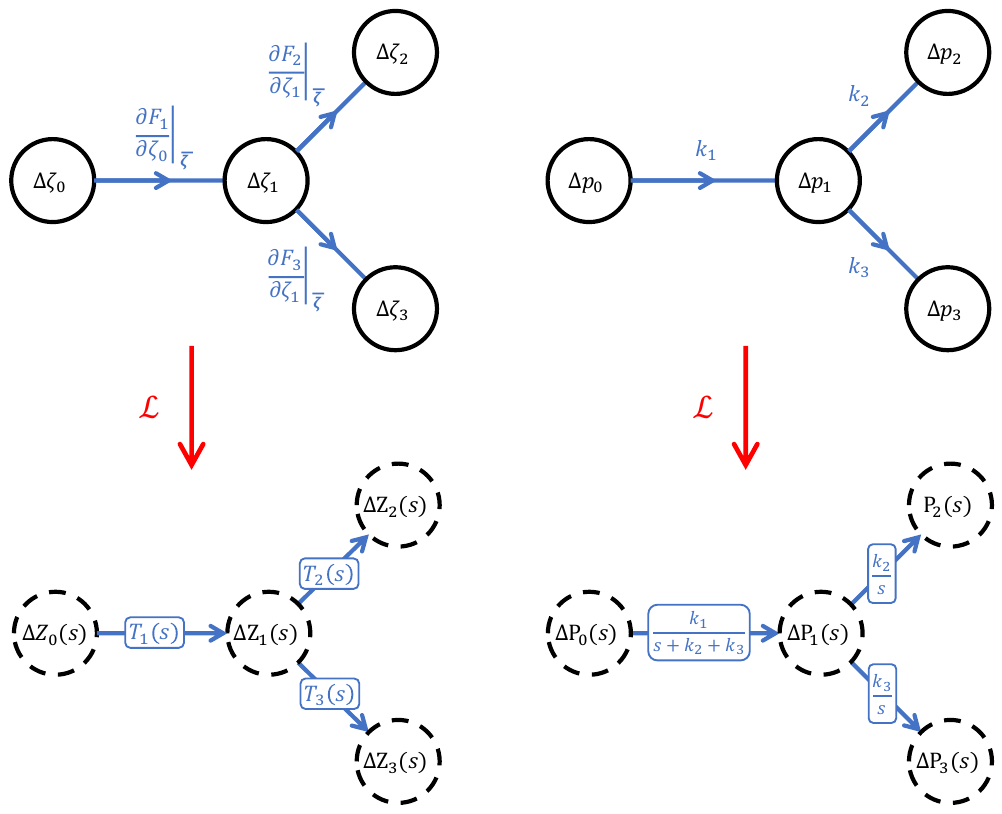}
    \textcolor{RoyalBlue}{
    \begin{equation*}
        T_{1}(s) 
        = \frac{\left.\frac{\partial F_1}{\partial u_1}\right|_{\overline{\boldsymbol{\zeta}}}}
        {s- \left( \left.\frac{\partial F_1}{\partial x_1}\right|_{\overline{\boldsymbol{\zeta}}} + \left.\frac{\partial F_3}{\partial x_1}\right|_{\overline{\boldsymbol{\zeta}}}\right)}
        \quad \text{,} \quad
        T_{2}(s) 
        = \frac{\left.\frac{\partial F_2}{\partial x_1}\right|_{\overline{\boldsymbol{\zeta}}}}
        {s-\left.\frac{\partial F_2}{\partial x_2}\right|_{\overline{\boldsymbol{\zeta}}}} 
        \quad \text{,} \quad
        T_{3}(s) 
        = \frac{\left.\frac{\partial F_3}{\partial x_2}\right|_{\overline{\boldsymbol{\zeta}}}}
        {s-\left.\frac{\partial F_3}{\partial x_3}\right|_{\overline{\boldsymbol{\zeta}}}} 
    \end{equation*}
    }
    \caption{The Laplace transform converts networks of biochemical concentrations to networks of changes in biochemical concentrations. This is illustrated for the chemical reaction example from Figure \ref{fig:lctiODE}.}
    \label{fig:LaplaceNetwork}
\end{figure}

Transforming into a network of perturbations is a natural graphical convention because perturbations are the natural language of oscillations and, hence, the natural language of periodic forcing which we are studying. An illustrative example of this transformation from one network into another is shown in Figure \ref{fig:LaplaceNetwork}, where we build upon the  chemical reaction system of Figure \ref{fig:lctiODE}. This example highlights the essence of the Laplace transform --- we are now able to perform math with pictures instead of equations. Nowhere in this process did we need to start with the system of lctiODEs in Equation \eqref{lctiODEexamp} and apply the integral transform in Equation \eqref{Ltransform} to solve for the ratio in Equation \eqref{TF}. It was all done visually. 

Transforming into this final network of perturbations is a natural graphical choice, as perturbations form the fundamental language of oscillations—and thus, of periodic forcing, which is the focus of our study. An illustrative example of this transformation is shown in Figure \ref{fig:LaplaceNetwork}, which builds upon the chemical reaction system in Figure \ref{fig:lctiODE}. This example captures the core utility of the Laplace transform in biochemical circuits: \hl{the Laplace transform allows us to perform math with biochemical pictures instead of equations}. At no point was it necessary to start with the system of lctiODEs in Equation \eqref{lctiODEexamp}, apply the integral transform in Equation \eqref{Ltransform} and solve for the ratio in Equation \eqref{TF}. The entire transformation was carried out visually.

\subsection{Unidirectional Interactions and Timescale Decoupling}

When one species (A) regulates another (B), the conversion from A to B can be either conserved or unconserved. When conserved, the rate of gain in B corresponds exactly to a rate of loss in A --- the net flow is balanced at each instant in time, reflecting a direct conversion of the upstream into the downstream species. When unconserved, A does not experience any corresponding loss and appears unaffected by the regulation it exerts on B. These two cases of define the nature of interaction between a pair of species. \textbf{Bidirectional coupling} refers to the conserved case, where the dynamics of one species are intrinsically linked to the other through reciprocal effects. \textbf{Unidirectional coupling} refers to the unconserved case, where one species influences another without being influenced in return. This apparent asymmetry is not arbitrary. It is a consequence of our choice of timescale on which the system is observed. Specifically, it arises when this choice of timescale is larger than that of the dynamics of A. If A depletes and recovers more rapidly than the resolution of observation, its transient changes become effectively invisible every time we measure its population, creating the illusion that it has not changed. In this way, the choice of timescale can ``decouple'' the symmetry between an interaction between two species and render it effectively unidirectional. We can modify the example in Figure 1 to create another example illustrating this unidirectional behaviour:
\begin{equation} \label{lctiODEexamp_uni}
    \frac{dP_0}{dt} = 0, \quad
    \frac{dP_1}{dt} = k_1 P, \quad
    \frac{dP_2}{dt} = k_2 P_1, \quad
    \frac{dP_3}{dt} = k_3 P_1 .
\end{equation}

\begin{figure}[ht]
    \centering
    \includegraphics[width=0.4\linewidth]{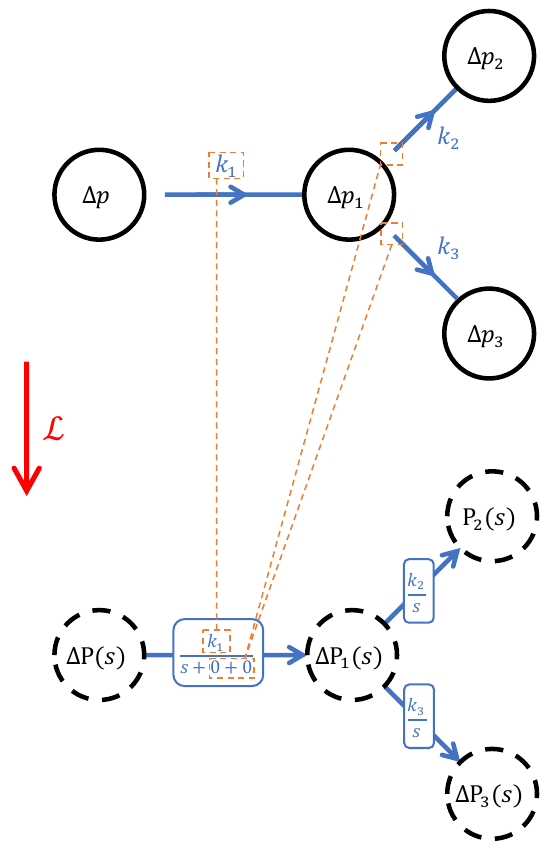}
    \caption{(Top) Biochemical network illustrating unidirectional coupling as described by the system in Equation \eqref{lctiODEexamp_uni}. Arrows disconnected from source nodes represent this effective one-way influence. (Bottom) Corresponding Laplace-transformed network. Note that the transfer function between $\Delta P(s)$ and $\Delta P_1(s)$ no longer contains constant terms in the denominator, compared to its original in Figure \ref{fig:LaplaceNetwork}, reflecting the new absence of outgoing rates from $\Delta p$.}
    \label{fig:UniCoupling_Eg}
\end{figure}

To visually represent such unidirectional regulation (or asymmetric interactions), we adopt a notational convention: the regulatory arrow is intentionally drawn disconnected from the source node. This emphasises that the interaction occurs in only one direction, such that the source does not deplete and the sink gains. The equivalent representation of Equation \ref{lctiODEexamp_uni} is illustrated in Figure \ref{fig:UniCoupling_Eg} (top). Its graphical Laplace transform, using the rules from Figure \eqref{fig:LaplaceNetwork}, is illustrated in Figure \ref{fig:UniCoupling_Eg} (bottom).

\subsection{Transfer function stability --- characterising containment and resonance in biochemical outputs} \label{stability}
The stability of a transfer function determines the boundedness of a system’s output in response to a sinusoidal input. Since both the numerator and denominator of Equation \eqref{TF2} are polynomials in $s$, by the fundamental theorem of algebra, the transfer function can further be expressed in the following factored form:

\begin{equation}
    T(s) = K\cdot \frac{(s-\zeta_1)(s-\zeta_2)...(s-\zeta_\alpha)}{(s-\rho_1)(s-\rho_2)...(s-\rho_\beta)} \text{ .}
    \label{eq:TF_factored}
\end{equation}

One can eliminate coinciding pole-zero pairs in Equation \eqref{eq:TF_factored} to obtain a reduced form that contains only the ``observable'' poles and zeros, that is:

\begin{equation}
    T(s) = \frac{(s-z_1)(s-z_2)...(s-z_m)}{(s-p_1)(s-p_2)...(s-p_n)} = K \cdot \frac{\prod_{i=1}^m\left(s-z_i\right)}{\prod_{j=1}^n\left(s-p_j\right)}: z_i \ne p_j \quad \forall i, j \text{ .}
    \label{TF_observable}
\end{equation}

From Equation \eqref{TF_observable}, three stability classifications can be made --- each listed in Table \ref{tab:stability} with their conditions. 

\begin{table}[H]
    \centering
    \begin{tabular}{||c|c||}
        \hline
        \textbf{Stability} & \textbf{Condition} \\ 
        \hline
        Asymptotically stable & $\forall i, \Re\left\{p_i\right\}<0$ \\
        Marginally stable & $\left(\forall i: \Re\left\{p_i\right\} \leq 0\right) \wedge\left(\exists j: \Re\left\{p_j\right\}=0\right)$ \\
        Unstable & $\exists i: \Re\left\{p_i\right\}>0$ \\ 
        \hline
    \end{tabular}
    \vspace{3pt}
    \caption{Conditions of stability of a transfer function. Here, $\left\{p_i\right\}$ denotes the set of all poles.}
    \label{tab:stability}
\end{table}

\begin{shaded}
    If a biochemical system is asymptotically stable, then a bounded sinusoidal input will produce a bounded sinusoidal output. On the contrary, if a biochemical system is unstable, the same bounded sinusoidal input will produce an unbounded sinusoidal output. Finally, if a biochemical system is marginally stable, it possesses a resonance frequency equal to the imaginary part of each unique imaginary pole. A bounded sinusoidal input at each of these frequencies will produce a sinusoidal output that grows unbounded with time.
\end{shaded}

With our notation in Table \ref{tab:stability}, this final statement of marginal stability states that: \textbf{if a system is marginally stable then it has a resonant frequency at $\omega=\Im\left\{p_j\right\}$ for each pole in the set $\left\{p_j\right\}$, and a bounded sinusoidal input of frequency $\omega=\Im\left\{p_j\right\}$ will produce a sinusoidal output that grows toward infinite amplitude as time appraoches infinity}. To show this, suppose the $j^{th}$ pole of Equation \eqref{TF_observable} satisfies the marginally stable condition in Table \ref{tab:stability}. That is, $p_j = 0+i\omega$. The transfer function can now be explicitly written as:

\begin{equation}
    T(s) = \frac{(s-z_1)(s-z_2)...(s-z_m)}{(s-p_1)(s-p_2)...(s-p_j)...(s-p_n)} = \frac{(s-z_1)(s-z_2)...(s-z_m)}{(s-p_1)(s-p_2)...(s-i\omega)...(s-p_n)} \text{ .}
    \label{TF_marginal}
\end{equation}

Now consider sinusoidal input $x(t)=\sin{(\omega t)}$, or equivalently:

\begin{equation}
    X(s) = \frac{\omega}{(s+i\omega)(s-i\omega)} \text{ .}
\end{equation}

By convolution, the output is:

\begin{equation}
    Y(s) = T(s)X(s) = \frac{(s-z_1)(s-z_2)...(s-z_m)\cdot\omega}{(s-p_1)(s-p_2)...(s+i\omega)(s-i\omega)^2...(s-p_n)} \text{ .}
    \label{eq:convolved}
\end{equation}

Equation \eqref{eq:convolved} describes the full frequency profile of the output in response to $\sin{(\omega t)}$, over all time. To see the asymptotic behaviour due to periodic forcing of $\sin{(\omega t)}$ over time, we must zoom into the limit of $s \rightarrow i\omega$ in Equation \eqref{eq:convolved}. Zooming in yields:

\begin{equation}
    Y(s) \sim \frac{1}{(s-i\omega)^2} \text{ .}
\end{equation}

In the time domain, this is equivalent to:

\begin{equation}
    y(t) = \mathscr{L}^{-1} \left\{Y(s)\right\} = te^{i\omega t} = t\left(\cos{(\omega t)} + i \sin{(\omega t)}\right) \text{ ,}
\end{equation}

which exhibits unbounded growth as $t \rightarrow \infty$. Alternatively, one could compute the full inverse Laplace transform of Equation \eqref{eq:convolved} to obtain all time-domain components explicitly, where it becomes evident that the long-term behaviour is governed by that of the second-order pole at $s = i\omega$, with all other terms decaying over time.

\textbf{N.B.} Although a pole at $s = 0+0i$ is technically classified as marginally stable, it is irrelevant to our context of resonance frequencies in biochemical networks. The resonance associated with an $s = 0 + 0i$ pole occurs at $\omega = 0$ --- a zero-frequency oscillation, or more precisely, a constant signal. As such, a biochemical network whose only purely imaginary poles are at the origin does not possess a resonance frequency. What this does imply, however, is that the system can respond to constant inputs, even in the absence of dynamic (oscillatory) forcing.

\subsection{Angular frequency response functions --- how biochemical networks at equilibrium filter oscillations} \label{sec:AngFreqResp}
\begin{figure}[ht]
    \centering
    \includegraphics[width=0.65\linewidth]{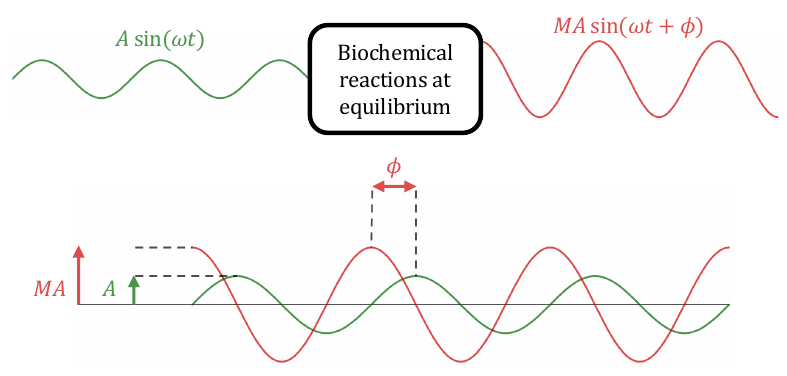}
    \caption{Amplitude modulation and phase shift induced by a biochemical network driven at angular frequency $\omega_0$, conveyed by the angular frequency response function $T(i\omega_0)$.}
    \label{fig:inout}
\end{figure}

Recall that the transfer function $T(s=\sigma+i\omega)$ captures the action of biochemical network on exponentially modulated sinusoids of modulation rates $\sigma$ and oscillation frequencies $\omega$. We can see what this action looks like in the special case of unmodulated sinusoids ($\sigma=0$) by evaluating the action $T(s)$ at $s=0+i\omega$. This resulting function is purely dependent on angular frequency and, so, will be referred to as the (angular) \textbf{frequency-response function} $T(i\omega)$ hereon. Each unique angular frequency $\omega_0$ corresponds to a unique phasor $T(i\omega_0)$. Each phasor \textit{encodes} two pieces of information: the amplitude modulation and the phase shift induced by the network at that frequency. That is, for a biochemical input oscillation $A\sin{(\omega_0 t)}$, $T(i\omega_0)$ tells us how the biochemical network amplitude modulates and phase shifts the signal into a new biochemical output oscillation $M(\omega_0)A\sin{(\omega_0 t+\phi(\omega_0))}$ --- see Figure \ref{fig:inout}.

These \textit{encodings} are respectively contained in the phasor distance from the origin $|T(i\omega_0)|=M(\omega_0)$ and its argument with the positive real axis $\angle T(i\omega_0)=\phi(\omega_0)$, respectively. $\phi(\omega_0)$ further informs us of the time delay between input and output, given by:

\begin{equation}
    \Delta t_0 = \frac{\phi(\omega_0)}{\omega_0} \text{ .}
    \label{eq:timedelay}
\end{equation}

\begin{figure}[ht]
    \centering
    \includegraphics[width=0.45\linewidth]{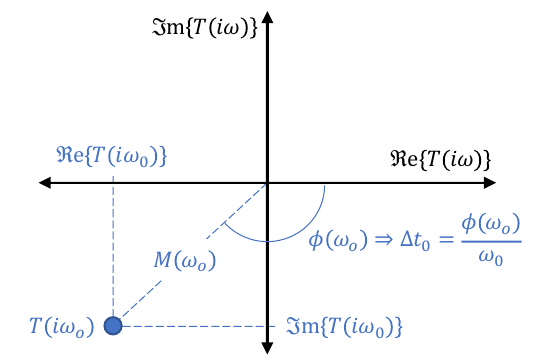}
    \caption{Phasor plot encoding the amplitude modulation and phase shift of a biochemical network at equilibrium on an incoming pure sinusoid of angular frequency $\omega_0$.}
    \label{fig:PhasorPlot}
\end{figure}

We can visualise this with a \textbf{phasor plot}, as shown in Figure \ref{fig:PhasorPlot}. The phasor’s motion through the complex plane as frequency evolves traces the full frequency-response function $T(i\omega)$, capturing how amplitude and phase shift vary across different input frequencies.

We can explicitly extract the magnitude frequency-response function $M$ and the phase frequency-response function $\phi$. These are defined, respectively, as:
\begin{subequations}
    \begin{alignat}{4}
        M(\omega) &=& \big| &T(i\omega) \big| &=& \sqrt{\Re \big\lbrace T(i\omega) \big\rbrace^2 + \Im \big\lbrace T(i\omega) \big\rbrace ^2} \label{mag} \\
        \phi(\omega) &=& \hspace{2pt} \arg\big\lbrace& T(i\omega)\big\rbrace &=& \arctantwo \Big( \Im \big\lbrace T(i\omega) \big\rbrace, \Re \big\lbrace T(i\omega) \big\rbrace \Big)  \label{phi}
    \end{alignat}
\end{subequations}

Alternatively, we can obtain a more descriptive form from the reduced factored transfer function in Equation \eqref{TF_observable}.

\begin{subequations}
    \begin{align}
        M(\omega) &= 
        |K| \cdot 
        \frac{\prod_{i=1}^n \sqrt{\Big[\Re(z_i)\Big]^2+\Big(\omega-\Big[\Im(z_i)\Big]\Big)^2}}
        {\prod_{j=1}^m \sqrt{\Big[\Re(p_i)\Big]^2+\Big(\omega-\Big[\Im(p_i)\Big]\Big)^2}}
        \label{mag2} \\
        \phi(\omega) &= 
        \arg(K) + \sum_{i=1}^n \arctantwo\Big(\omega-\Im(z_i),-\Re(z_i)\Big)-\sum_{j=1}^m \arctantwo\Big(\omega-\Im(p_i),-\Re(p_i)\Big)
        \label{phi2}
    \end{align}
\end{subequations}

The plot of $M(\omega)$ and $\phi(\omega)$ against $\omega$ are called the Bode magnitude and Bode phase plots, respectively. 

\subsection{Break frequencies} \label{sec:breakfreq}
The reduced factored form in Equation \eqref{TF_observable} also reveals so-called \textbf{break frequencies}. Break frequencies reveal changes in trends of amplitude modulation and phase shift, useful in conjunction with Bode plots.

\begin{figure}[ht]
    \centering
    \includegraphics[width=0.7\linewidth]{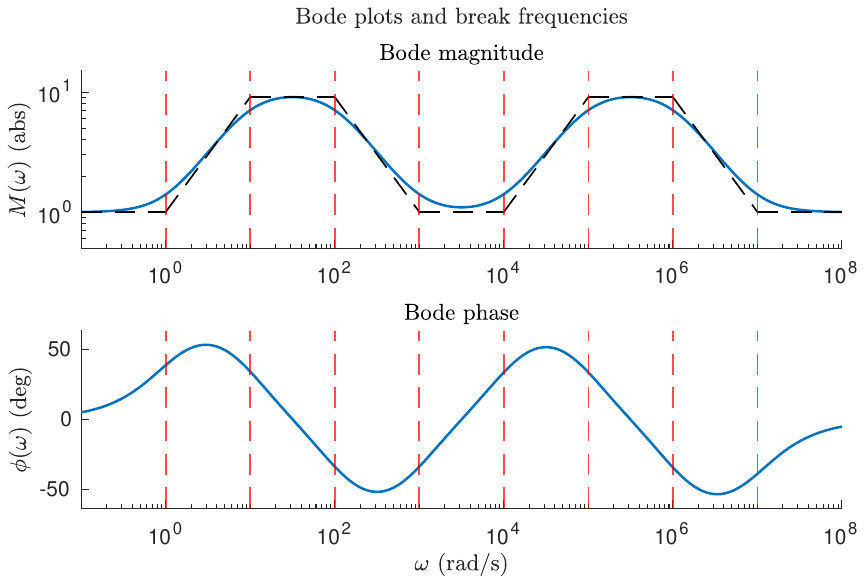}
    {\small
    \[
    TF(s) = \frac{(s + 10^0)(s + 10^3)(s + 10^4)(s + 10^7)}{(s + 10^1)(s + 10^2)(s + 10^5)(s + 10^6)}
    \]}
    \caption{
        Bode plots for the transfer function shown above. Break frequencies appear as visible corners where the slope of the amplitude modulation transitions between asymptotic regimes.
    }
    \label{fig:breakfreq}
\end{figure}

In Bode magnitude plots, each zero introduces a frequency at which the amplitude modulation factor gains an order of dependence on frequency, shifting from $M(\omega) \propto \omega^n$ to $M(\omega) \propto \omega^{n+1}$. This manifests as a unit increase in the slope of $M(\omega)$ on a conventional log-log scaled Bode plot. Conversely, each pole introduces a frequency at which the amplitude modulation factor loses an order of dependency on frequency, shifting from $M(\omega) \propto \omega^n$ to $M(\omega) \propto \omega^{n-1}$. Here, there is a unit decrease in the log-log slope instead. These zero- and pole-frequencies, at which transition from one asymptotic slope to another occurs, are the break frequencies. If the asymptotic slope segments of the plot were extended, they would intersect at visible “corners.” These corners not only visualise the transition toward new asymptotic behaviour but also mark the exact location of each break frequency. An example is shown in Figure \ref{fig:breakfreq} (top panel).

In Bode phase plots, each zero and pole contributes a sigmoidal-like segment of phase transition --- a rising sigmoid for a zero and a descending sigmoid for a pole. The break frequency acts as a qualitative guide to the centre of these transitions, but the precise shape, steepness, and symmetry depend on the structure and order of the transfer function, including factors such as damping. For first-order systems, the break frequency aligns exactly with the inflection point of the phase curve, where the rate of change is greatest. In contrast, for higher-order systems, the phase transitions become more contorted. In these cases, the inflection point no longer necessarily coincides with the break frequency. Nonetheless, break frequencies remain useful as milestones --- they often mark where a phase transition begins to level off toward its local asymptote. These qualitative segmental ``midpoints'' are illustrated in Figure \ref{fig:breakfreq} (bottom panel).

As will become apparent later, the values of poles and zeros (and thus their break frequencies), are determined by the physical rate constants of our biochemical network. Control over physical rate constants allows for control over amplitude modulation and phase shift. 

\subsection{Asymptotic phase shifts}
Closely associated with the reduced factored form in Equation \eqref{TF_observable} is the effective order of a transfer function in the high frequency limit. For sufficiently high frequencies, Equation \eqref{TF_observable} approaches the ratio $s^m / s^n$. This leads to an effective system order of n - m, and the corresponding asymptotic phase shift approaches:

\begin{equation}
    \phi(\omega \to \infty) = \frac{\pi}{2} (n-m) \text{ .}
    \label{phi_lim}
\end{equation}

This $\pi(n-m)/2$ value is also the asymptotic phase shift of a biochemical cascade containing n - m downstream species. 

\subsection{Block diagram algebra --- how biochemical networks at equilibrium work together}
Transfer functions are modular and obey rules known as block diagram algebra. Of the extensive list of rules, three will be most relevant for motif architectures of interest in this paper. Considering two transfer functions $T_1(s)$ and $T_2(s)$, these rules are highlighted in Table \ref{blockalg}.
\begin{table}[ht]
    \centering
    \renewcommand{\arraystretch}{1.5}
    \begin{tabular}{|c|c|}
        \hline
        \textbf{Rule} & \textbf{Collective Transfer Function} \\ 
        \hline\hline
        Series & $T_{II}(s)T_{I}(s)$ \\
        \hline
        Parallel & $T_I(s)+T_{II}(s)$ \\
        \hline
        Closed loop & $\dfrac{T_I(s)}{1 \mp T_I(s)T_{II}(s)}$ \\[7pt]
        \hline
    \end{tabular}
    \vspace{3pt}
    \caption{Block diagram algebra rules for series, parallel and closed-loop architectures involving two transfer functions $T_I(s)$ and $T_{II}(s)$.}
    \label{blockalg}
\end{table}

\begin{figure}[ht]
    \centering
    \includegraphics[width=0.8\linewidth]{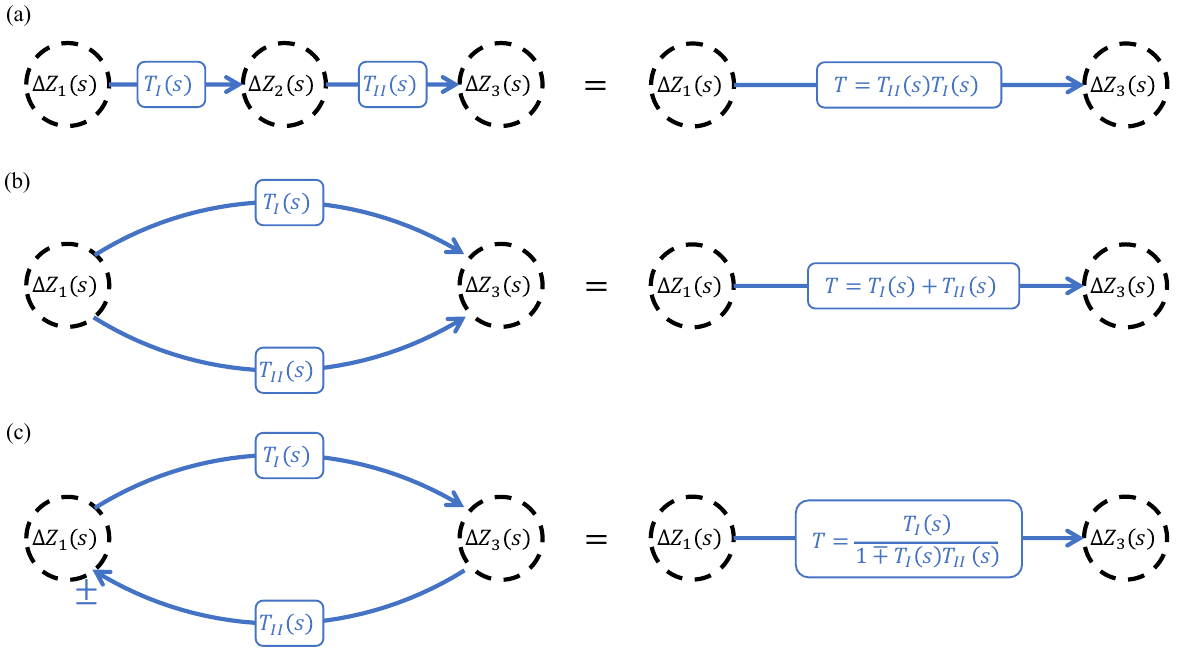}
    \caption{Block diagram algebra: (a) series rule, (b) parallel rule and (c) closed-loop rule.}
    \label{fig:BlockDiaAlg}
\end{figure}

%% file: 4_Discussion.tex
\section{Discussion}

The following table lists the collection of motifs which we will study. They have been ordered in order of increasing ``complexity''. Here, adding a node is less complex than adding an edge which is less complex than adding a feedback loop. 

\begin{table}[ht]
\centering
\setlength{\extrarowheight}{4pt}
\begin{tabular}{| c | c | c | c | c |}
    \hline
    \textbf{Category} & \textbf{Nodes} & \textbf{Edges} & \textbf{Feedbacks} & \textbf{No. of motif variants} \\
    \hline
    Simple regulation (SR) & 2 & 1 & 0 & 2 \\
    \hline
    Cascade (Cas)          & 3 & 2 & 0 & 4 \\
    \hline
    Feedforward loop (FFL) & 3 & 3 & 0 & 16 \\
    \hline
    Forced autoregulation (fAR) & 1 & 2 & 1 & 4 \\
    \hline
    Forced double feedback loop (fdPFL) & 2 & 3 & 1 & 8 \\
    \hline
    Autonomous autoregulation (aAR) & 1 & 1 & 1 & 2 \\
    \hline
    Autonomous double feedback loop (adPFL) & 2 & 2 & 1 & 4 \\
    \hline
\end{tabular}
\vspace{3pt}
\caption{Network motifs to be discussed in this section, grouped by category and listed in order of complexity. The final column indicates the number of distinct motif variants within each category. The final two autonomous architectures are discussed in Appendices B and C, respectively.}
\end{table} 

%% file: 4a_SimpleRegulation.tex
\subsection{Simple regulation}
\subsubsection{General architecture}
Simple regulation (SR) is the most basic network motif category. In SR, one biochemical species A regulates another B. This is described in general by the equation:

\begin{equation} \label{SR_genODE}
    \frac{d B}{d t}=f_1(A)-f_2(B)
\end{equation}

Here, $f_1$ describes the production rate of concentration B as a function of concentration A, while $f_2$ describes the rate of loss in concentration B as a function of itself. This interaction is often illustrated by a network of two nodes (each representing a different species) connected by a directed edge (indicating the ``direction of regulation''). If the edge points away from A and towards B, then A is said to ``simply regulate'' B. 
\begin{figure}[H]
    \centering
    \includegraphics[scale=0.7]{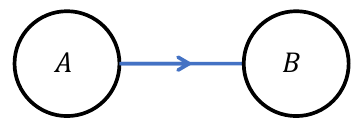}
    \caption{Simple regulation motif architecture.}
    \label{fig:ar_gen}
\end{figure}

\begin{figure}[H]
    \centering
    \includegraphics[scale=0.7]{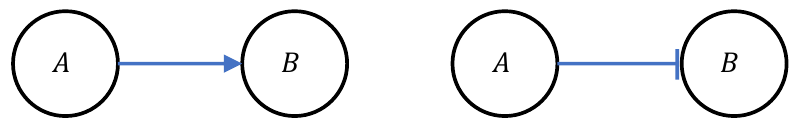}
    \caption{Positive SR (left) and negative SR (right).}
    \label{fig:SR_cases}
\end{figure}

Regulation often occurs in two modes: promotion and repression. In promotion, increasing A increases the production rate of B.  In repression, increasing A decreases the production rate of B. As such, the choice of regulation mode dictates the functional form of $f_1$. In diagrams, promotional interaction is often represented with a positive (sharp) arrow while repressive interaction is often represented with a negative (blunt) arrow, as shown in Figure \ref{fig:SR_cases}. 

Before we discuss positive and negative SR, let us foreshadow what we expect their properties to look like with our knowledge of the general SR architecture. 

We first would like to obtain a description of their dynamics near equilibrium. 
Recall that the basic idea behind this framework is the oscillation of inputs about a mean value once the system has reached an equilibrium due to sustained forcing of that mean input value. As such, suppose that after sufficiently long time, population of B reaches equilibrium value $B_e$ due to sustained input $A_e$. That is, the equilibrium coordinate of the system is defined by the pair of numbers $(A_e,B_e)$. 
We wish to “zoom” into the local linear dynamics about this equilibrium point $(B_e,A_e)$ so that we can study oscillations about this point. In this linear regime local to $(A_e,B_e)$, Equation \eqref{SR_genODE} can be approximated by the Taylor expansion:

\begin{equation}
    \left.\frac{d B}{d t} \sim \frac{d B}{d t}\right|_{\left(B_e, A_e\right)}+\left.\left(\frac{\partial}{\partial A}\left(\frac{d B}{d t}\right)\right)\right|_{\left(B_e, A_e\right)} \cdot\left(A-A_e\right)+\left.\left(\frac{\partial}{\partial B}\left(\frac{d B}{d t}\right)\right)\right|_{\left(B_e, A_e\right)} \cdot\left(B-B_e\right) \text{ .}
    \label{eq:sr_taylor1}
\end{equation}

But at equilibrium $(A_e,B_e)$, concentration $B$ does not change. Therefore:

\begin{equation}
    \left.\frac{d B}{d t}\right|_{\left(B_e, A_e\right)}=0
\end{equation}

Equation \eqref{eq:sr_taylor1} reduces to:

\begin{equation}
    \left.\frac{d B}{d t} \sim\left(\frac{\partial}{\partial A}\left(\frac{d B}{d t}\right)\right)\right|_{\left(B_e, A_e\right)} \cdot\left(A-A_e\right)+\left.\left(\frac{\partial}{\partial B}\left(\frac{d B}{d t}\right)\right)\right|_{\left(B_e, A_e\right)} \cdot\left(B-B_e\right) \text{ .}
    \label{eq:sr_taylor2}
\end{equation}

The region of points within which we will oscillate is defined by: 

\begin{equation}
    B=B_e+\Delta B \quad \text { and } \quad A=A_e+\Delta A \text{ ,}
    \label{eq:SR_B_A}
\end{equation}

where $\Delta B$ and $\Delta A$ are sufficiently small distances from $(B_e,A_e)$ such that our linear approximation still holds. Performing this change of variables, from Equation \eqref{eq:SR_B_A} into Equation \eqref{eq:sr_taylor2}, yields:

\begin{equation*}
    \frac{d (B_e + \Delta B)}{d t} \sim \left.\left(\frac{\partial}{\partial A}\left(\frac{d B}{d t}\right)\right)\right|_{\left(B_e, A_e\right)} \cdot\left(A_e+\Delta A-A_e\right)+\left.\left(\frac{\partial}{\partial B}\left(\frac{d B}{d t}\right)\right)\right|_{\left(B_e, A_e\right)} \cdot\left(B_e + \Delta B-B_e\right)
\end{equation*}
\begin{equation*}
    \Rightarrow \frac{d B_e}{d t} + \frac{d \Delta B}{d t} \sim \left.\left(\frac{\partial}{\partial A}\left(\frac{d B}{d t}\right)\right)\right|_{\left(B_e, A_e\right)} \cdot\left(\Delta A\right)+\left.\left(\frac{\partial}{\partial B}\left(\frac{d B}{d t}\right)\right)\right|_{\left(B_e, A_e\right)} \cdot\left(\Delta B\right)
\end{equation*}
\begin{shaded}
\begin{equation}
    \Rightarrow \frac{d \Delta B}{d t} \sim \left.\left(\frac{\partial}{\partial A}\left(\frac{d B}{d t}\right)\right)\right|_{\left(B_e, A_e\right)} \cdot\left(\Delta A\right)+\left.\left(\frac{\partial}{\partial B}\left(\frac{d B}{d t}\right)\right)\right|_{\left(B_e, A_e\right)} \cdot\left(\Delta B\right) \label{linODE_SRgen}
\end{equation}
\begin{equation*}
    = J_1\cdot\Delta A + J_2\cdot\Delta B \quad , \quad J_1,J_2=\text{constants .}
\end{equation*}
    Equation \eqref{linODE_SRgen} describes the dynamics between fluctuations in input species $A$, measured relative to its equilibrium value $A_e$, and fluctuations in output species $B$, measured relative to its equilibrium value $B_e$. Specifically, it states that, for a general SR circuit architecture at steady state, the instantaneous rate of change of output fluctuations is proportional to the sum of two physical quantities at that same point in time: the amplitude of the input fluctuation and the amplitude of the corresponding output fluctuation. We can expect the near equilibrium dynamics of our positive and negative SR motifs to also be sums of linear combinations of input perturbations and output perturbations.
\end{shaded}
We now derive the corresponding transfer function structure. To do this, we require the frequency domain information from Equation \eqref{linODE_SRgen}. This is done by converting it into the frequency domain via a Laplace transform to both sides of the equation, yielding:

\begin{align}
    \mathcal{L}\left\{\frac{d(\Delta B)}{d t}\right\}(s)
    &=
    \mathcal{L}\left\{
    \left.\left(\frac{\partial}{\partial A}\left(\frac{d B}{d t}\right)\right)\right|_{\left(B_e, A_e\right)} \cdot\left(\Delta A\right)
    +
    \left.\left(\frac{\partial}{\partial B}\left(\frac{d B}{d t}\right)\right)\right|_{\left(B_e, A_e\right)} \cdot\left(\Delta B\right)
    \right\}(s) \nonumber\\
    \Rightarrow s \mathcal{L}\{\Delta B\}(s)-\left.\Delta B\right|_{t=0}
    &=
    \left.\left(\frac{\partial}{\partial A}\left(\frac{d B}{d t}\right)\right)\right|_{\left(B_e, A_e\right)} \cdot \mathcal{L}\{\Delta A\}(s)- \left.\left(\frac{\partial}{\partial B}\left(\frac{d B}{d t}\right)\right)\right|_{\left(B_e, A_e\right)} \cdot \mathcal{L}\{\Delta B\}(s)
    \label{eq:SR_sLB}
\end{align}

But $\Delta B|_{t=0}=0$, since the system is initially in a state of rest at equilibrium and so the oscillation deviation $\Delta B$ from $B_e$ is initially zero. Equation \eqref{eq:SR_sLB}, therefore, becomes:

\begin{equation*}
    s \mathcal{L}\{\Delta B\}(s)=\left.\left(\frac{\partial}{\partial A}\left(\frac{d B}{d t}\right)\right)\right|_{\left(B_e, A_e\right)} \cdot \mathcal{L}\{\Delta A\}(s)- \left.\left(\frac{\partial}{\partial B}\left(\frac{d B}{d t}\right)\right)\right|_{\left(B_e, A_e\right)} \cdot \mathcal{L}\{\Delta B\}(s)
\end{equation*}
\begin{shaded}
    \begin{equation}
        \Rightarrow \frac{\mathcal{L}\{\Delta B\}(s)}{\mathcal{L}\{\Delta A\}(s)}=\frac{\left.\left(\dfrac{\partial}{\partial A}\left(\dfrac{d B}{d t}\right)\right)\right|_{\left(B_e, A_e\right)}}{s+\left.\left(\dfrac{\partial}{\partial B}\left(\dfrac{d B}{d t}\right)\right)\right|_{\left(B_e, A_e\right)}} = \dfrac{J_1}{s+J_2} \quad , \quad J_1,J_2=\text{constants .}
        \label{TF_SRgen}
    \end{equation}
    Equation \eqref{TF_SRgen} gives the functional form of the transfer function for the general SR architecture described by Equation \eqref{SR_genODE} and illustrated in Figure \ref{fig:ar_gen}. Notice that it shares the same form as Equation \eqref{eq:TF_rate_note}. We can expect our positive and negative SR transfer functions to be of functional form $\text{constant}/(s+\text{constant})$.
\end{shaded}

\subsection{Positive simple regulation}
With this foresight, we are now ready to proceed to the first variant of SR: positive SR. We analyse positive SR in the context of a specific model --- in fact, the most common model in the biological literature --- yet the results retain their generality. Changing the model does not affect the functional form of the near-equilibrium dynamics or the resulting transfer functions; it only changes the values of the fluctuation coefficients (previously denoted by $J$ in Equations \eqref{linODE_SRgen} and \eqref{TF_SRgen}). Using the most common model has the added benefit of allowing readers to readily connect our results with existing studies in the literature.

\subsubsection{Dynamics in time: a common model}
In positive SR, the production rate of $B$ monotonically increases with $A$ until an asymptotic maximum rate is reached. This is a consequence of the physics of molecular binding. This monotonic increase, described by $f_1(A)$, is often modelled as a positive Hill function \cite{alon_introduction_2019} of form:

\begin{equation}
    f_1(A)=\alpha \frac{A^n}{A^n+k^n} \quad \text{,} \quad \alpha,k,n \geq 0\text{ .}
\end{equation}

As $B$ increases, the loss rate of $B$ also increases proportionally. This is a cumulative effect from molecular mechanisms including degradation and dephosphorylation. This proportional loss rate, captured by $f_2(B)$, is often modelled as a linear function:

\begin{equation}
    f_2(B)=c B \quad \text{,} \quad c \geq 0\text{ .}
\end{equation} 

Combining these two models gives the common model for positive SR in biological literature:

\begin{equation}
    \frac{d B}{d t}=\alpha \frac{A^n}{A^n+k^n}-c B \text{ .}
    \label{eq:SRpos}
\end{equation}

\subsubsection{Dynamics in time: general near-equilibrium properties}
From Equation \eqref{eq:SRpos}, the local production and local degradation rates are: 

\begin{equation}
    \left.\left(\frac{\partial}{\partial A}\left(\frac{d B}{d t}\right)\right)\right|_{\left(B_e, A_e\right)}=\frac{A_e^{n-1} \alpha k^n n}{\left(A_e^n+k^n\right)^2}=m\left(A_e, \alpha, k, n,\right) \in \mathbb{R}^{+}
\end{equation}

and 

\begin{equation}
    \left.\left(\frac{\partial}{\partial B}\left(\frac{d B}{d t}\right)\right)\right|_{\left(B_e, A_e\right)}=-c \in \mathbb{R}^{-}
\end{equation}

respectively. Local to equilibrium $(A_e,B_e)$, Equation \eqref{eq:SRpos} has Taylor approximation

\begin{equation*}
    \left.\frac{d B}{d t} \sim\left(\frac{\partial}{\partial A}\left(\frac{d B}{d t}\right)\right)\right|_{\left(B_e, A_e\right)} \cdot\left(A-A_e\right)+\left.\left(\frac{\partial}{\partial B}\left(\frac{d B}{d t}\right)\right)\right|_{\left(B_e, A_e\right)} \cdot\left(B-B_e\right)
\end{equation*}
\begin{equation}
    = m\cdot(A-A_e) - c\cdot(B-B_e)
\end{equation}

Realising that our variables of interest are the relative concentrations of $A$ and $B$ with respect to the new equilibrium — equivalently, the fluctuations $\Delta A$ and $\Delta B$ rather than the absolute concentrations $A$ and $B$ — we perform a change of variables, setting $A = A_e+\Delta A$ and $B=B_e+\Delta B$ to arrive at:

\begin{align}
    \frac{d\left(B_e+\Delta B\right)}{d t} & \sim m \cdot\left(A_e+\Delta A-A_e\right)-c \cdot\left(B_e+\Delta B-B_e\right) \nonumber\\
    \Rightarrow \frac{d B_e}{d t}+\frac{d(\Delta B)}{d t} & \sim m \cdot(\Delta A)-c \cdot(\Delta B) \nonumber\\
    \Rightarrow \frac{d(\Delta B)}{d t} & \sim m \cdot \Delta A-c \cdot \Delta B \label{eq:SR_dDeltaB}
\end{align}
\begin{shaded}
Equation \eqref{eq:SR_dDeltaB} describes the dynamics between fluctuations in input A centred about equilibrium value $A_e$ and fluctuations in output B centred about equilibrium value $B_e$.
\end{shaded}

\subsubsection{Transfer function}
To extract the frequency domain information from Equation \eqref{eq:SR_dDeltaB}, we must first convert it into the frequency domain. This is done by applying the Laplace transform to both sides of the equation, yielding:

\begin{align}
    \mathcal{L}\left\{\frac{d(\Delta B)}{d t}\right\}(s)&=\mathcal{L}\{m \cdot \Delta A-c \cdot \Delta B\}(s) \nonumber\\
    \Rightarrow s \mathcal{L}\{\Delta B\}(s)-\left.\Delta B\right|_{t=0}&=m \mathcal{L}\{\Delta A\}(s)-c \mathcal{L}\{\Delta B\}(s)
    \label{eq:SRpos_sLB}
\end{align}

Recall that, initially, the system is settled into steady state and there are no perturbations in $A$ to cause perturbations in $B$. Therefore, $\Delta B|_{t=0}=0$, giving:

\begin{align}
s \mathcal{L}\{\Delta B\}(s)&=m \mathcal{L}\{\Delta A\}(s)-c \mathcal{L}\{\Delta B\}(s) \nonumber \\
\Rightarrow(s+c) \mathcal{L}\{\Delta B\}(s)&=m \mathcal{L}\{\Delta A\}(s) \nonumber \\
\Rightarrow \frac{\mathcal{L}\{\Delta B\}(s)}{\mathcal{L}\{\Delta A\}(s)}&=\frac{m}{s+c} \text{ .}
\end{align}

\begin{shaded}
    The transfer function for positive simple regulation is:
    \begin{equation}
        T F_{S R^{+}}(s)=\frac{m}{s+c} \hspace{1em},\hspace{1em} c\in \mathbb{R}^+ \text{ .}
        \label{eq:TF_SR+}
    \end{equation}
\end{shaded}

\subsubsection{Motif stability and asymptotic response}

Recall from Section \ref{stability} that the stability of a transfer function determines the boundedness of its outputs. $TF_{SR^+}$ has a single pole at $s = -c$. When degradation is present, $c > 0$ and this pole is strictly negative-real. From Table \ref{tab:stability}, the motif is asymptotically stable. In this case, all biochemical oscillations fed into the motif produce bounded biochemical output oscillations --- no unbounded growth can occur. When degradation is absent, $c = 0$ and the pole shifts to the origin. Here, the motif is marginally stable but the corresponding resonance frequency is $\omega = 0$ and thus does not constitute a meaningful resonance frequency. Again, all sinusoidal input oscillations still produce bounded output oscillations.
\begin{shaded}
The positive SR motif is asymptotically stable when degradation is present and marginally stable when degradation is absent. This motif possesses no resonance frequencies. It strictly converts all biochemical input oscillations into bounded biochemical output oscillations.
\end{shaded}

\subsubsection{Slow forcing response}
To analyse the dominant behaviour of the positive SR motif under low-frequency input oscillations, we consider the limit as $|s| \to 0$. In this regime, the transfer function approaches:

\begin{equation}
    \lim _{s \rightarrow 0} T F_{S R^{+}} \frac{\displaystyle\lim _{s \rightarrow 0}(m)}{\displaystyle\lim _{s \rightarrow 0}(s+c)}=\frac{m}{c}+0 i \text{ .}
    \label{eq:SR+_slow}
\end{equation}

This limiting complex number makes a distance of $m/c$  units from the origin and an angle of $0^\circ$ from the positive real axis, on the complex plane. These two quantities imply that low frequency oscillations of A are amplitude modulated by a factor of $m/c$ and phase shifted by an angle of $0^\circ$ when transformed into oscillations of B. For example, a low frequency A oscillation of amplitude $1$ will be amplified into an in-phase same frequency B oscillation of amplitude $m/c$. 
\begin{shaded}
    The positive SR motif behaves like a quasi-static amplifier in the low-frequency limit.
\end{shaded}

\subsubsection{Rapid forcing response}
Conversely, to analyse the dominant behaviour of  positive SR motif under high-frequency input oscillations, we consider the limit as $|s| \to \infty$. Here, the transfer function approaches:

\begin{subequations}
\begin{align}
    \lim_{s \rightarrow \infty} TF_{SR^+} 
    & = \frac{\displaystyle\lim_{s \rightarrow \infty} (m)}{\displaystyle\lim_{s \rightarrow \infty} (s + c)} \label{eq:SR_highfreq_a} \\
    & = \frac{m}{\infty + c} = 0 + 0i \label{eq:SR_highfreq_b}
\end{align}
\end{subequations}

This complex number makes a distance of $0$ units with the origin on the complex plane. Therefore, in the high frequency limit, A oscillations are completely suppressed causing B to not be produced. We can deduce this high-frequency regime suppression rate by inspecting Equation \eqref{eq:SR_highfreq_a}. For sufficiently large values of $s$,

\begin{equation}
    \frac{m}{s+c} \sim \frac{m}{s} \text{ .}
    \label{SRpos_order}
\end{equation}

$m/s$ is the effective transfer function of our motif when input A starts operating at large enough frequencies. An order of $1/s$ implies amplitude modulation roll-off at 1 order of magnitude per decade of frequency, that is 10 abs/decade. Raising the input frequency of A by a factor of 10 will dampen the B output by a factor of 10. 

We can also deduce the asymptotic phase shift from this $m/s$ effective transfer function. Employing Equation \eqref{phi_lim} onto $m/s$, the asymptotic phase shift is $-90^\circ$. Therefore, rapid oscillations of A will set oscillations of B one quarter of a cycle behind.

\begin{shaded}
    At sufficiently high frequencies, the positive SR motif attenuates input signals at a rate of 10 abs/decade. Under increasingly rapid oscillations of A, the response in B is driven toward zero amplification and phase quadrature with output lagging behind input (if it remains detectable at all after such dampening).
\end{shaded}

\subsubsection{Frequency response functions}
Recall, from Section \ref{sec:AngFreqResp}, that the magnitude frequency-response and phase frequency-response functions can be directly obtained from a transfer function by applying Equations \eqref{mag} and \eqref{phi}.

\begin{shaded}
The complete analytical frequency-response functions describing how much the positive SR motif amplitude modulates and phase shifts incoming oscillations at any given frequency $\omega$ are, therefore:
\begin{subequations}
\begin{align}
M_{S R^{+}}(\omega) & = \frac{m}{\sqrt{\mathrm{c}^2+\omega^2}} \label{M_SR+}\\
\phi_{S R^{+}}(\omega) & = - \arctan \left(\frac{c}{\omega} \right) \label{phi_SR+}
\end{align}
\end{subequations}
\end{shaded}

\begin{figure}[ht]
    \centering
    \includegraphics[width=0.9\linewidth]{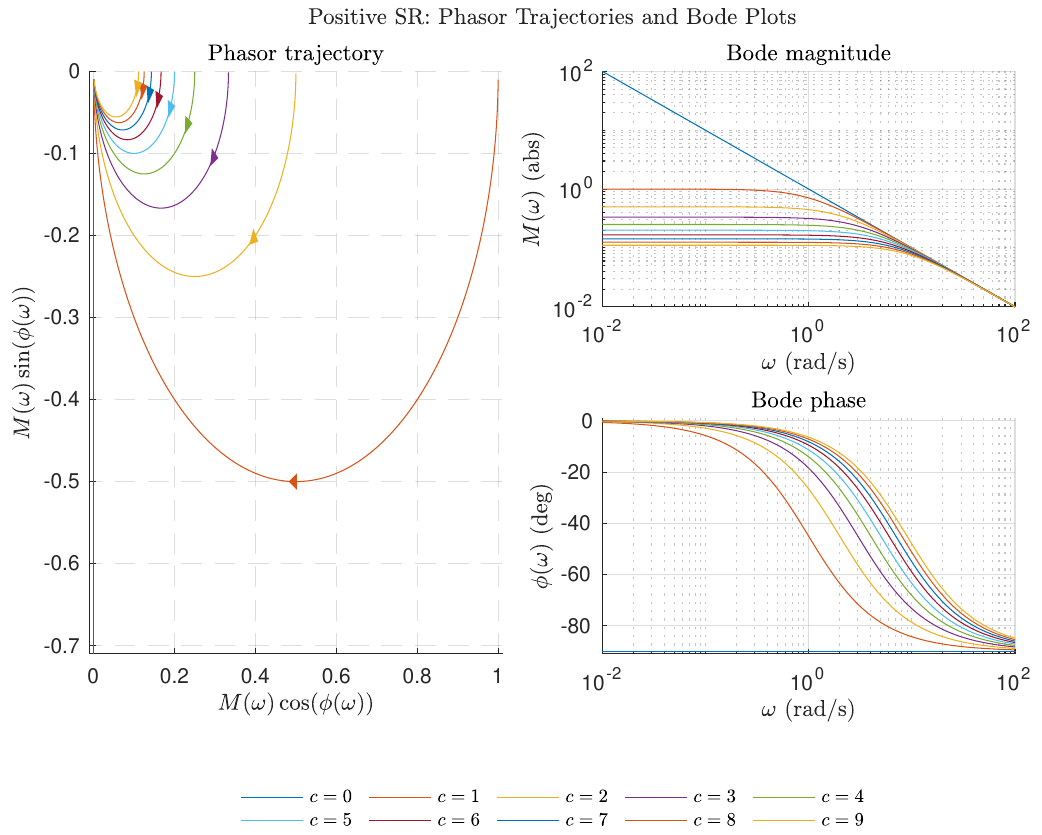}
    \caption{Phasor trajectories and Bode plots illustrating the angular frequency response functions of the positive SR motif.}
    \label{fig:SRpos_w}
\end{figure}

Although not necessary for the positive SR motif, since its full frequency response is already neatly described by Equations \eqref{M_SR+} and \eqref{phi_SR+}, it is helpful -- while the system remains simple -- to build familiarity with their counterpart phasor and Bode plots. These plots offer offer an alternative means of identifying key frequency response properties. By seeing how the same information is reflected both analytically and visually, we develop confidence in using these plots to extract insight—an essential skill as we move toward more complex network architectures, where the analytical frequency-response functions often become too unwieldy to be useful. In such cases, phasor trajectories and Bode plots become more essential tools. For the positive SR motif, these plots are shown in Figure \ref{fig:SRpos_w}.

To obtain the phasor trajectories, we evaluate Equation \eqref{eq:TF_SR+} at $s=i\omega$. This gives:

\begin{equation}
    T F_{S R^{+}}(i \omega) = \frac{m}{i \omega+c} = \frac{m c}{c^2+\omega^2} + i \left( -\frac{m c}{c^2+\omega^2} \right)
    \label{eq:SRpos_phasor}
\end{equation}

Each pair of $(m,c)$ uniquely defines a positive SR system. Allowing $\omega$ to vary while fixing $m$ and $c$ generates a trajectory describing the frequency-evolution of the amplitude modulation $M$ and phase shift $\phi$ for that particular system. By plotting the trajectories associated with each $(m,c)$ combination on the same set of axes, we can visualise a family of phasor plots describing the different frequency response functions that the positive SR motif can exhibit. Without loss of generality, let us examine the effect of varying degradation constant $c$ by considering 10 different systems defined by the following set of $(m,c)$ pairs:

\begin{equation*}
    \{(1,0),(1,1),(1,2),(1,3),(1,4),(1,5),(1,6),(1,7),(1,8),(1,9)\} \text{ .}
\end{equation*} 

Substituting these values into Equation \eqref{eq:SRpos_phasor} give the family of phasor trajectories in Figure \ref{fig:SRpos_w} (left panel). All positive SR systems exhibit a semi-circular trajectory, rotating clockwise in the fourth quadrant, from $(m/c,0)$ at $\omega = 0$ to $(0,0)$ at $\omega=\infty$. The only exception is when $c=0$. Here, the curve traces vertically upwards along the negative imaginary axis towards the origin. We can still think of this path as a clockwise semi-ellipse, however, now with an infinitely large radius.

\begin{shaded}
    The phasor plot tells us:\\
    All positive SR phasor trajectories are clockwise semicircular arcs in the fourth quadrant, with start and end points:
    \begin{equation}
        \lim _{\omega \rightarrow 0} \bigg( T F_{S R^{+}}(i \omega) \bigg)=\left(\frac{m}{c},-\frac{0}{c^2}\right) \text { and } \lim _{\omega \rightarrow \infty} \bigg( T F_{S R^{+}}(i \omega) \bigg)=(0,0)
    \end{equation}
    where we have employed a slight abuse of notation in the expression $-0/c^2$ when $c=0$ to mean $-\infty$. 
\end{shaded}

From Section \ref{sec:breakfreq}, the purely real-valued pole at $s=-c$ reveals that the positive SR motif has a break frequency at $\omega_{\text{break}}=c$. This is observed in the asymptotic corners on our Bode magnitude plots and points of inflection on Bode phase plots, in the top- and bottom-right panels of Figure \ref{fig:SRpos_w} respectively. The amplitude modulation curves exhibit constant order before and first-order roll-off after this break frequency. The phase modulation curves show a robust sigmoidal transition from being in phase at low frequencies to being in phase quadrature at high frequencies, defined by an inflection point at the same break frequency. These two features together are indicative of what is known as a first-order low-pass filter in control systems. Increasing $c$ increases the break frequency, and hence both the corner and inflection points, observed by vertical dilation of Bode magnitude curves of factor $1/c$ and horizontal translation of Bode phase curves by $c$ units.

\begin{shaded}
    The Bode plots tell us:\\
    Positive SR is a conventional first order low-pass filter with break frequency $\omega_{\text{break}}=c$. \textcolor{red}{A biological example is illustrated in Figure \ref{fig:Biology_SR}}
\end{shaded}

\subsection{Negative simple regulation}
\subsubsection{Dynamics in time: a common model}
Unlike positive SR, in negative SR, the production rate of B decreases with A until rate reaches an asymptotic minimum. Due to the similar physical mechanisms to promotion, the profile of this rate is symmetric to that of positive SR. As such, $f_1$ is often modelled as a decreasing Hill function --- the symmetric counterpart to the increasing Hill function. That is:

\begin{equation}
    f_1(A)=\alpha \frac{k^n}{A^n+k^n} \text{ ,}
\end{equation}

where $\alpha$, $k$ and $n$ are positive constants. The degradation mechanism remains the same and so:

\begin{equation}
    f_2(B)=cB \text{ ,}
\end{equation}

where c is a positive constant. 
\begin{shaded}
The corresponding dynamical equation of the system is therefore:
\begin{equation}
    \frac{d B}{d t}=\alpha \frac{k^n}{A^n+k^n}-c B \text{ .}
    \label{eq:SRneg}
\end{equation}
\end{shaded}

\subsubsection{Dynamics in time: general near-equilibrium properties} \label{sec:SRneg_DynamEquil}
Following the same linearisation logic of positive SR, we arrive at the linearised ODE for negative SR as follows:

\begin{equation}
    \left.\left(\frac{\partial}{\partial A}\left(\frac{d B}{d t}\right)\right)\right|_{\left(B_e, A_e\right)}=-\frac{A_e^{n-1} \alpha k^n n}{\left(A_e^n+k^n\right)^2}=-m\left(A_e, \alpha, k, n,\right) \in \mathbb{R}^{-}
\end{equation}

and 

\begin{equation}
    \left.\left(\frac{\partial}{\partial B}\left(\frac{d B}{d t}\right)\right)\right|_{\left(B_e, A_e\right)}=-c \in \mathbb{R}^{-} \text{ .}
\end{equation}

Therefore:

\begin{equation}
    \frac{d(\Delta B)}{d t} \sim-m \cdot \Delta A-c \cdot \Delta B \text{ .}
    \label{eq:SRneg_dDeltaB}
\end{equation}

\begin{shaded}
Like Equation \eqref{eq:SR_dDeltaB}, here, Equation \eqref{eq:SRneg_dDeltaB} describes the dynamics between input A centred about equilibrium value $A_e$ and output B centred about equilibrium value $B_e$.

\textbf{N.B.} \textit{The only difference between Equation \eqref{eq:SR_dDeltaB} and Equation \eqref{eq:SRneg_dDeltaB} is the negative sign in front of $m(A_e,\alpha,k,n,)$ --- the local production rate of species B due to A. Previously in positive SR, for each unit of increase in A near equilibrium, $m$ units of B were produced. Now in negative SR, for each unit of increase in A near equilibrium, $m$ units of B are lost. \textbf{This ``symmetry'' of production rates will present itself later in the symmetry between transfer functions of positive SR and negative SR.}}
\end{shaded}

\subsubsection{Transfer function}
The corresponding transfer function is then derived similarly to what was done in positive SR:

\begin{align}
    \mathcal{L}\left\{\frac{d(\Delta B)}{d t}\right\}(s)&=\mathcal{L}\{-m \cdot \Delta A-c \cdot \Delta B\}(s) \nonumber\\
    \Rightarrow s \mathcal{L}\{\Delta B\}(s)-\left.\Delta B\right|_{t=0}&=-m \mathcal{L}\{\Delta A\}(s)-c \mathcal{L}\{\Delta B\}(s)
    \label{eq:SRneg_sLB}
\end{align}

but $\left.\Delta B\right|_{t=0}=0$, so:

\begin{align}
    s \mathcal{L}\{\Delta B\}(s)&=-m \mathcal{L}\{\Delta A\}(s)-c \mathcal{L}\{\Delta B\}(s) \nonumber \\
    \Rightarrow \frac{\mathcal{L}\{\Delta B\}(s)}{\mathcal{L}\{\Delta A\}(s)}&=-\frac{m}{s+c} \text{ .}
\end{align}

\begin{shaded}
    The transfer function for negative simple regulation is:
    \begin{equation}
        T F_{S R^{-}}(s)=-\frac{m}{s+c} \hspace{1em},\hspace{1em} c\in \mathbb{R}^+ \text{ .}
        \label{eq:TF_SR-}
    \end{equation}
\end{shaded}

\begin{shaded}
    \textbf{Remark:} \textit{Notice that $T F_{S R^{-}}(s)$ is just the negative of $T F_{S R^{+}}(s)$, and vice versa. That is:
    \begin{equation}
        T F_{S R^{-}}(s) = -T F_{S R^{+}}(s) \text{ .}
    \end{equation}
    This will become important in the collective picture of SR later.}
\end{shaded}

\subsubsection{Motif stability and asymptotic response}
$T F_{S R^{-}}(s)$ has the same single pole at $s=-c$, as was the case for $T F_{S R^{+}}(s)$. Therefore, like positive SR, negative SR is asymptotically stable when $c>0$ and marginally stable when $c=0$.
\begin{shaded}
    \textbf{Remark:} \textit{Changing the mode of regulation from promotion to inhibition does not affect the stability and time-constants of simple regulation. However, as we will see later, what this change in regulation mode does affect is the system’s phase modulation, which is a more subtly hidden behaviour and more difficult to identify in the time domain.}
\end{shaded}

\subsubsection{Slow forcing response}
In the limit of low frequencies, that is $|s|\to0$:

\begin{equation}
    \lim _{s \rightarrow 0} T F_{S R^{-}}=-\frac{m}{c}+0 i \text{ .}
\end{equation}

As was for positive SR in Equation \eqref{eq:SR+_slow}, this limiting complex number makes distance $m/c$ units from the origin. Therefore, low frequency (and constant) signals are amplitude modulated by factor $m/c$. Unlike positive SR however, the complex number makes an angle of $180^\circ$ with the positive real axis. Inputs and outputs are, therefore, completely out of phase with each other at low frequencies. 

\begin{shaded}
    The negative SR motif also behaves like a quasi-static amplifier in the low-frequency limit.
\end{shaded}

\subsubsection{Rapid forcing response}
In the limit of high frequencies, that is $|s|\to\infty$:

\begin{subequations}
\begin{align}
    \lim_{s \rightarrow \infty} TF_{SR^-} 
    & = -\frac{\displaystyle\lim_{s \rightarrow \infty} (m)}{\displaystyle\lim_{s \rightarrow \infty} (s + c)} \label{eq:SRneg_highfreq_a} \\
    & = -\frac{m}{\infty + c} = 0 + 0i \label{eq:SRneg_highfreq_b}
\end{align}
\end{subequations}

Just like positive SR, this limiting number creates zero distance with the origin --- implying the negative SR motif suppresses high frequency signals. We can deduce this high-frequency regime suppression rate from Equation \eqref{eq:SRneg_highfreq_a}. For sufficiently large $s$,

\begin{equation}
    - \frac{m}{s+c} \sim \frac{m}{s} \text{ .}
\end{equation}

Again, just like positive SR in the high frequency regime, the transfer function of negative SR reduces to order $1/s$ and so exhibits roll-off at 1 order of magnitude per decade of frequency. However, unlike positive SR, the asymptotic phase shift from Equation \eqref{phi_lim} returns $+90^\circ$. Instead of lagging behind its input by a quarter-cycle, as was the case with positive SR, the output now leads by a quarter-cycle. For the same high frequency input, the outputs of positive and negative SR are always exactly $180^\circ$ out of phase with each other. This property becomes particularly useful when assembling SR motifs in series and parallel, where we can exploit properties such as superposition of amplitudes to cause destructively interference between biochemical outputs or convolution to shift signal phases in discrete units of quarter-cycles.

\begin{shaded}

    At sufficiently high frequencies, the negative SR motif attenuates input signals at a rate of 10 abs/decade. Under increasingly rapid oscillations of A, the response in B is driven toward zero amplification and phase quadrature with output leading input (if it remains detectable at all after such dampening).
    
    \textbf{N.B.} \textit{Changing the mode of regulation in SR, from promotion to inhibition, does not affect the roll-off rate but does reverse the asymptotic phase relationship. This a consequence of the ``symmetry'' between the two motifs hinted at earlier in section \eqref{sec:SRneg_DynamEquil}.}
\end{shaded}

\subsubsection{Frequency response functions}
From Equations \eqref{mag} and \eqref{phi}, the angular frequency response functions of negative SR have the exact analytical forms:
\begin{shaded}
\begin{subequations}
\begin{align}
M_{S R^{-}}(\omega) & = \frac{m}{\sqrt{\mathrm{c}^2+\omega^2}} \label{M_SR-}\\
\phi_{S R^{-}}(\omega) & = \pi - \arctan \left(\frac{c}{\omega} \right) \label{phi_SR-}
\end{align}
\end{subequations}

Equations \eqref{M_SR-} and \eqref{phi_SR-} fully describe how the negative SR motif performs amplitude modulation and phase shift on oscillations of angular frequency $\omega$.
\end{shaded}

\begin{figure}[ht]
    \centering
    \includegraphics[width=0.9\linewidth]{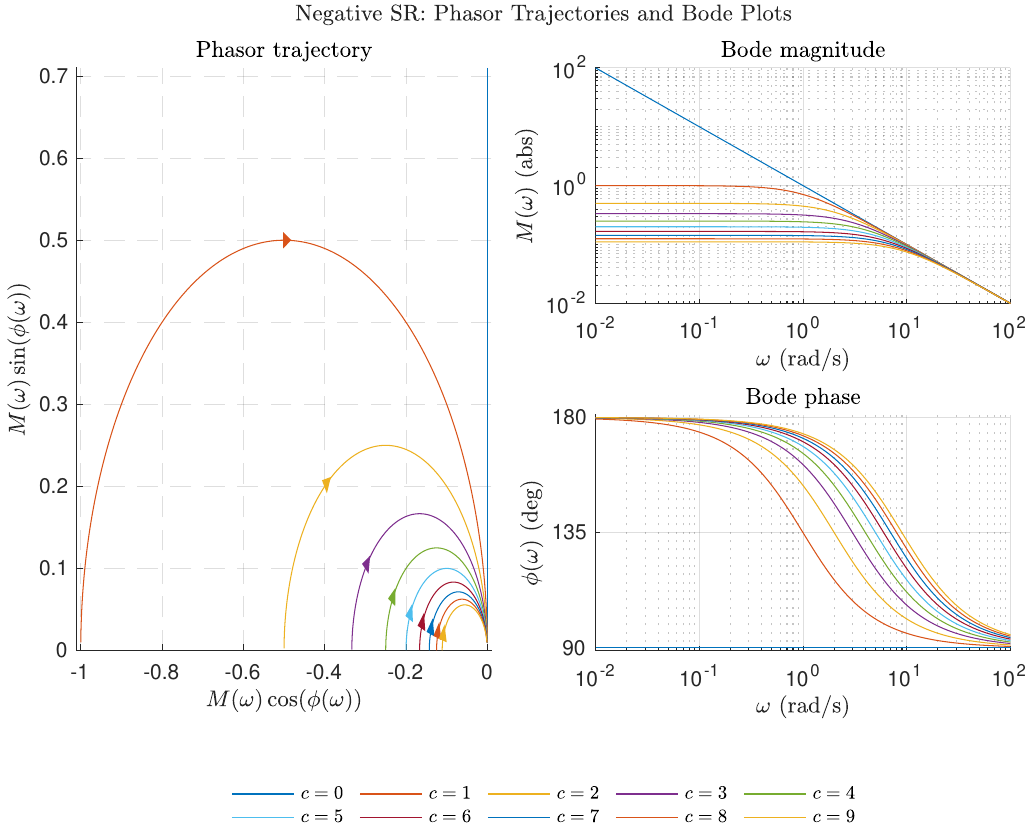}
    \caption{Phasor trajectories and Bode plots illustrating the angular frequency response functions of the negative SR motif.}
    \label{fig:SRneg_w}
\end{figure}

These relationships are plotted for representative parameters in Figure \ref{fig:SRneg_w} (right panels). The magnitude and phase curves exhibit identical shapes as those of positive SR. However, while the phase shifts in positive SR transition from $\phi = 0^\circ$ to $\phi = -90^\circ$, those of negative SR instead transition from $\phi = 180^\circ$ to $\phi = 90^\circ$. At any given frequency, the amplitude modulations of the two systems are indistinguishable. However, the phase difference between their outputs remains fixed at $180^\circ$, providing a clear distinguishing feature. Thus, under periodic forcing at equilibrium, a positive and a negative SR motif of the same $m$ and $c$ parameters yield identical amplitude profiles and can only be distinguished by their phase responses.

These two separate properties can be captured together in a single phasor plot. The corresponding phasor trajectories are described by Equation \eqref{eq:SRneg_phasor} and plotted in Figure \ref{fig:SRneg_w} (left). These negative SR phasor trajectories are just those of positive SR under a reflection about the line $\Re(s)=\Im(s)$. All negative SR systems exhibit a semi-circular trajectory, rotating clockwise in the second quadrant from $(-m/c,0)$ at $\omega=0$ to $(0,0)$ at $\omega=\infty$, with an exception when $c=0$. In this case, the phasor trajectory traces vertically downward along the positive imaginary axis towards the origin. Again, we can still think of this path as a clockwise semi-circle, however, now with an infinitely large radius. These semi-circular geometries show how $M(\omega)$ curves are strictly constrained into monotonically decreasing functions and how the second quadrant bounds $\phi(\omega)$ curves between $180^\circ$ and $90^\circ$.

\begin{equation}
    T F_{S R^{-}}(i \omega) = -\frac{m}{i \omega+c} = -\frac{m c}{c^2+\omega^2} + i \left( \frac{m c}{c^2+\omega^2} \right)
    \label{eq:SRneg_phasor}
    \end{equation}

Like $TF_{\text{SR}^+}$, the pole of $TF_{\text{SR}^-}$ also occurs at $s=-c$ and so the break frequency remains at $\omega_{\text{break}}=c$. This is observed in the asymptotic corners of the Bode magnitude curves and inflection points of the Bode phase curves in \ref{fig:SRneg_w}.

\begin{shaded}
    All negative SR phasor trajectories are clockwise semicircular arcs in the second quadrant, with start and end points:
    \begin{equation}
        \lim _{\omega \rightarrow 0} \bigg( T F_{S R^{-}}(i \omega) \bigg)=\left(-\frac{m}{c},\frac{0}{c^2}\right) \text { and } \lim _{\omega \rightarrow \infty} \bigg( T F_{S R^{-}}(i \omega) \bigg)=(0,0)
    \end{equation}
    where we have employed a slight abuse of notation in the expression $0/c^2$ when $c=0$ to mean $\infty$.
\end{shaded}

\subsection{Collective picture of simple regulation: the square}
\begin{figure}[ht]
    \centering
    \includegraphics[width=1\linewidth]{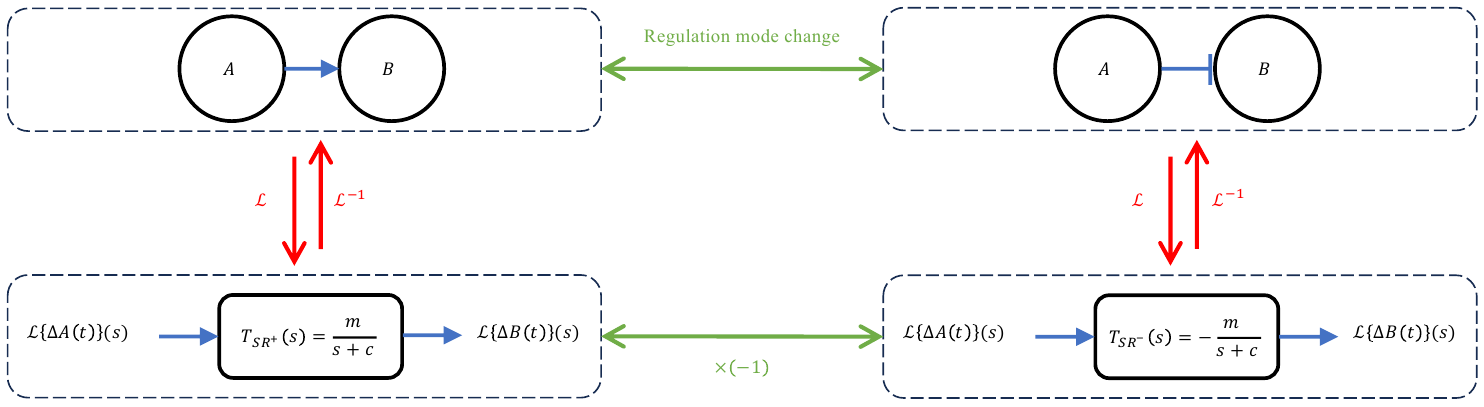}
    \caption{Collective picture of SR motifs: the relationship of positive and negative modes in time and frequency.}
    \label{fig:SR_collective}
\end{figure}

We can relate all four pictures of simple regulation to each other via the diagram shown in Figure \ref{fig:SR_collective}. The Laplace and inverse transforms are drawn in red. These allow us to move between the time domain and frequency domain pictures of simple regulation. The change in regulation mode in time and, equivalently, change in transfer function sign in frequency are drawn in green. These allow us to move between the two types of simple regulation modes. 

%% file: 4b_Cascades.tex
\subsection{Cascade}
\subsubsection{General architecture}
\begin{figure}[ht]
    \centering
    \includegraphics[width=0.75\linewidth]{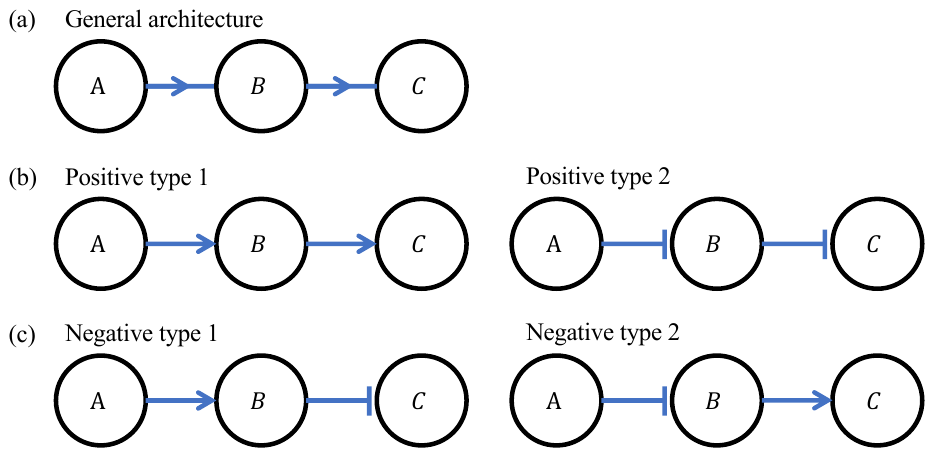}
    \caption{(a) General architecture of a cascade. (b) The two different forms of the positive cascade motif. (c) The two different forms of the negative cascade motif.}
    \label{fig:Cascade_archs}
\end{figure}

Cascades are linear sequences of biochemical species in which each species consecutively regulates the next. They can be viewed as chains of simple regulation (SR) motifs, linked end-to-end. While cascades may involve any number of species, the term cascade motif in systems biology typically refers to a three-species arrangement --- the simplest extension of a single two-species SR motif. This three-species cascade structure is usually depicted in literature --- whether in signal transduction, gene regulation, or other contexts --- as Figure \ref{fig:Cascade_archs}a, and is described by the following system of equations:

\begin{subequations}
   \begin{align}
   \dfrac{dB}{dt} &= f_1(A) - g_1(B) \\
   \dfrac{dC}{dt} &= f_2(B) - g_2(C)
   \end{align}
   \label{cascade_odes}
\end{subequations} 

A cascade composed of two identical modes of regulation is referred to as a positive cascade, as shown in Figure \ref{fig:Cascade_archs}b. Conversely, a cascade composed of two opposing modes of regulation is referred to as a negative cascade, as illustrated in Figure \ref{fig:Cascade_archs}c.

\subsubsection{Dynamics in time: a common model}
Like SR, promoting and repressing modes of regulation in cascades are commonly modelled using increasing and decreasing Hill functions, respectively, each superimposed with linear degradation. The corresponding models for the four cascades are summarised in Table \ref{tab:cascade-motifs}.

\begin{table}[H]
\renewcommand{\arraystretch}{2}
\centering
\begin{tabular}{|c|c|c|c|c|}
\hline
\textbf{Motif} & $f_1(A)$ & $g_1(B)$ & $f_2(B)$ & $g_2(C)$ \\
\hline \hline
Positive type 1 & $\alpha_1 \dfrac{A^{n_1}}{A^{n_1}+k_1^{n_1}}$ & $\gamma_1 B$ & $\alpha_2 \dfrac{B^{n_2}}{B^{n_2}+k_2^{n_2}}$ & $\gamma_2 C$ \\
\hline
Positive type 2 & $\alpha_1 \dfrac{k_1^{n_1}}{A^{n_1}+k_1^{n_1}}$ & $\gamma_1 B$ & $\alpha_2 \dfrac{k_2^{n_2}}{B^{n_2}+k_2^{n_2}}$ & $\gamma_2 C$ \\
\hline
Negative type 1 & $\alpha_1 \dfrac{A^{n_1}}{A^{n_1}+k_1^{n_1}}$ & $\gamma_1 B$ & $\alpha_2 \dfrac{k_2^{n_2}}{B^{n_2}+k_2^{n_2}}$ & $\gamma_2 C$ \\
\hline
Negative type 2 & $\alpha_1 \dfrac{k_1^{n_1}}{A^{n_1}+k_1^{n_1}}$ & $\gamma_1 B$ & $\alpha_2 \dfrac{B^{n_2}}{B^{n_2}+k_2^{n_2}}$ & $\gamma_2 C$ \\
\hline
\end{tabular}
\vspace{3pt}
\caption{Activation and degradation terms for different cascade motif types.}
\label{tab:cascade-motifs}
\end{table}

\subsubsection{Dynamics in time: general near-equilibrium properties}

Near equilibrium $(A_e,B_e,C_e)=\overline{\textbf{e}}$, the dynamics from Equation \eqref{cascade_odes} approach their linear regime

\begin{subequations}
   \begin{align}
   \dfrac{dB}{dt} 
   & \sim \left. \frac{d B}{d t}\right|_{\overline{\mathbf{e}}}
   + \left.\left(\frac{\partial}{\partial A}\left(\frac{d B}{d t}\right)\right)\right|_{\overline{\mathbf{e}}} \cdot (A - A_e)
   + \left.\left(\frac{\partial}{\partial B}\left(\frac{d B}{d t}\right)\right)\right|_{\overline{\mathbf{e}}} \cdot (B - B_e)
   \\
   \dfrac{dC}{dt} &
   \sim \left. \frac{d C}{d t}\right|_{\overline{\mathbf{e}}}
   + \left.\left(\frac{\partial}{\partial B}\left(\frac{d C}{d t}\right)\right)\right|_{\overline{\mathbf{e}}} \cdot (B - B_e)
   + \left.\left(\frac{\partial}{\partial C}\left(\frac{d C}{d t}\right)\right)\right|_{\overline{\mathbf{e}}} \cdot (C - C_e)
   \end{align}
\end{subequations}

and so

\begin{subequations}
   \begin{align}
   \dfrac{d\Delta B}{dt} 
   & \approx \left.\left(\frac{\partial}{\partial A}f_1(A)\right)\right|_{\overline{\mathbf{e}}} \cdot \Delta A
   + \left.\left(\frac{\partial}{\partial B} (-g_1(B))\right)\right|_{\overline{\mathbf{e}}} \cdot \Delta B
   \\
   \dfrac{d\Delta C}{dt} 
   & \approx \left.\left(\frac{\partial}{\partial B}f_2(B)\right)\right|_{\overline{\mathbf{e}}} \cdot \Delta B
   + \left.\left(\frac{\partial}{\partial C}(-g_2(C))\right)\right|_{\overline{\mathbf{e}}} \cdot \Delta C
   \end{align}
   \label{eq:cascade_lctiode}
\end{subequations}

The coefficients in Equation \eqref{eq:cascade_lctiode} are given in Table \ref{tab:cascade-coefs}, for each motif form.

\begin{table}[H]
\renewcommand{\arraystretch}{2.5}
\centering
\begin{tabular}{|c|c|c|c|c|}
\hline
\textbf{Motif} &
$\left.\left(\dfrac{\partial}{\partial A}f_1(A)\right)\right|_{\overline{\mathbf{e}}}$ & 
$\left.\left(\dfrac{\partial}{\partial B}(-g_1(B))\right)\right|_{\overline{\mathbf{e}}}$ & 
$\left.\left(\dfrac{\partial}{\partial B}f_2(B)\right)\right|_{\overline{\mathbf{e}}}$ & 
$\left.\left(\dfrac{\partial}{\partial C}(-g_2(C))\right)\right|_{\overline{\mathbf{e}}}$
\\
\hline \hline
Positive type 1 
& 
$+\dfrac{A_e^{n_1-1} \alpha_1 k_1^{n_1} n_1}{\left(A_e^{n_1}+k_1^{n_1}\right)^2}$
& $-\gamma_1$ 
& 
$+\dfrac{B_e^{n_2-1} \alpha_2 k_2^{n_2} n_2}{\left(B_e^{n_2}+k_2^{n_2}\right)^2}$
& $-\gamma_2$ \\
\hline
Positive type 2 
& 
$-\dfrac{A_e^{n_1-1} \alpha_1 k_1^{n_1} n_1}{\left(A_e^{n_1}+k_1^{n_1}\right)^2}$
& $-\gamma_1$ 
& 
$-\dfrac{B_e^{n_2-1} \alpha_2 k_2^{n_2} n_2}{\left(B_e^{n_2}+k_2^{n_2}\right)^2}$
& $-\gamma_2$ \\
\hline
Negative type 1 
& 
$+\dfrac{A_e^{n_1-1} \alpha_1 k_1^{n_1} n_1}{\left(A_e^{n_1}+k_1^{n_1}\right)^2}$
& $-\gamma_1$ 
& 
$-\dfrac{B_e^{n_2-1} \alpha_2 k_2^{n_2} n_2}{\left(B_e^{n_2}+k_2^{n_2}\right)^2}$
& $-\gamma_2$ \\
\hline
Negative type 2 
& 
$-\dfrac{A_e^{n_1-1} \alpha_1 k_1^{n_1} n_1}{\left(A_e^{n_1}+k_1^{n_1}\right)^2}$
& $-\gamma_1$ 
& 
$+\dfrac{B_e^{n_2-1} \alpha_2 k_2^{n_2} n_2}{\left(B_e^{n_2}+k_2^{n_2}\right)^2}$
& $-\gamma_2$ \\
\hline
\end{tabular}
\vspace{3pt}
\caption{Coefficients of each motif in linear regime.}
\label{tab:cascade-coefs}
\end{table}

\subsubsection{Transfer function}
\begin{figure}[ht]
    \centering
    \includegraphics[width=0.8\linewidth]{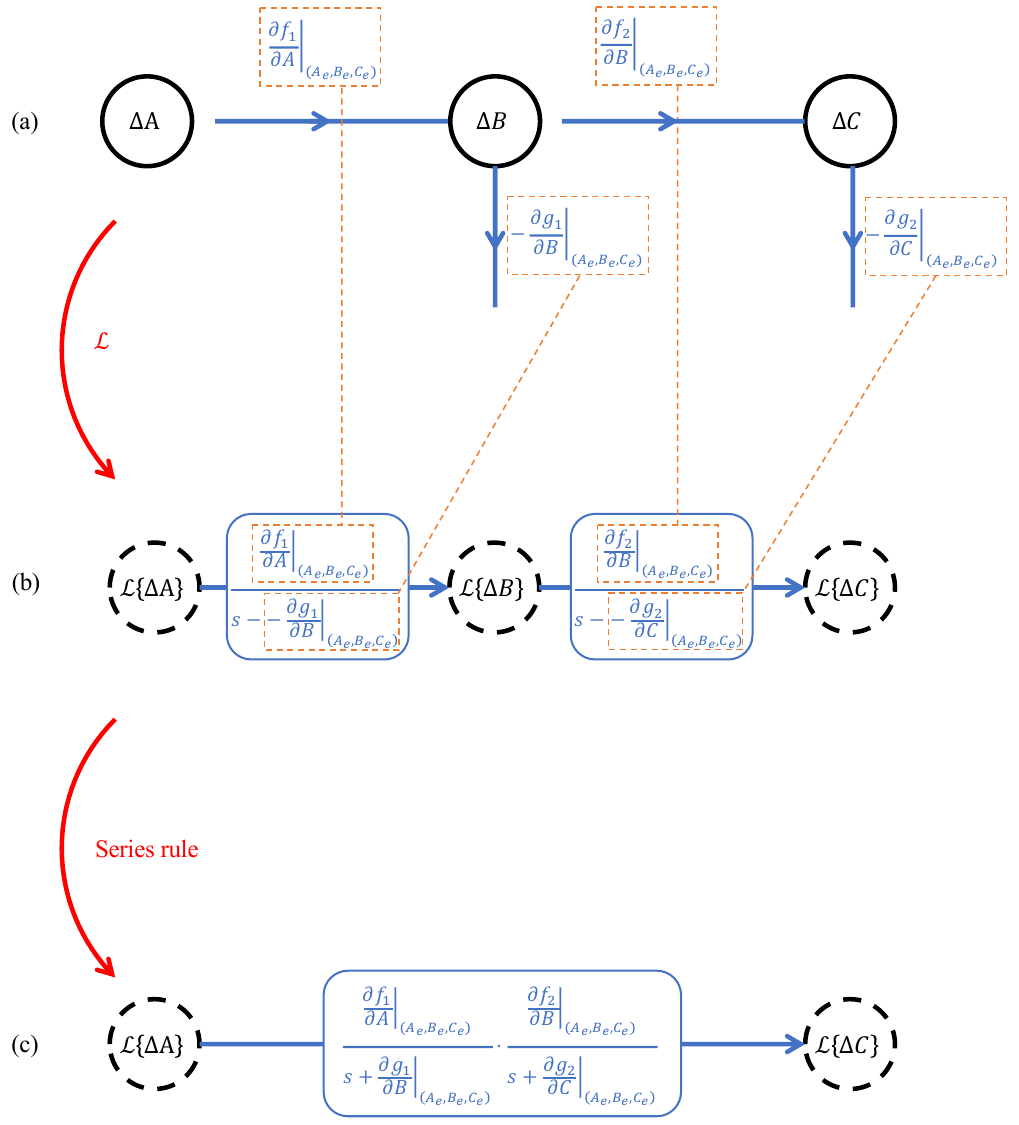}
    \caption{(a) Reaction network at equilibrium. (b) Laplace transform of network. (c) Reduced network using block diagram algebra.}
    \label{fig:Cascade_schema}
\end{figure}

We derive the transfer functions using the visual approach discussed in section \ref{sec:mathwithpictures} and highlighted in Figure \ref{fig:Laplace_zeta_Z}(c). This process begins by drawing system of lctiODEs in Equation \eqref{eq:cascade_lctiode} to its equivalent biochemical network at equilibrium in Figure \ref{fig:Cascade_schema}(a). With this picture, we can perform a visual Laplace transform that takes us to the frequency domain equivalent in Figure \ref{fig:Cascade_schema}(b). The final transfer function is obtained by employing the series rule, highlighted in Figure \ref{fig:BlockDiaAlg}. This returns Figure \ref{fig:Cascade_schema}(c) containing the transfer function from input oscillations of species A to output oscillations of species C. Defining

\begin{equation*}
    \dfrac{A_e^{n_1-1} \alpha_1 k_1^{n_1} n_1}{\left(A_e^{n_1}+k_1^{n_1}\right)^2}=m_1(A_e,\alpha_1,k_1,n_1) \in \mathbb{R}^+ \quad \text{and} \quad \dfrac{B_e^{n_2-1} \alpha_2 k_2^{n_2} n_2}{\left(A_e^{n_2}+k_2^{n_2}\right)^2}=m_2(B_e,\alpha_2,k_2,n_2) \in \mathbb{R}^+ \text{  ,}
\end{equation*}

the transfer functions for each motif are listed in Table \ref{tab:cascade_tf}. Notice how both positive motifs and both negative motifs have the same transfer function. This is why, in the literature, you will sometimes see the positive and negative cascade motifs be define as their type 1 forms.

\begin{table}[H]
\renewcommand{\arraystretch}{2}
\centering
\begin{tabular}{|c|c|c|c|c|}
\hline
\textbf{Motif} & Positive type 1 & Positive type 2 & Negative type 1 & Negative type 2 \\
\hline
\textbf{Transfer Function} & 
$+\dfrac{m_1 m_2}{(s + \gamma_1)(s + \gamma_2)}$ & 
$+\dfrac{m_1 m_2}{(s + \gamma_1)(s + \gamma_2)}$ & 
$-\dfrac{m_1 m_2}{(s + \gamma_1)(s + \gamma_2)}$ & 
$-\dfrac{m_1 m_2}{(s + \gamma_1)(s + \gamma_2)}$ \\
\hline
\end{tabular}
\vspace{3pt}
\caption{Transfer functions for each cascade motif type.}
\label{tab:cascade_tf}
\end{table}

\begin{shaded}
    For both their types 1 and 2 variants, the positive and negative cascades have general transfer function:
    \begin{equation}
    T_{\text{Cas}^\pm}=\pm\dfrac{m_1 m_2}{(s + \gamma_1)(s + \gamma_2)} = \pm T_{\text{Cas}} \text{ .}
    \end{equation}
\end{shaded}

\subsubsection{Motif stability and asymptotic response}
All four forms of the cascade motif have poles at $s=-\gamma_1,-\gamma2$. Since when $\gamma_1,\gamma2$ are both positive, that is when both B and C degrade, all poles are negative-real valued and the motifs are asymptotically stable. Oscillations of A about equilibrium $A_e$ cannot cause unbounded oscillations in C about equilibrium $C_e$. It takes only one species to stop degrading to render the entire cascade marginally stable. 

It is also worthwhile to explicitly connect our pole-inspection method to the underlying mechanics of second-order systems, in order to clarify why resonance is impossible in cascades, since we will later rely solely on this pole-inspection technique in more complex networks. It is well known from electrical engineering that, a general second order system of form:

\begin{equation*}
    \frac{1}{s^2+2 \xi \omega_n s+\omega_n^2}
\end{equation*}

where $\omega_n$ is the natural frequency of the system and $\xi$ is the damping constant, has resonant frequency:

\begin{equation*}
    \omega_r=\omega_n \sqrt{1-2 \xi^2}
\end{equation*}

Our cascade transfer functions have form:

\begin{equation*}
    \pm \dfrac{m_1 m_2}{(s + \gamma_1)(s + \gamma_2)} \text{ .}
\end{equation*}

Therefore, the natural frequency and damping constant of a positive cascade are:

\begin{equation*}
    \omega_n=\sqrt{\gamma_1 \gamma_2} \quad \text {and} \quad \xi=\frac{1}{2} \frac{\gamma_1+\gamma_2}{\gamma_1 \gamma_2}
\end{equation*}

respectively. As such, the resonant frequency of a cascade would be: 

\begin{equation*}
    \omega_r=\sqrt{c_1 c_2-\frac{\left(c_1+c_2\right)^2}{2}}\text{ .}
\end{equation*}

But, since $\gamma_1\geq0$ and $\gamma_2\geq0$, it must be that $\omega_r$ cannot be positive real. Therefore, by the nature of the cascade architecture, resonance is impossible. 

\subsubsection{Slow and rapid forcing responses}
Taking the limits as $|s|\rightarrow0$ and $|s|\rightarrow\infty$, we can deduce the following table. 

\begin{table}[H]
\renewcommand{\arraystretch}{2}
\centering
\begin{tabular}{|c|c|c|c|c|c|c|}
\hline
\textbf{Motif} & 
\textbf{\makecell{Slow Forcing\\Limit $s \to 0$}} & 
\makecell{Quasistatic\\Gain Limit} & 
\makecell{Low-Frequency\\Phase Shift\\Limit} & 
\textbf{\makecell{Rapid Forcing\\Limit $s \to \infty$}} & 
\makecell{Asymptotic\\Gain Limit} &
\makecell{High-Frequency\\Phase Shift\\Limit}\\
\hline
\textbf{\makecell{Positive\\type 1}} & $+\dfrac{m_1 m_2}{\gamma_1 \gamma_2} + 0i$ & $\dfrac{m_1 m_2}{\gamma_1 \gamma_2}$ & \makecell{$0^\circ$ (asym. stable)\\$-90^\circ$ (marg. stable)} & $0 + 0i$ & $0$ & $-180^\circ$\\
\hline
\textbf{\makecell{Positive\\type 2}} & $+\dfrac{m_1 m_2}{\gamma_1 \gamma_2} + 0i$ & $\dfrac{m_1 m_2}{\gamma_1 \gamma_2}$ & \makecell{$0^\circ$ (asym. stable)\\$-90^\circ$ (marg. stable)} & $0 + 0i$ & $0$ & $-180^\circ$\\
\hline
\textbf{\makecell{Negative\\type 1}} & $-\dfrac{m_1 m_2}{\gamma_1 \gamma_2} + 0i$ & $\dfrac{m_1 m_2}{\gamma_1 \gamma_2}$ & \makecell{$180^\circ$ (asym. stable)\\$90^\circ$ (marg. stable)} & $0 + 0i$ & $0$ & $0^\circ$\\
\hline
\textbf{\makecell{Negative\\type 2}} & $-\dfrac{m_1 m_2}{\gamma_1 \gamma_2} + 0i$ & $\dfrac{m_1 m_2}{\gamma_1 \gamma_2}$ & \makecell{$180^\circ$ (asym. stable)\\$90^\circ$ (marg. stable)} & $0 + 0i$ & $0$ & $0^\circ$\\
\hline
\end{tabular}
\vspace{4pt}
\caption{Frequency response characteristics of the four cascade motifs in both low- and high-frequency forcing regimes.}
\label{tab:casc_freq_response}
\end{table}

\subsubsection{Frequency response functions}
\begin{figure}[ht]
    \centering
    \includegraphics[width=0.85\linewidth]{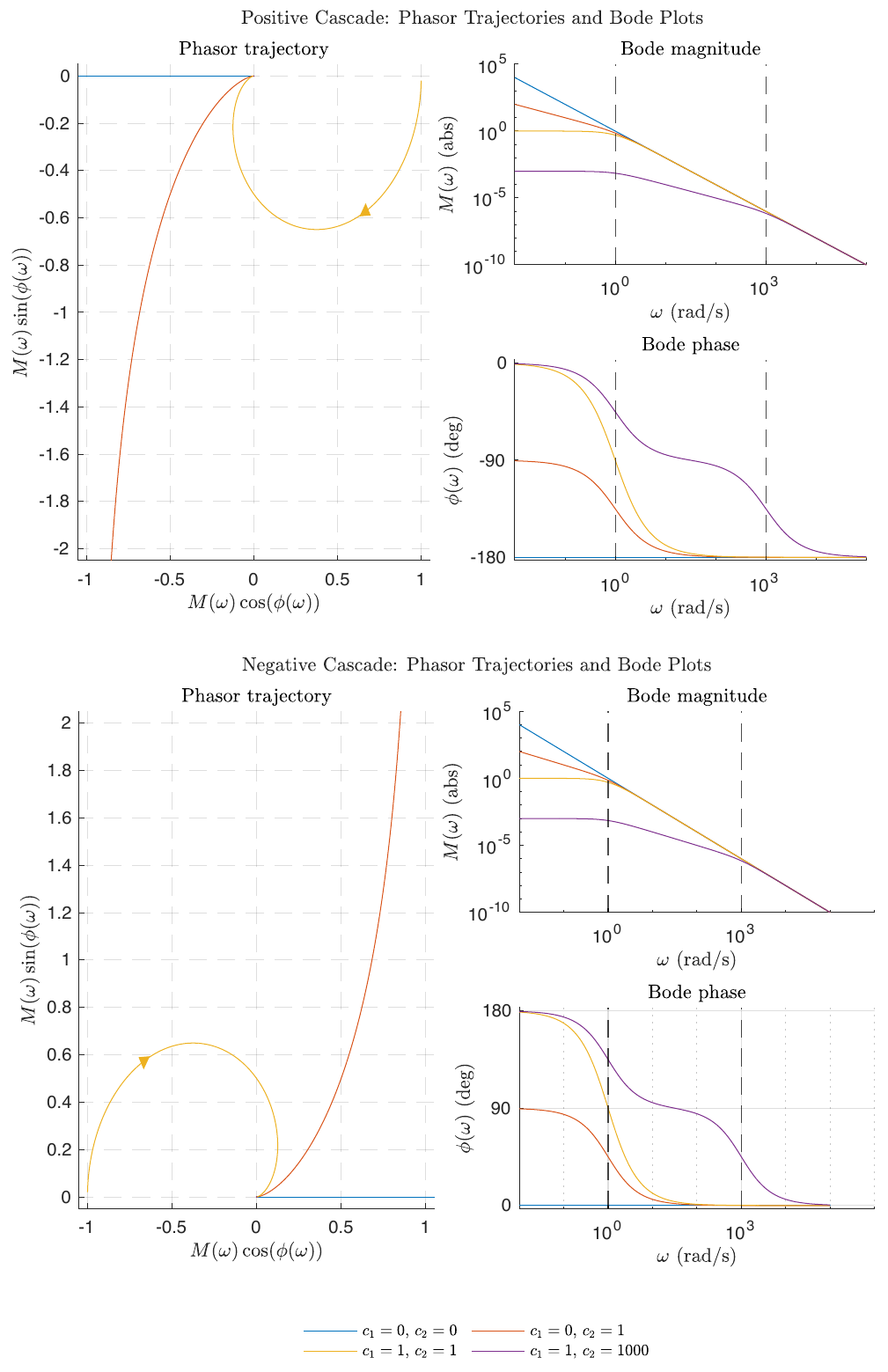}
    \caption{Phasor trajectories and Bode plots of the positive and negative cascade motifs (types 1 and 2). Without loss of generality, local production rates $m_1$,$m_2$ are set to unity and degradation rates $\gamma_1$,$\gamma_2$ are varied.}
    \label{fig:Cascade_PhasorBode}
\end{figure}

With poles $p_1=\gamma_1+0i$ and $p_2=\gamma_2+0i$, the corresponding analytical frequency response functions are

\begin{subequations}
    \begin{align}
        M(\omega) &= 
        \frac{|m_1m_2|}
        {\sqrt{\left( \gamma_1^2+\omega^2 \right)\left( \gamma_2^2+\omega^2 \right)}}
        \label{cascadepos_mag2} \\
        \phi(\omega) &= 
        - \arctantwo\left(\omega,-\gamma_1\right) - \arctantwo\left(\omega,-\gamma_2\right)
        \label{cascadepos_phi2}
    \end{align}
\end{subequations}

for the positive cascade motifs, and 

\begin{subequations}
    \begin{align}
        M(\omega) &= 
        \frac{|m_1m_2|}
        {\sqrt{\left( \gamma_1^2+\omega^2 \right)\left( \gamma_2^2+\omega^2 \right)}}
        \label{cascadeneg_mag2} \\
        \phi(\omega) &= 
        \pi - \arctantwo\left(\omega,-\gamma_1\right) - \arctantwo\left(\omega,-\gamma_2\right)
        \label{cascadeneg_phi2}
    \end{align}
\end{subequations}

for the negative cascade motifs. The relevant phasor and Bode plots are shown in Figure \ref{fig:Cascade_PhasorBode} for the positive and negative cascade motifs, respectively. Only a single set of plots are presented for each motif, as their type 1 and type 2 variants share identical transfer functions and therefore exhibit identical frequency responses. Break frequencies in all plots occur at $\omega=\gamma_1,\gamma_2$.

All Bode magnitude curves are upper bounded by the case with absent degradation mechanisms (blue) and lower bounded by the case with maximal degradation presence (purple). This reveals that cascade filtering is weakest when its species do not degrade and strongest when all possible degradation pathways are active at their highest rates. 

The Bode phase curves suggest that, by selecting the number of degradation mechanisms to be included, we can control the low-frequency phase shift to be either $0^\circ$, $-90^\circ$ or $-180^\circ$ for positive cascades, and $180^\circ$, $90^\circ$ or $0^\circ$ for negative cascades. Unlike the magnitude response, this low-frequency phase limit depends only on the presence of degradation mechanisms and not their specific rates. However, the rate of transition between low- and high-frequency phase shift limits does become slower as the degradation rates increase. 

By tuning the degradation rates, we control the positioning of break frequencies and thereby create an intermediate window of width $|\gamma_1 - \gamma_2|$ where transition dynamics emerge. This is most clearly illustrated by the purple curve in Figure~\ref{fig:Cascade_PhasorBode}, where a widened $|\gamma_1 - \gamma_2|$ interval produces a distinct region of order 1 roll-off and phase quadrature—previously inaccessible when not all degradation mechanisms were present, or impractically narrow when the gap was too small. The purple curve thus realises the cascade motif’s full potential: exhibiting order 0 behaviour at low frequencies, order 1 dynamics in the intermediate regime, and order 2 filtering at high frequencies. A special case occurs when both degradation rates are equal,  eliminating the intermediate frequency window. This is shown in yellow. As a result, the system jumps directly from zeroth- to second-order behaviour at this single break frequency. This is particularly useful in scenarios where intermediate timescale dynamics must be preserved when sweeping frequencies in an experiment.

These magnitude and phase properties, which depend on our choice of degradation rates, arise directly from basic phasor multiplication—where magnitudes are multiplied and phases are summed.

\subsection{Collective picture of cascades: the cube}
\begin{figure}[ht]
    \centering
    \includegraphics[width=1\linewidth]{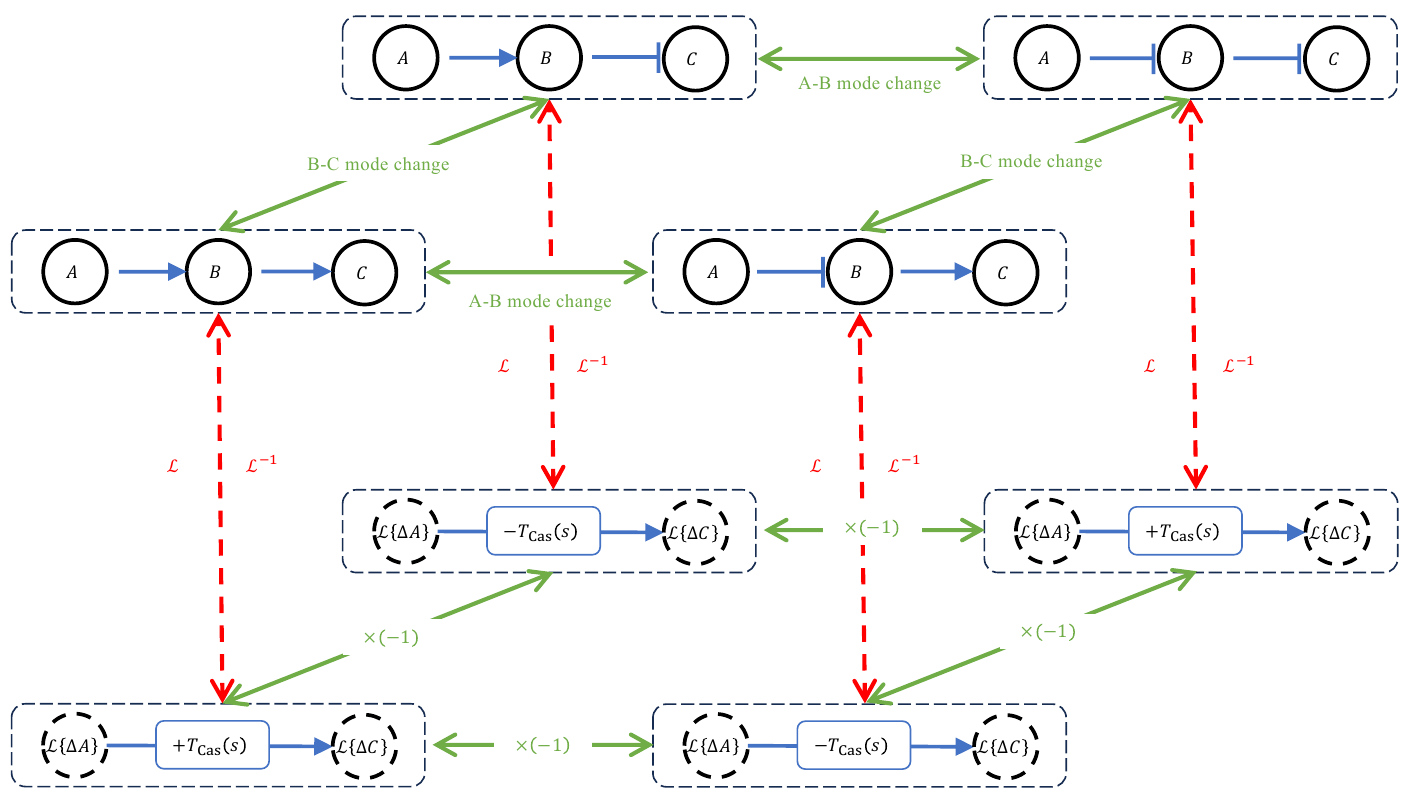}
    \caption{Collective picture of cascade motifs: the relationship of positive and negative cascades in time and frequency.}
    \label{fig:Cas_collective}
\end{figure}

The four cascade types are interrelated by switching the regulatory mode of each edge. Collectively, their time-domain and frequency-domain representations span a cube, as illustrated in Figure \ref{fig:Cas_collective}.

%% file: 4c_Feedforward.tex
\subsection{Feedforward loop}
\subsubsection{General architecture}
\begin{figure}[ht]
    \centering
    \includegraphics[width=0.7\linewidth]{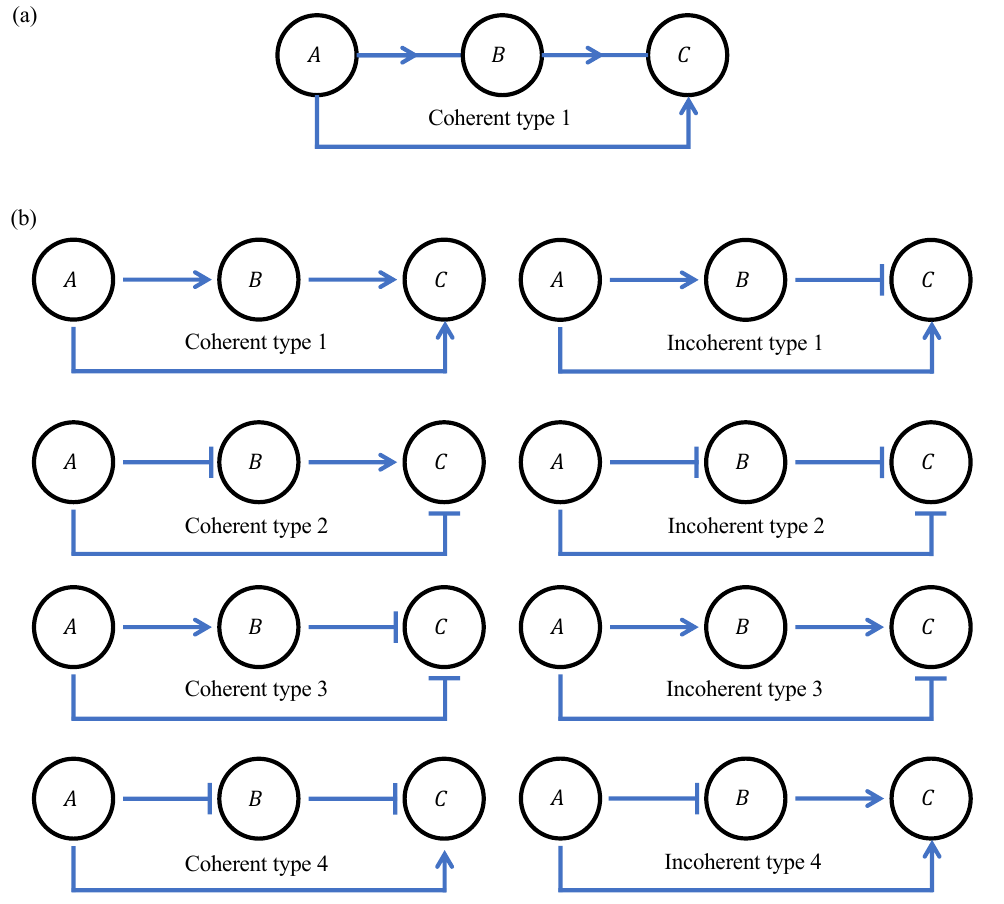}
    \caption{(a) General architecture of a feedforward loop. (b) All eight motif variants.}
    \label{fig:FFL_archs}
\end{figure}

Feedforward loops (FFLs) are extensions of cascades, distinguished by an additional regulatory path from the first to the last species. Like cascades, they involve three biochemical species --- one acting as a pure source and another as a pure sink. This results in two regulatory paths from input to output, as seen in the general structure drawn in Figure \ref{fig:FFL_archs}(a). FFLs are described by the following system of equations:

\begin{subequations}
   \begin{align}
   \dfrac{dB}{dt} &= f_1(A) - g_1(B) \\
   \dfrac{dC}{dt} &= f_2(A,B) - g_2(C)
   \end{align}
   \label{ffl_odes}
\end{subequations}

Each regulatory edge in Figure \ref{fig:FFL_archs}(a) assumes one of two regulation modes --- either promotion or repression. Given three edges, a total of eight regulation combinations are possible. These eight combinations form the eight FFL variants, illustrated in Figure \ref{fig:FFL_archs}(b). Regulation modes are also assigned sign values --- with promotion being +1 and repression being -1. The sign of a linear path is the product of its constituent signs. As such, any two linear paths can only ever be of the same sign (a $+1$ pair) or of opposing signs (a $\pm1$ pair). In the former case, the pair of paths are said to be coherent. In the latter case, they are said to be incoherent. Each of the eight FFLs can be classified as either coherent or incoherent.

\subsubsection{Dynamics in time: a common model}
Like the motifs discussed previously, FFLs share the same Hill-like mechanisms for upstream-dependent production and linear-like mechanisms for self-dependent degradation. However, unlike those motifs, the FFL is the first architecture in which a single species (C) is regulated by two upstream species (A and B) --- captured in $f_2(A,B)$. This dual dependence warrants first understanding the underlying mechanism, and then selecting an appropriate modelling approach to represent the combined influence of A and B. In the literature, two approaches are often used: individual dependence, in which binding of either A or B is sufficient to induce production of C, and mutual dependence, where both A and B must bind simultaneously \cite{mangan_structure_2003}. In the former, the gain rate of C is modelled as the sum of individual contributions from A and B (OR logic). In the latter, it is modelled as their product (AND logic). Under these two logic models, Equation \eqref{ffl_odes} becomes Equation \eqref{ffl_odes2}. Their flux terms for the eight FFL motifs are summarised in Table \ref{tab:FFL_motifs}.

\begin{shaded}
    \begin{subequations}
    \begin{align}
    \dfrac{dB}{dt} &= f_1(A) - g_1(B) \\
    \dfrac{dC}{dt} &=
    \begin{cases}
    f_{2,A}(A) + f_{2,B}(B) - g_2(C) & \text{(OR \hspace{0.75em}logic)} \label{ffl_odes_cases}\\
    f_{2,A}(A) \hspace{2pt} \cdot \hspace{3pt} f_{2,B}(B) - g_2(C) & \text{(AND logic)}
    \end{cases}
    \end{align}
    \label{ffl_odes2}
    \end{subequations}
\end{shaded}

\begin{table}[ht]
\renewcommand{\arraystretch}{2}
\centering
\begin{tabular}{|c|c|c|c|c|c|}
\hline
\textbf{Motif} & $f_1(A)$ & $g_1(B)$ & $f_{2A}$ & $f_{2B}$ & $g_2(C)$ \\
\hline \hline
\makecell{Coherent\\type 1} 
& $\alpha_1 \dfrac{A^{n_1}}{A^{n_1}+k_1^{n_1}}$ 
& $\gamma_1 B$ 
& $\alpha_{2,A} \dfrac{A^{n_{2,A}}}{A^{n_{2,A}}+k_{2,A}^{n_{2,A}}}$ 
& $\alpha_{2,B} \dfrac{B^{n_{2,B}}}{B^{n_{2,B}}+k_{2,B}^{n_{2,B}}}$
& $\gamma_2 C$ \\
\hline
\makecell{Coherent\\type 2} 
& $\alpha_1 \dfrac{k_1^{n_1}}{A^{n_1}+k_1^{n_1}}$ 
& $\gamma_1 B$ 
&  $\alpha_{2,A} \dfrac{k_{2,A}^{n_{2,A}}}{A^{n_{2,A}}+k_{2,A}^{n_{2,A}}}$ 
& $\alpha_{2,B} \dfrac{B^{n_{2,B}}}{B^{n_{2,B}}+k_{2,B}^{n_{2,B}}}$
& $\gamma_2 C$ \\
\hline
\makecell{Coherent\\type 3} 
& $\alpha_1 \dfrac{A^{n_1}}{A^{n_1}+k_1^{n_1}}$ 
& $\gamma_1 B$ 
& $\alpha_{2,A} \dfrac{k_{2,A}^{n_{2,A}}}{A^{n_{2,A}}+k_{2,A}^{n_{2,A}}}$ 
& $\alpha_{2,B} \dfrac{k_{2,B}^{n_{2,B}}}{B^{n_{2,B}}+k_{2,B}^{n_{2,B}}}$ 
& $\gamma_2 C$ \\
\hline
\makecell{Coherent\\type 4} 
& $\alpha_1 \dfrac{k_1^{n_1}}{A^{n_1}+k_1^{n_1}}$ 
& $\gamma_1 B$ 
& $\alpha_{2,A} \dfrac{A^{n_{2,A}}}{A^{n_{2,A}}+k_{2,A}^{n_{2,A}}}$ 
& $\alpha_{2,B} \dfrac{k_{2,B}^{n_{2,B}}}{B^{n_{2,B}}+k_{2,B}^{n_{2,B}}}$ 
& $\gamma_2 C$ \\
\hline
\makecell{Incoherent\\type 1} 
& $\alpha_1 \dfrac{A^{n_1}}{A^{n_1}+k_1^{n_1}}$ 
& $\gamma_1 B$ 
& $\alpha_{2,A} \dfrac{A^{n_{2,A}}}{A^{n_{2,A}}+k_{2,A}^{n_{2,A}}}$ 
& $\alpha_{2,B} \dfrac{k_{2,B}^{n_{2,B}}}{B^{n_{2,B}}+k_{2,B}^{n_{2,B}}}$  
& $\gamma_2 C$ \\
\hline
\makecell{Incoherent\\type 2} 
& $\alpha_1 \dfrac{k_1^{n_1}}{A^{n_1}+k_1^{n_1}}$ 
& $\gamma_1 B$ 
& $\alpha_{2,A} \dfrac{k_{2,A}^{n_{2,A}}}{A^{n_{2,A}}+k_{2,A}^{n_{2,A}}}$ 
& $\alpha_{2,B} \dfrac{k_{2,B}^{n_{2,B}}}{B^{n_{2,B}}+k_{2,B}^{n_{2,B}}}$ 
& $\gamma_2 C$ \\
\hline
\makecell{Incoherent\\type 3} 
& $\alpha_1 \dfrac{A^{n_1}}{A^{n_1}+k_1^{n_1}}$ 
& $\gamma_1 B$ 
&  $\alpha_{2,A} \dfrac{k_{2,A}^{n_{2,A}}}{A^{n_{2,A}}+k_{2,A}^{n_{2,A}}}$ 
& $\alpha_{2,B} \dfrac{B^{n_{2,B}}}{B^{n_{2,B}}+k_{2,B}^{n_{2,B}}}$
& $\gamma_2 C$ \\
\hline
\makecell{Incoherent\\type 4} 
& $\alpha_1 \dfrac{k_1^{n_1}}{A^{n_1}+k_1^{n_1}}$ 
& $\gamma_1 B$ 
& $\alpha_{2,A} \dfrac{A^{n_{2,A}}}{A^{n_{2,A}}+k_{2,A}^{n_{2,A}}}$ 
& $\alpha_{2,B} \dfrac{B^{n_{2,B}}}{B^{n_{2,B}}+k_{2,B}^{n_{2,B}}}$
& $\gamma_2 C$ \\
\hline
\end{tabular}
\vspace{3pt}
\caption{Activation and degradation terms for different cascade motif types. All $\alpha,k,n$ terms are positive-real and $\gamma$ terms are nonnegative-real.}
\label{tab:FFL_motifs}
\end{table}

\subsubsection{Dynamics in time: general near-equilibrium properties}
Before writing the exact analytical form, it is fruitful to establish that, regardless of the choice in Equation \ref{ffl_odes_logic}, the structure of their dynamical equations collapse into the same form in the linear regime. In this regime, Equation \eqref{ffl_odes2} becomes:

\begin{subequations}
   \begin{align}
   \dfrac{d\Delta B}{dt} 
   & \approx \left.\frac{\partial f_1}{\partial A}\right|_{\overline{\mathbf{e}}} \cdot \Delta A
   + \left.\frac{\partial(-g_1)}{\partial B}\right|_{\overline{\mathbf{e}}} \cdot \Delta B
   \\
   \dfrac{d\Delta C}{dt} 
   & \approx 
   \left.\frac{\partial f_2}{\partial A}\right|_{\overline{\mathbf{e}}} \cdot \Delta A
   +\left.\frac{\partial f_2}{\partial B}\right|_{\overline{\mathbf{e}}} \cdot \Delta B
   + \left.\frac{\partial(-g_2)}{\partial B}\right|_{\overline{\mathbf{e}}} \cdot \Delta C \\
   & =
   \begin{cases}\label{ffl_odes_logic}
   \left.\dfrac{\partial f_{2,A}}{\partial A}\right|_{\overline{\mathbf{e}}} \cdot \Delta A
   +\left.\dfrac{\partial f_{2,B}}{\partial B}\right|_{\overline{\mathbf{e}}} \cdot \Delta B
   + \left.\dfrac{\partial(-g_2)}{\partial C}\right|_{\overline{\mathbf{e}}} \cdot \Delta C 
   & \text{(OR \hspace{0.75em}logic)}\\
   \left(\left.\dfrac{\partial f_{2,A}}{\partial A}f_{2,B}\right)\right|_{\overline{\mathbf{e}}} \cdot \Delta A
   +\left(f_{2,A}\left.\dfrac{\partial f_{2,B}}{\partial B}\right)\right|_{\overline{\mathbf{e}}} \cdot \Delta B
   + \left.\dfrac{\partial(-g_2)}{\partial C}\right|_{\overline{\mathbf{e}}} \cdot \Delta C
   & \text{(AND logic)} 
    \end{cases}
   \end{align}
   \label{eq:ffl_lctiode}
\end{subequations}

Regardless of the logic choice in Equation \eqref{ffl_odes_logic} the structure of their equations near equilibrium, are the same, having form:
\begin{shaded}
    \begin{subequations}
   \begin{align}
   \dfrac{dB}{dt} &= \lambda_{1,A} \cdot \Delta A + \lambda_{1,B} \cdot \Delta B \\
   \dfrac{dC}{dt} &= \lambda_{2,A} \cdot \Delta A + \lambda_{2,B} \cdot \Delta B  + \lambda_{2,C} \cdot \Delta C
   \end{align}
   \label{ffl_form}
\end{subequations}
\end{shaded}
where $\lambda_{1,A}$ to $\lambda_{2,C}$ are all real constants. The associated $\lambda$ terms for each of FFL motif under each logic assumption is summarised in Table \ref{tab:ffl_ORAND_lctiODE}, with the following definitions: 

\begin{alignat}{2}
    &\hspace{1.6em} \frac{A_e^{n_1-1} \alpha_1 k_1^{n_1} n_1}{\left(A_e^{n_1}+k_1^{n_1}\right)^2}
    &&= m_1\left(A_e, \alpha_1, k_1, n_1\right) \hspace{3.4em} \in \mathbb{R}^{+} 
    \\
    &\frac{A_e^{n_{2,A}-1} \alpha_{2,A} k_{2,A}^{n_{2,A}} n_{2,A}}{\left(A_e^{n_{2,A}}+k_{2,A}^{n_{2,A}}\right)^2}
    &&= m_{2,A}\left(A_e, \alpha_{2,A}, k_{2,A}, n_{2,A}\right) \in \mathbb{R}^{+} 
    \\
    &\frac{B_e^{n_{2,B}-1} \alpha_{2,B} k_{2,B}^{n_{2,B}} n_{2,B}}{\left(B_e^{n_{2,B}}+k_{2,B}^{n_{2,B}}\right)^2}
    &&= m_{2,B}\left(B_e, \alpha_{2,B}, k_{2,B}, n_{2,B}\right) \in \mathbb{R}^{+}
\end{alignat}

\begin{table}[ht]
    \renewcommand{\arraystretch}{2.3}
    \centering
    \resizebox{\textwidth}{!}{ 
    \begin{tabular}{|c|c|c|c|c|c|c|c|}
    \hline \textbf{Motif} & $\lambda_{1, A}$ & $\lambda_{1, B}$ & \makecell{$\lambda_{2, A}$\\(OR CASE)} & \makecell{$\lambda_{2, B}$\\(OR CASE)} & \makecell{$\lambda_{2, A}$\\(AND CASE)} & \makecell{$\lambda_{2, B}$\\(AND CASE)} & $\lambda_{2, C}$ \\
    \hline \hline
    \makecell{Coherent\\type 1} 
    & $+m_1$ 
    & $-\gamma_1$ 
    & $+m_{2, A}$ 
    & $+m_{2, B}$ 
    & $+m_{2, A} \cdot \alpha_{2,B} \dfrac{B_e^{n_{2,B}}}{B_e^{n_{2,B}}+k_{2,B}^{n_{2,B}}}$ 
    & $+\alpha_{2,A} \dfrac{A_e^{n_{2,A}}}{A_e^{n_{2,A}}+k_{2,A}^{n_{2,A}}} \cdot m_{2, B}$ 
    & $-\gamma_2$ \\
    \hline \makecell{Coherent\\type 2} 
    & $-m_1$ 
    & $-\gamma_1$ 
    & $-m_{2, A}$ 
    & $+m_{2, B}$ 
    & $-m_{2, A} \cdot \alpha_{2,B} \dfrac{B_e^{n_{2,B}}}{B_e^{n_{2,B}}+k_{2,B}^{n_{2,B}}}$ 
    & $+\alpha_{2,A} \dfrac{k_{2,A}^{n_{2,A}}}{A_e^{n_{2,A}}+k_{2,A}^{n_{2,A}}} \cdot m_{2, B}$ 
    & $-\gamma_2$ \\
    \hline \makecell{Coherent\\type 3} 
    & $+m_1$ 
    & $-\gamma_1$ 
    & $-m_{2, A}$ 
    & $-m_{2, B}$ 
    & $-m_{2, A} \cdot \alpha_{2,B} \dfrac{k_{2,B}^{n_{2,B}}}{B_e^{n_{2,B}}+k_{2,B}^{n_{2,B}}}$ 
    & $-\alpha_{2,A} \dfrac{k_{2,A}^{n_{2,A}}}{A_e^{n_{2,A}}+k_{2,A}^{n_{2,A}}} \cdot m_{2, B}$ 
    & $-\gamma_2$ \\
    \hline \makecell{Coherent\\type 4} 
    & $-m_1$ 
    & $-\gamma_1$ 
    & $+m_{2, A}$ 
    & $-m_{2, B}$ 
    & $+m_{2, A} \cdot \alpha_{2,B} \dfrac{k_{2,B}^{n_{2,B}}}{B_e^{n_{2,B}}+k_{2_B}^{n_{2,B}}}$ 
    & $-\alpha_{2,A} \dfrac{A_e^{n_{2,A}}}{A_e^{n_{2,A}}+k_{2,A}^{n_{2,A}}} \cdot m_{2, B}$ 
    & $-\gamma_2$ \\
    \hline \makecell{Incoherent\\type 1} 
    & $+m_1$ 
    & $-\gamma_1$ 
    & $+m_{2, A}$ 
    & $-m_{2, B}$ 
    & $+m_{2, A} \cdot \alpha_{2,B} \dfrac{k_{2,B}^{n_{2,B}}}{B_e^{n_{2,B}}+k_{2,B}^{n_{2,B}}}$ 
    & $-\alpha_{2,A} \dfrac{A_e^{n_{2,A}}}{A_e^{n_{2,A}}+k_{2,A}^{n_{2,A}}} \cdot m_{2, B}$ 
    & $-\gamma_2$ \\
    \hline \makecell{Incoherent\\type 2} 
    & $-m_1$ 
    & $-\gamma_1$ 
    & $-m_{2, A}$ 
    & $-m_{2, B}$ 
    & $-m_{2, A} \cdot \alpha_{2,B} \dfrac{k_{2,B}^{n_{2,B}}}{B_e^{n_{2,B}}+k_{2,B}^{n_{2,B}}}$ 
    & $-\alpha_{2,A} \dfrac{k_{2,A}^{n_{2,A}}}{A_e^{n_{2,A}}+k_{2,A}^{n_{2,A}}} \cdot m_{2, B}$ 
    & $-\gamma_2$ \\
    \hline \makecell{Incoherent\\type 3} 
    & $+m_1$ 
    & $-\gamma_1$ 
    & $-m_{2, A}$ 
    & $+m_{2, B}$ 
    & $-m_{2, A} \cdot \alpha_{2,B} \dfrac{B_e^{n_{2,B}}}{B_e^{n_{2,B}}+k_{2,B}^{n_{2,B}}}$ 
    & $+\alpha_{2,A} \dfrac{k_{2,A}^{n_{2,A}}}{A_e^{n_{2,A}}+k_{2,A}^{n_{2,A}}} \cdot m_{2, B}$ 
    & $-\gamma_2$ \\
    \hline \makecell{Incoherent\\type 4} 
    & $-m_1$ 
    & $-\gamma_1$ 
    & $+m_{2, A}$ 
    & $+m_{2, B}$ 
    & $+m_{2, A} \cdot \alpha_{2,B} \dfrac{B_e^{n_{2,B}}}{B_e^{n_{2,B}}+k_{2,B}^{n_{2,B}}}$ 
    & $+\alpha_{2,A} \dfrac{A_e^{n_{2,A}}}{A_e^{n_{2,A}}+k_{2,A}^{n_{2,A}}} \cdot m_{2, B}$ 
    & $-\gamma_2$ \\
    \hline
    \end{tabular}
    }
    \vspace{3pt}
    \caption{FFL lctiODE coefficients for both individual and mutual dependcy cases of C on A and B.}
    \label{tab:ffl_ORAND_lctiODE}
\end{table}

\subsubsection{Transfer function}
\begin{figure}
    \centering
    \includegraphics[width=0.75\linewidth]{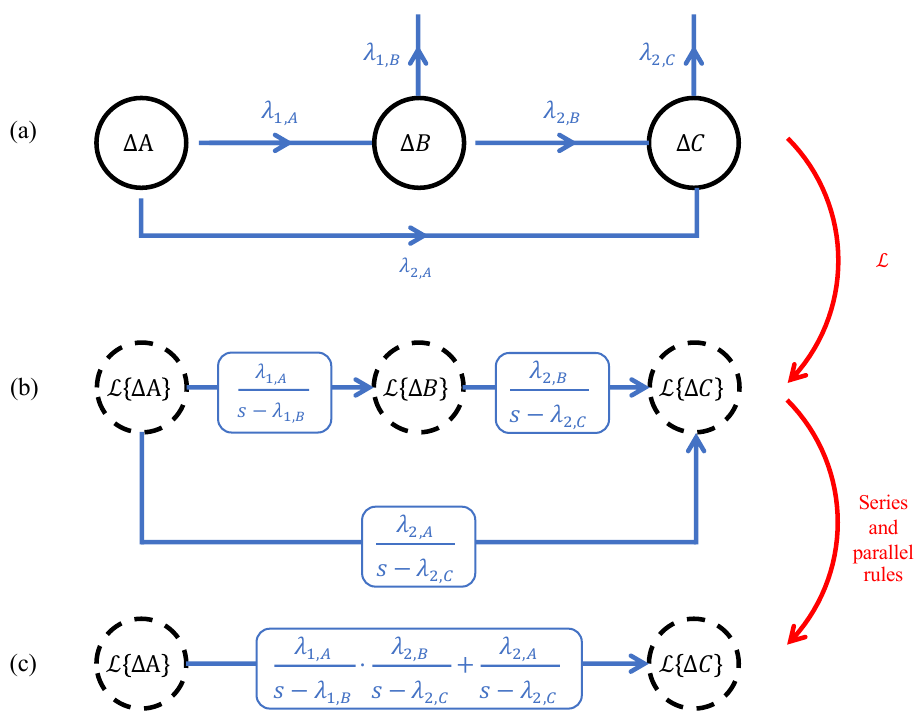}
    \caption{(a) Reaction network at equilibrium. (b) Laplace transform of network. (c) Reduced network using block diagram algebra.}
    \label{fig:FFL_schema}
\end{figure}

We again derive the FFL transfer functions visually. Beginning with our FFL lctiODEs in \eqref{eq:ffl_lctiode} --- the definition of the FFL architecture at equilibrium --- we can draw its biochemical network in Figure \ref{fig:FFL_schema}(a). The visual Laplace transform then produces the new picture in panel (b) that collapses, via block diagram algebra, into the transfer function of interest in panel (c). The transfer function between species A and C of all eight FFL motif types is, therefore:

\begin{shaded}
\begin{equation} \label{eq:FFL_tf}
    TF_{\text{FFL}} 
    = 
    \frac{\lambda_{1, A}\lambda_{2, B}+\lambda_{2, A}(s-\lambda_{1, B})}{(s-\lambda_{1, B})(s-\lambda_{2, C})} 
    = 
    \lambda_{2, A} 
    \cdot 
    \frac{s-\left( \lambda_{1, B} - \dfrac{\lambda_{1, A}\lambda_{2, B}}{\lambda_{2, A}}\right)}{\left(s-\lambda_{1, B}\right)\left(s-\lambda_{2, C}\right)}
    \text{ .}
\end{equation}
\end{shaded}

where $\lambda_{1, A},\lambda_{1, B},\lambda_{2, A},\lambda_{2, B}\in\mathbb{R}$. The evaluated expressions for each FFL type under the OR and AND assumptions are given in Table \eqref{tab:ffl_transfer_subs}.

\begin{table}[ht]
    \renewcommand{\arraystretch}{3.25}
    \centering
    \resizebox{\textwidth}{!}{
    \begin{tabular}{|c|c|c|}
    \hline
    \textbf{Motif} & \textbf{\makecell{Transfer Function\\(OR CASE)}} & \textbf{\makecell{Transfer Function\\(AND CASE)}} \\
    \hline\hline
    \makecell{Coherent\\type 1} 
    & \makecell{
    $ +\dfrac{m_1m_{2,B} + m_{2,A}(s + \gamma_1)}{(s + \gamma_1)(s + \gamma_2)}$ }
    & \makecell{
    $ +\dfrac{m_1 m_{2,B} \alpha_{2,A} \dfrac{A_e^{n_{2,A}}}{A_e^{n_{2,A}}+k_{2,A}^{n_{2,A}}} +  m_{2,A} \alpha_{2,B} \dfrac{B_e^{n_{2,B}}}{B_e^{n_{2,B}}+k_{2,B}^{n_{2,B}}} (s + \gamma_1)}{(s + \gamma_1)(s + \gamma_2)}$}\\
    \hline
    \makecell{Coherent\\type 2} 
    & \makecell{
    $ -\dfrac{m_1m_{2,B} + m_{2,A}(s + \gamma_1)}{(s + \gamma_1)(s + \gamma_2)}$ }
    & \makecell{
    $ -\dfrac{m_1 m_{2,B} \alpha_{2,A} \dfrac{k_{2,A}^{n_{2,A}}}{A_e^{n_{2,A}}+k_{2,A}^{n_{2,A}}} + m_{2,A} \alpha_{2,B} \dfrac{B_e^{n_{2,B}}}{B_e^{n_{2,B}}+k_{2,B}^{n_{2,B}}} (s + \gamma_1)}{(s + \gamma_1)(s + \gamma_2)}$}\\
    \hline
    \makecell{Coherent\\type 3} 
    & \makecell{
    $ -\dfrac{m_1m_{2,B} + m_{2,A}(s + \gamma_1)}{(s + \gamma_1)(s + \gamma_2)}$ }
    & \makecell{
    $ -\dfrac{m_1 m_{2,B} \alpha_{2,A} \dfrac{k_{2,A}^{n_{2,A}}}{A_e^{n_{2,A}}+k_{2,A}^{n_{2,A}}}  + m_{2,A} \alpha_{2,B} \dfrac{k_{2,B}^{n_{2,B}}}{B_e^{n_{2,B}}+k_{2,B}^{n_{2,B}}} (s + \gamma_1)}{(s + \gamma_1)(s + \gamma_2)}$}\\
    \hline
    \makecell{Coherent\\type 4} 
    & \makecell{
    $\displaystyle +\frac{m_1m_{2,B} + m_{2,A}(s + \gamma_1)}{(s + \gamma_1)(s + \gamma_2)}$ }
    & \makecell{
    $ + \dfrac{m_1 m_{2,B} \alpha_{2,A} \dfrac{A_e^{n_{2,A}}}{A_e^{n_{2,A}}+k_{2,A}^{n_{2,A}}} + m_{2,A} \alpha_{2,B} \dfrac{k_{2,B}^{n_{2,B}}}{B_e^{n_{2,B}}+k_{2,B}^{n_{2,B}}} (s + \gamma_1)}{(s + \gamma_1)(s + \gamma_2)}$}\\
    \hline
    \makecell{Incoherent\\type 1} 
    & \makecell{
    $ -\dfrac{m_1m_{2,B} - m_{2,A}(s + \gamma_1)}{(s + \gamma_1)(s + \gamma_2)}$ }
    & \makecell{
    $ - \dfrac{m_1 m_{2,B} \alpha_{2,A} \dfrac{A_e^{n_{2,A}}}{A_e^{n_{2,A}}+k_{2,A}^{n_{2,A}}} - m_{2,A} \cdot \alpha_{2,B} \dfrac{k_{2,B}^{n_{2,B}}}{B_e^{n_{2,B}}+k_{2,B}^{n_{2,B}}} (s + \gamma_1)}{(s + \gamma_1)(s + \gamma_2)}$}\\
    \hline
    \makecell{Incoherent\\type 2} 
    & \makecell{
    $ +\dfrac{m_1m_{2,B} - m_{2,A}(s + \gamma_1)}{(s + \gamma_1)(s + \gamma_2)}$ }
    & \makecell{
    $ + \dfrac{m_1 m_{2,B} \alpha_{2,A} \dfrac{k_{2,A}^{n_{2,A}}}{A_e^{n_{2,A}}+k_{2,A}^{n_{2,A}}}  - m_{2,A} \alpha_{2,B} \dfrac{k_{2,B}^{n_{2,B}}}{B_e^{n_{2,B}}+k_{2,B}^{n_{2,B}}} (s + \gamma_1)}{(s + \gamma_1)(s + \gamma_2)}$}\\
    \hline
    \makecell{Incoherent\\type 3} 
    & \makecell{
    $ +\dfrac{m_1m_{2,B} - m_{2,A}(s + \gamma_1)}{(s + \gamma_1)(s + \gamma_2)}$ } 
    & \makecell{
    $ +\dfrac{m_1 m_{2,B} \alpha_{2,A} \dfrac{k_{2,A}^{n_{2,A}}}{A_e^{n_{2,A}}+k_{2,A}^{n_{2,A}}} - m_{2,A} \alpha_{2,B} \dfrac{B_e^{n_{2,B}}}{B_e^{n_{2,B}}+k_{2,B}^{n_{2,B}}} (s + \gamma_1)}{(s + \gamma_1)(s + \gamma_2)}$}\\
    \hline
    \makecell{Incoherent\\type 4} 
    & \makecell{
    $ -\dfrac{m_1m_{2,B} - m_{2,A}(s + \gamma_1)}{(s + \gamma_1)(s + \gamma_2)}$ }
    & \makecell{
    $ -\dfrac{m_1 m_{2,B} \alpha_{2,A} \dfrac{A_e^{n_{2,A}}}{A_e^{n_{2,A}}+k_{2,A}^{n_{2,A}}} -  m_{2,A} \alpha_{2,B} \dfrac{B_e^{n_{2,B}}}{B_e^{n_{2,B}}+k_{2,B}^{n_{2,B}}} (s + \gamma_1)}{(s + \gamma_1)(s + \gamma_2)}$}\\
    \hline
    \end{tabular}
    }
    \vspace{3pt}
    \caption{Substituted transfer functions for each FFL motif type in both OR and AND logic cases.}
    \label{tab:ffl_transfer_subs}
\end{table}

\subsubsection{Motif stability and asymptotic response}
From Equation \eqref{eq:FFL_tf}, the transfer function contains two real valued poles at $s=\lambda_{1,B},\lambda_{2,C}=-\gamma_1,-\gamma_2$ --- which are both non-positive real numbers. As such, the system is asymptotically stable and does not possess resonant behaviour but, instead, exhibits break frequencies at these two pole values.

\subsubsection{Frequency response functions}
Previously, the transfer functions we analysed were relatively simple, with monotonic magnitude and phase responses and only a small number of parameters and model variants. This made it feasible to classify their behaviour across all frequencies by inspecting the low- and high-frequency limits. In contrast, the FFL architecture comprises 16 transfer function variants --- some of which exhibit qualitative changes in both asymptotic and intermediate behaviours depending on parameter choices. Given this added complexity, we now shift our focus to deriving general analytical expressions and identifying key behaviours characteristic of the architecture, rather than attempting to exhaustively catalogue all possible dynamics. 

$TF_{\text{FFL}}$ contains one zero at $z_1=\left|\lambda_{1,B}-\lambda_{1,A}\lambda_{2,B}/\lambda_{2,A}\right|$, and two poles at $p_1=\lambda_{1,B}$, $p_2=\lambda_{2,C}$. Recall from Section \ref{sec:breakfreq} that zeros mark the frequency at which the system gains an order of frequency dependence, while poles mark those at which an order is lost. If $z_1<p_1<p_2$ then our the frequency dependencies go from $\omega^0$ (at $\omega=0$) to $\omega^1$ (at $\omega=z_1$) to $\omega^0$ (at $\omega=p_1$) to $\omega^{-1}$ (at $\omega=p_2$), creating a local maximum of amplitude modulation at $\omega=p_1$. This is sometimes called a ``low pass filter with resonance''. Within the window $z_1<\omega<p_2$, there is an effective local band-pass. If $p_1<z_1<p_2$ then our the frequency dependencies go from $\omega^0$ (at $\omega=0$) to $\omega^{-1}$ (at $\omega=p_1$) to $\omega^0$ (at $\omega=z_1$) to $\omega^{-1}$ (at $\omega=p_2$), creating a staircase-like profile. If $p_1<p_2<z_1$ then our the frequency dependencies go from $\omega^0$ (at $\omega=0$) to $\omega^{-1}$ (at $\omega=p_1$) to $\omega^-2$ (at $\omega=p_2$) to $\omega^{-1}$ (at $\omega=z_1$), creating a decay with inflection at $\omega=z_1$. These amplitude modulation profiles were previously inaccessible by previous motifs due to the absence of the zero, which is a manifestation of biochemical signal interference between the $A\rightarrow C$ and $A\rightarrow C$ branches of regulation local to equilibrium. These results are summarised in Table \ref{tab:ffl_breaks}.

\begin{table}[ht]
\renewcommand{\arraystretch}{1.8}
\centering
\begin{tabular}{|c|c|c|}
\hline
\textbf{Ordering} & \textbf{Frequency Dependence of Amplitude Modulation} & \textbf{Behaviour} \\
\hline
$z_1 < p_1 < p_2$ & 
\makecell{
$\omega^{+1}$ at $\omega = z_1$ , $\omega^0$ at $\omega = p_1$ , $\omega^{-1}$ at $\omega = p_2$
} & 
Local band-pass \\
\hline
$p_1 < z_1 < p_2$ & 
\makecell{
$\omega^{-1}$ at $\omega = p_1$ , $\omega^0$ at $\omega = z_1$ , $\omega^{-1}$ at $\omega = p_2$
} & 
Staircase filtering \\
\hline
$p_1 < p_2 < z_1$ & 
\makecell{
$\omega^{-1}$ at $\omega = p_1$ , $\omega^{-2}$ at $\omega = p_2$ , $\omega^{-1}$ at $\omega = z_1$
} & 
Decay with inflection \\
\hline
\end{tabular}
\vspace{3pt}
\caption{Amplitude modulation profiles for different orderings of zero \( z_1 \) and poles \( p_1, p_2 \).}
\label{tab:ffl_breaks}
\end{table}

The general amplitude modulation and phase shift of FFLs are given by:

\begin{subequations}
    \begin{gather}
        M(\omega) = 
        \left|\lambda_{2, A}\right| \cdot 
        \frac{ \left|\lambda_{1,B}-\dfrac{\lambda_{1,A}\lambda_{2,B}}{\lambda_{2,A}}\right|}
        {\left(\sqrt{\left(\lambda_{1, B}\right)^2+\omega^2}\right)\left(\sqrt{\left(\lambda_{2, C}\right)^2+\omega^2}\right)}
        \\
        \phi(\omega) = 
        \arg\Bigg(\lambda_{2, A}\Bigg) + \arctantwo\Bigg(\omega,-\lambda_{1,B}+\dfrac{\lambda_{1,A}\lambda_{2,B}}{\lambda_{2,A}}\Bigg)- \arctantwo\Bigg(\omega,-\lambda_{1, B}\Bigg)
        - \arctantwo\Bigg(\omega,-\lambda_{2, C}\Bigg)
    \end{gather}
\end{subequations}

\subsection{Collective picture of FFLs}
For the set of coherent and set of incoherent FFLs under OR logic, one can construct a cube analogous to Figure \eqref{fig:Cas_collective} each. In each cube, type 1 and 4 motifs occupy opposite corners, while type 2 and 3 motifs occupy the remaining opposite corners. The same cannot be done, however, for our set of coherent and set of incoherent FFLs under AND logic. \textbf{This contrast highlights a design advantage of using OR logic interactions in achieving modularity at a scale smaller than a full FFL unit.}

%% file: 4d_Autoregulation.tex
\subsection{Autoregulation} \label{Sec:autoreg}
Here we encounter the first member of a larger family of architectures, known as feedback loops. A feedback loop that consists of only a single species is known as an autoregulation --- abbreviated AR from here on. Before proceeding, it is important to clarify what we mean by ``feedback loop'', and hence AR. For readers new to the intersection of biology and control theory, the terminology in the literature can often be inconsistent or imprecise. To remove ambiguity, we will classify feedback loops into two distinct forms: autonomous and forced. 

A \textbf{feedback loop} is a collection of two sets: a set of biochemical species, and a set of directed interactions among them such that there exists at least one directed path that begins and ends on the same species. In other words, the only requirement is that at least one species ultimately influences its own dynamics through a chain of interactions. This is why, in biological literature, circuits that are not strictly single cyclic loops are nevertheless referred to as ``feedback loops''. A circuit may encompass multiple interconnected loops or contain interaction paths branching out from a central loop, yet still be called a ``feedback loops''. By convention, the nomenclature typically identifies one cyclic path of interest and appends the term ``feedback loop'', even though a broader network of interactions is involved. 

An \textbf{autonomous feedback loop} is a feedback loop in which the set of directed interactions do not allow for any pure sources to exist. The system’s population dynamics arise solely from these internal interactions, with no driving inputs from the external environment. In the language of ctiODEs, such a system is described by:

\begin{equation}
\frac{dx_i}{dt} = F_i\big(x_1(t), x_2(t), \dots, x_n(t)\big),
\label{autonFB}
\end{equation}

where there exists a sequence $(i_0, i_1, \dots, i_m)$ with $i_m = i_0$ satisfying

\begin{equation}
    \frac{\partial F_{i_{j+1}}}{\partial x_{i_j}} \neq 0
\quad , \quad \forall j \text{ .}
\label{loopcondition}
\end{equation}

Equation \eqref{autonFB} represents a closed system, with the additional condition that at least one interaction path begins and ends at the same species.

A \textbf{forced feedback loop} is an autonomous feedback loop where one species within the cyclic loop is regulated by an external species. Without loss of generality, let us denote this external species $u_1$ and define the internal species coupled to this external source $x_1$. The ctiODE description then becomes:
\begin{equation}
    \begin{aligned}
    \frac{dx_1}{dt} &= F_1(u_1,x_1,x_2,...,x_n) \\
    \frac{dx_i}{dt} &= F_i(x_1,x_2,...,x_n) \quad , \quad x>1
    \end{aligned}
    \label{forcedFB}
\end{equation}
such that there exists a sequence $(i_0,i_1,...,i_m)$ with $i_m=i_0=1$ that satisfies condition \eqref{loopcondition}.

In biological literature, the autonomous definition predominates, whereas in control-oriented treatments of biology, the forced definition is more common \cite{elowitz_biological_nodate,shin_linear_2010}. This is because, in control theory, a ``feedback loop'' refers to a very specific system architecture that necessarily involves an external input. In the biochemical context, this external input translates into the presence of an upstream species regulating one within the loop --- the same architecture captured by our definition of ``forced'' feedback loops in Equation \eqref{forcedFB}. This subtle difference in definitions between the two fields can create confusion when attempting to assign dynamical descriptions to architecture names. In this section, we focus on the control-theoretic formulation --- \textbf{forced AR}. For a complementary treatment of the autonomous counterpart, we refer the reader to Appendix B; which highlights why transfer functions do not exist in the traditional sense for autonomous ARs, and how the closest we can get to such a quantity is by introducing the idea of an ``intrinsic transfer function''. \textit{By introducing such a quantity, modularity of autonomous feedback loops becomes permissible in what was once not allowed.}

\subsubsection{General architecture}
In forced AR (fAR), the general dynamics of an autoregulated species $A$ obey:

\begin{equation} \label{AR_forced_general}
    \frac{dA}{dt} = f_1(A') + f_2(A) - g_1(A) \text{ ,}
\end{equation}

where the primed symbol $A'$ denotes an arbitrary upstream species to $A$. This naming convention has been chosen solely to maintain consistency with our discussion of autonomous AR later in Appendix B. $f_1$ and $f_2$ describe the production rates arising from upstream forcing and autoregulatory mechanisms, respectively, while $g_1$ represents the population's loss rate. 

\begin{figure}[ht]
    \centering
    \includegraphics[scale=0.7]{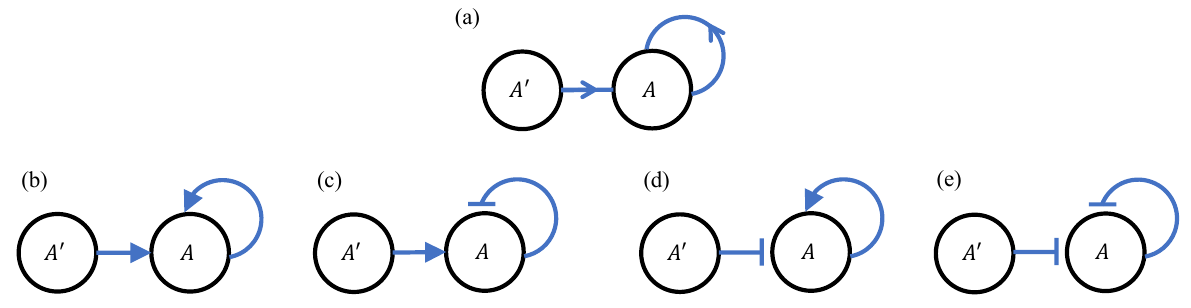}
    \caption{General architecture of forced AR (a) and its four special cases: positively forced positive AR (b), positively forced negative AR (c), negatively forced positive AR (d) and negatively forced negative AR (e).}
    \label{fig:AR_archs_forced}
\end{figure}

We depict Equation \eqref{AR_forced_general} with a target node, $A$, receiving two inputs: one from its upstream node $A'$ and the other from itself (Figure \ref{fig:AR_archs_forced}a). Like all interactions up till now, both the upstream forcing and autoregulatory mechanisms can operate in one of two modes: promotion or repression, denoted by a positive ($+$) and negative ($-$) sign, respectively. This yields four distinct realisations of our forced AR architecture: positively-forced positive-AR (f$^+$AR$^+$), positively-forced negative-AR (f$^+$AR$^-$), negatively-forced positive-AR (f$^-$AR$^+$) and negatively-forced negative-AR (f$^-$AR$^-$); illustrated in Figures \ref{fig:AR_archs_forced}b to e, respectively.

\subsubsection{Dynamics in time: a common model} \label{sec:forcedAR_model}
Assuming positive Hill kinetics for promotion and negative Hill kinetics for repression, the functional form of each rate in Equation \eqref{AR_forced_general} for each architecture variant is given in Table \ref{tab:forced_AR_terms}.

\begin{table}[ht]
\renewcommand{\arraystretch}{2}
\centering
\begin{tabular}{|c|c|c|c|}
\hline
\textbf{Architecture} & $f_1(A')$ & $f_2(A)$ & $g_1(A)$ \\
\hline \hline
\makecell{Positively-forced\\positive-AR} 
& $\alpha_1 \dfrac{{A'}^{n_1}}{{A'}^{n_1} + k_1^{n_1}}$
& $\alpha_2 \dfrac{A^{n_2}}{A^{n_2} + k_2^{n_2}}$
& $\gamma A$ \\
\hline
\makecell{Positively-forced\\negative-AR} 
& $\alpha_1 \dfrac{{A'}^{n_1}}{{A'}^{n_1} + k_1^{n_1}}$
& $\alpha_2 \dfrac{k_2^{n_2}}{A^{n_2} + k_2^{n_2}}$
& $\gamma A$ \\
\hline
\makecell{Negatively-forced\\positive-AR} 
& $\alpha_1 \dfrac{k_1^{n_1}}{{A'}^{n_1} + k_1^{n_1}}$
& $\alpha_2 \dfrac{A^{n_2}}{A^{n_2} + k_2^{n_2}}$
& $\gamma A$ \\
\hline
\makecell{Negatively-forced\\negative-AR} 
& $\alpha_1 \dfrac{k_1^{n_1}}{{A'}^{n_1} + k_1^{n_1}}$
& $\alpha_2 \dfrac{k_2^{n_2}}{A^{n_2} + k_2^{n_2}}$
& $\gamma A$ \\
\hline
\end{tabular}
\vspace{3pt}
\caption{Production and degradation terms for the four forced autoregulation architectures. All $\alpha, k, n$ terms are positive-real, and $\gamma$ terms are nonnegative-real.}
\label{tab:forced_AR_terms}
\end{table}

\subsubsection{Dynamics in time: general near-equilibrium properties}
Near equilibrium $(A'_e,A_e)=\overline{\textbf{e}}$, the dynamics of our general forced AR architecture in Equation \eqref{AR_forced_general} is given converges into:

\begin{shaded}
\begin{equation}
    \frac{d \Delta A}{d t} \sim \left.\frac{\partial f_1}{\partial A'}\right|_{\overline{\textbf{e}}} \cdot \Delta A' + \left( \left.\frac{\partial f_2}{\partial A}\right|_{\overline{\textbf{e}}} -\left.\frac{\partial g_1}{\partial A}\right|_{\overline{\textbf{e}}} \right) \cdot \Delta A\text{ .}
\label{eq:ar_forced_taylor1}
\end{equation}

Equation \eqref{eq:ar_forced_taylor1} tells us that, at equilibrium, the perturbation speed of species $A$ in a forced AR motif is simply the a linear combination of two physical quantities: the perturbation of its upstream precursor and the perturbation of itself at that point in time. 
\end{shaded}

Returning to the representative models for each forced AR variant in Table \ref{tab:forced_AR_terms}, their corresponding coefficients near equilibrium are given in Table \ref{tab:forced_AR_coeff}.

\begin{table}[ht]
\renewcommand{\arraystretch}{2.5}
\centering
\begin{tabular}{|c|c|c|c|}
\hline
\textbf{Architecture} & 
$\left.\dfrac{\partial f_1}{\partial A'}\right|_{\overline{\mathbf{e}}}$ & 
$\left.\dfrac{\partial f_2}{\partial A}\right|_{\overline{\mathbf{e}}}$ & 
$\left.\dfrac{\partial g_1}{\partial A}\right|_{\overline{\mathbf{e}}}$ \\
\hline \hline
\makecell{Positively-forced\\positive-AR} 
& $\displaystyle
+ \frac{(A'_e)^{n_1 - 1} \,\alpha_1 \,k_1^{n_1} \,n_1}{\Big((A'_e)^{n_1} + k_1^{n_1}\Big)^2}$
& $\displaystyle
+ \frac{(A_e)^{n_2 - 1} \,\alpha_2 \,k_2^{n_2} \,n_2}{\Big((A_e)^{n_2} + k_2^{n_2}\Big)^2}$
& $\gamma$ \\
\hline
\makecell{Positively-forced\\negative-AR} 
& $\displaystyle
+ \frac{(A'_e)^{n_1 - 1} \,\alpha_1 \,k_1^{n_1} \,n_1}{\Big((A'_e)^{n_1} + k_1^{n_1}\Big)^2}$
& $\displaystyle
- \frac{(A_e)^{n_2 - 1} \,\alpha_2 \,k_2^{n_2} \,n_2}{\Big((A_e)^{n_2} + k_2^{n_2}\Big)^2}$
& $\gamma$ \\
\hline
\makecell{Negatively-forced\\positive-AR} 
& $\displaystyle
- \frac{(A'_e)^{n_1 - 1} \,\alpha_1 \,k_1^{n_1} \,n_1}{\Big((A'_e)^{n_1} + k_1^{n_1}\Big)^2}$
& $\displaystyle
+ \frac{(A_e)^{n_2 - 1} \,\alpha_2 \,k_2^{n_2} \,n_2}{\Big((A_e)^{n_2} + k_2^{n_2}\Big)^2}$
& $\gamma$ \\
\hline
\makecell{Negatively-forced\\negative-AR} 
& $\displaystyle
- \frac{(A'_e)^{n_1 - 1} \,\alpha_1 \,k_1^{n_1} \,n_1}{\Big((A'_e)^{n_1} + k_1^{n_1}\Big)^2}$
& $\displaystyle
- \frac{(A_e)^{n_2 - 1} \,\alpha_2 \,k_2^{n_2} \,n_2}{\Big((A_e)^{n_2} + k_2^{n_2}\Big)^2}$
& $\gamma$ \\
\hline
\end{tabular}
\vspace{3pt}
\caption{lctiODE coefficients for all four forced AR architectures. All $\alpha,k,n$ terms are positive real, and $\gamma$ terms are nonnegative real.}
\label{tab:forced_AR_coeff}
\end{table}

Defining

\begin{equation*}
    \frac{(A'_e)^{n_1 - 1} \,\alpha_1 \,k_1^{n_1} \,n_1}{\Big((A'_e)^{n_1} + k_1^{n_1}\Big)^2} = m_1\left(A'_e, \alpha_1, k_1, n_1,\right) \in \mathbb{R}^{+} \quad \text{and} \quad \frac{(A_e)^{n_2 - 1} \,\alpha_2 \,k_2^{n_2} \,n_2}{\Big((A_e)^{n_2} + k_2^{n_2}\Big)^2} = m_2\left(A_e, \alpha_2, k_2, n_2,\right) \in \mathbb{R}^{+} \text{ ,}
\end{equation*}
\begin{shaded}
the linear dynamics of all four forced AR representative models can be summarised by the following relationship:
\begin{equation}
    \textbf{f$^{\textcolor{red}{\pm}}$AR$^{\textcolor{teal}{\pm}}$:} 
    \quad 
    \frac{d\Delta A}{dt} \sim \textcolor{red}{\pm} m_1 \Delta A' 
    + 
    (\textcolor{teal}{\pm} m_2 - \gamma)\Delta A
    \label{eq:f^+-AR^+-linear}
\end{equation}
\end{shaded}

\subsubsection{Transfer function}

Applying the Laplace transform to Equation \eqref{eq:ar_forced_taylor1} and noting zero initial fluctuation in $A$ from equilibrium, we can solve for the following transfer function:

\begin{equation}
    \text{TF}_{\text{fAR}}=\dfrac{\mathcal{L}\{\Delta A\}(s)}{\mathcal{L}\{\Delta A'\}(s)}
    = 
    \dfrac{\left.\dfrac{\partial f_1}{\partial A'}\right|_{\overline{\mathbf{e}}}}{s - \left.\dfrac{\partial f_2}{\partial A}\right|_{\overline{\textbf{e}}} +\left.\dfrac{\partial g_1}{\partial A}\right|_{\overline{\textbf{e}}}} \text{ .}
    \label{fAR_TFs_gen}
\end{equation}

Notice that this is shares the same $J_1/(s+J_2)$ transfer function structure of positive SR, with the only difference being that $J_2$ is now influenced by the value of the local autoregulatory rate $\partial f_2/\partial A|_{\overline{\textbf{e}}}$. The four forced AR models have transfer function:
\begin{shaded}
\begin{equation}
    \text{TF}_{\text{f}^{\textcolor{red}{\pm}} \text{AR}^{\textcolor{teal}{\pm}}} = \frac{{\textcolor{red}{\pm}} m_1}{s{\textcolor{teal}{\mp}} m_2 + \gamma} \text{ .}
    \label{fAR_TFs}
\end{equation}
\end{shaded}
Notice that, with a slight rearrangement of this equation, the closed loop rule from Table \ref{blockalg} reemerges:

\begin{equation}
    \text{TF}_{\text{f}^{\textcolor{red}{\pm}} \text{AR}^{\textcolor{teal}{\pm}}} = \dfrac{{\textcolor{red}{\pm}} \dfrac{m_1}{s+\gamma}}{1 {\textcolor{teal}{\mp}} \left( \dfrac{m_1}{s+\gamma} \right) \left( \dfrac{m_2}{m_1} \right)} \text{ .}
    \label{fAR_block}
\end{equation}

Equation \eqref{fAR_block} represents a closed loop consisting of a forward path with SR transfer function $\pm {m_1}/({s+\gamma})$ and a feedback path of constant strength ${m_2}/{m_1}$, corresponding to the ratio between local autoregulatory and local simple regulatory production rates. Consequently, in control-oriented treatments of systems biology, autoregulation is often defined as a closed feedback loop comprising a simple regulation branch and a constant feedback branch, as illustrated in the following Figure \ref{fig:forcedAR_block}. Since the forward SR branch implicitly assumes the presence of a driving species $A'$, one can see why a biological-oriented reader who is accustomed to the idea of autonomous autoregulation may find this confusing.

\begin{figure}[ht]
    \centering
    \includegraphics[width=0.4\linewidth]{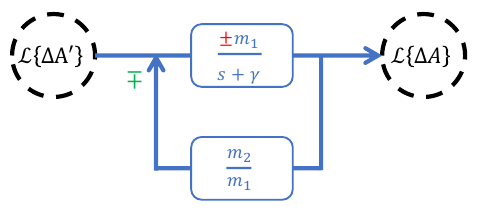}
    \caption{In control-oriented treatments of biology, autoregulation is often described as a closed feedback loop between a simple regulation branch and a constant feedback branch.}
    \label{fig:forcedAR_block}
\end{figure}

\subsubsection{Motif stability and asymptotic response}
From Equations \eqref{fAR_TFs_gen} and \eqref{fAR_TFs}, forced autoregulation possesses a single pole at frequency equal to

\begin{equation*}
    s = \left.\dfrac{\partial f_2}{\partial A}\right|_{\overline{\textbf{e}}} - \left.\dfrac{\partial g_1}{\partial A}\right|_{\overline{\textbf{e}}} = 
    \\
    \begin{cases}
        +m_2 - \gamma \quad, \quad \text{if AR mode is positive}\\
        -m_2 - \gamma \quad, \quad \text{if AR mode is negative}
    \end{cases}
\end{equation*}

Notice that this frequency value depends on our choice of autoregulatory mode and not forcing mode. The stability of forced autoregulation is independent of whether its upstream driver activates or represses the autonomous loop. 

Forced positive AR can exhibit all three stability modes. When self-production is slower than self-degradation ($\gamma > m_2$), degradation dominates and counteracts the driving force from $A'$, preventing unbounded growth and yielding \textit{asymptotic stability}. When the two self processes are exactly balanced ($\gamma = m_2$), their effects cancel, allowing any displacement of $A$ from equilibrium to persist at its displaced level, corresponding to \textit{marginal stability}. When self-production is faster than self-degradation ($\gamma < m_2$), only amplification effects remain, leading to unbounded growth and thus \textit{instability}.

\begin{center}
\begin{tabular}{ccc}
\toprule
\textbf{Condition} & \textbf{Pole Location} & \textbf{Stability of fAR$^+$ Motif} \\
\midrule
$\gamma > m_2$ & $s = m_2 - \gamma < 0$ & Asymptotically stable \\
$\gamma = m_2$ & $s = 0$ & Marginally stable \\
$\gamma < m_2$ & $s = m_2 - \gamma > 0$ & Unstable \\
\bottomrule
\end{tabular}
\end{center}

Forced negative AR can only be asymptotically stable. This is because the autoregulatory rate constant $-m_2$ is always negative, so it consistently acts as a counteracting flux opposing the drive from $A'$.

\subsubsection{Slow and high limit forcing response}
Taking the limits as $|s|\rightarrow0$ and $|s|\rightarrow\infty$, we can deduce the Table \ref{tab:AR_freq_response_forced}. 

\begin{table}[ht]
\renewcommand{\arraystretch}{2}
\centering
\begin{tabular}{|c|c|c|c|c|c|c|}
\hline
\textbf{Motif} & 
\textbf{\makecell{Slow Forcing\\Limit $s \to 0$}} & 
\makecell{Quasistatic\\Gain Limit} & 
\makecell{Low-Frequency\\Phase Shift\\Limit} & 
\textbf{\makecell{Rapid Forcing\\Limit $s \to \infty$}} & 
\makecell{Asymptotic\\Gain Limit} &
\makecell{High-Frequency\\Phase Shift\\Limit}\\
\hline
\textbf{\makecell{f$^+$AP$^+$}} 
& $\dfrac{+m_1}{\gamma-m_2} + 0i$ 
& $ \dfrac{m_1}{\left|\gamma-m_2\right|} $
& \makecell{$0^\circ$ (asym. stab.)
\\ $-90^\circ$ (marg. stab.)
\\ $-180^\circ$ (unstable)}
& $0 + 0i$ 
& $0$ 
& $-90^\circ$\\
\hline
\textbf{\makecell{f$^-$AP$^+$}} 
& $\dfrac{-m_1}{\gamma-m_2} + 0i$ 
& $ \dfrac{m_1}{\left|\gamma-m_2\right|} $
& \makecell{$180^\circ$ (asym. stab.)
\\ $90^\circ$ (marg. stab.)
\\ $0^\circ$ (unstable)}
& $0 + 0i$ 
& $0$ 
& $+90^\circ$\\
\hline
\textbf{\makecell{f$^+$AP$^-$}} 
& $\dfrac{+m_1}{\gamma+m_2} + 0i$ 
& $\dfrac{m_1}{\gamma+m_2}$ 
& $0^\circ$ 
& $0 + 0i$ 
& $0$ 
& $-90^\circ$\\
\hline
\textbf{\makecell{f$^-$AP$^-$}} 
& $\dfrac{-m_1}{\gamma+m_2} + 0i$ 
& $\dfrac{m_1}{\gamma+m_2}$ 
& $180^\circ$ 
& $0 + 0i$ 
& $0$ 
& $+90^\circ$\\
\hline
\end{tabular}
\vspace{3pt}
\caption{Frequency response characteristics of forced AR motifs in both low- and high-frequency forcing regimes.}
\label{tab:AR_freq_response_forced}
\end{table}

\subsubsection{Frequency response functions}
\begin{figure}[ht]
    \centering
    \includegraphics[width=1\linewidth]{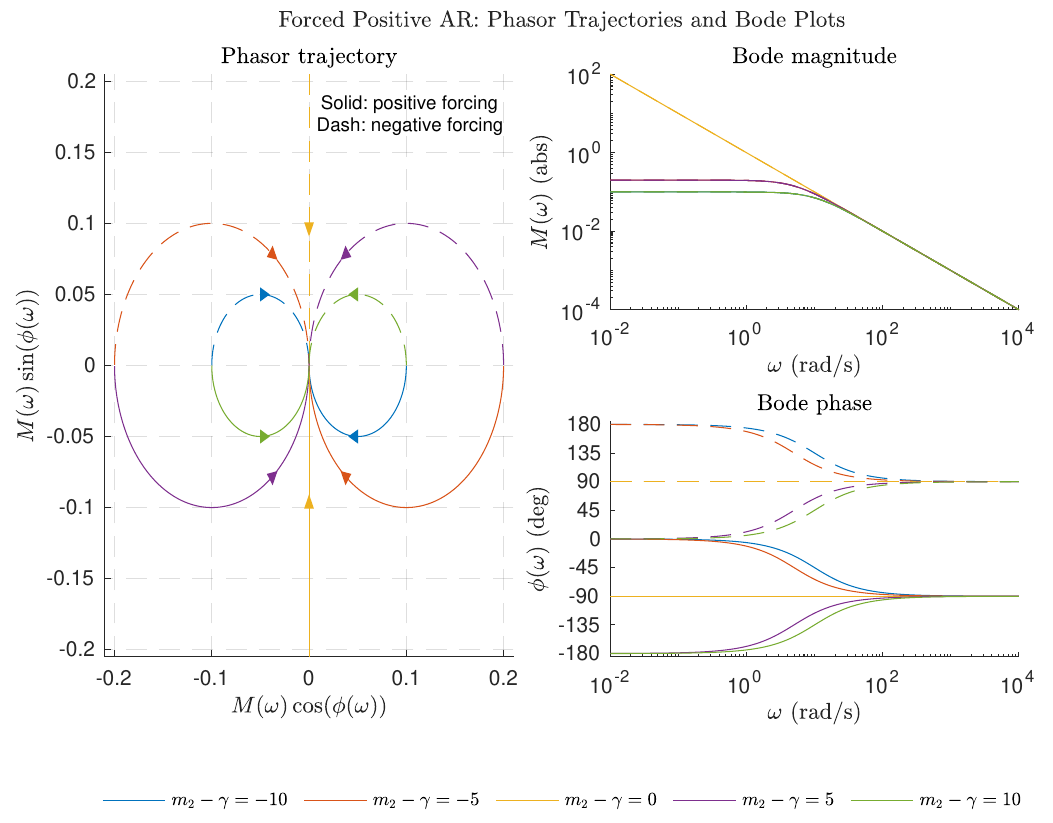}
    \caption{Phasor and Bode plots for forced AR, with positive and negative forcing modes denoted by solid and dashed lines, respectively. The blue-green and orange-purple pairs share identical modulus profiles, as seen in their Phasor trajectories, and so overlap each other when plotted as Bode magnitude curves. Without loss of generality, we assign $m_1=1$ for illustrative purposes.}
    \label{fig:placeholder}
\end{figure}

The analytical amplitude modulation and phase shift frequency-response functions are
\begin{subequations}
    \begin{align}
        M_{f^\pm AR^{+}}(\omega) &= 
        \frac{m_1}
        {\sqrt{\left( (\gamma-m_2)^2 +\omega^2 \right)}}
        \\
        \phi_{f^+ AR^{+}}(\omega) &= 
        - \arctantwo\left(\omega,\gamma-m_2\right)
        \\
        \phi_{f^- AR^{+}}(\omega) &= 
        180^\circ- \arctantwo\left(\omega,\gamma-m_2\right)
    \end{align}
\end{subequations}

for the pair of forced positive AR motifs, and 
\begin{subequations}
    \begin{align}
        M_{f^\pm AR^{-}}(\omega) &= 
        \frac{m_1}
        {\sqrt{\left( (\gamma+m_2)^2 +\omega^2 \right)}}
        \\
        \phi_{f^+ AR^{-}}(\omega) &= 
        - \arctantwo\left(\omega,\gamma+m_2\right)
        \\
        \phi_{f^- AR^{-}}(\omega) &= 
        180^\circ- \arctantwo\left(\omega,\gamma+m_2\right)
    \end{align}
\end{subequations}

for the pair of forced negative AR motifs. The corresponding phasor and Bode plots for positive autoregulation (AR) are shown in Figure \ref{fig:placeholder}. Note that, since the poles the two negative AR motifs are strictly non-negative, their frequency response functions are subsets of those for the two positive AR motifs when $\gamma \geq m_2$. All four motif architectures have a single break frequency equal to their poles.. These frequency response plots are shown in Figure \ref{fig:placeholder}.

%% file: 4e_Feedback.tex
\subsection{Double feedback loops}
In this section, we consider two-species feedback loops or double feedback loops (dFBLs) in the control-oriented formulation. For completeness, the autonomous counterparts are presented in Appendix C.

\subsubsection{General architecture}
\begin{figure} [ht]
    \centering
    \includegraphics[width=0.4\linewidth]{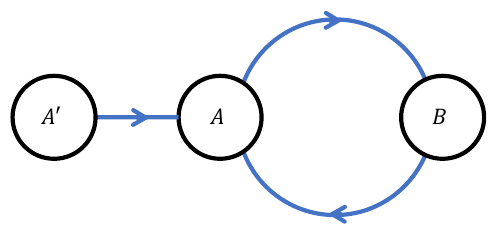}
    \caption{Forced two-species feedback loop architecture.}
    \label{fig:dFBL}
\end{figure}

Forced dFBLs (fdFBLs) are mutual regulations between two biochemical species $A$ and $B$, driven by an upstream source $A'$. This is described by Equation \eqref{dfbl_odes} and depicted in Figure \ref{fig:dFBL}.
\begin{subequations}
   \begin{align}
   \dfrac{dA}{dt} &= f_{A'}(A') + f_{B,A}(B) - g_A(A) \\
   \dfrac{dB}{dt} &= f_{A,B}(A) - g_B(B)
   \end{align}
   \label{dfbl_odes}
\end{subequations}

\subsubsection{Dynamics in time: a common model}
Each flux term in Equation \eqref{dfbl_odes} can take one of two regulation modes: promotion ($+$) or repression ($-$). We specify the regulation modes of an fFBL architecture in the order: $f_{A'}(A')$, $f_{A,B}(B)$, and $f_{B,A}(A)$. That is, a positively forced dFBL in which $A$ represses $B$ and $B$ promotes $A$ is denoted f$^{+}$dFBL$^{-+}$. A total of eight combinations are possible. Under the common Hill function assumptions, the forms of each of these terms in each motif variants are listed in Table \ref{tab:dfbl_motifs}.

\begin{table}[ht]
\renewcommand{\arraystretch}{2}
\centering
\begin{tabular}{|c||c|c|c|c|c|}
\hline
\textbf{Motif} 
& $f_{A'}(A')$ 
& $f_{B,A}(B)$ 
& $g_1(A)$ 
& $f_{A,B}(A)$ 
& $g_2(B)$ \\
\hline\hline
f$^{+}$FBL$^{++}$ 
& $H^+(\alpha_{A'},k_{A'},n_{A'};A')$ 
& $H^+(\alpha_{B,A},k_{B,A},n_{B,A};B)$ 
& $\gamma_A A$ 
& $H^+(\alpha_{A,B},k_{A,B},n_{A,B};A)$ 
& $\gamma_B B$ \\
\hline
f$^{+}$FBL$^{--}$ 
& $H^+(\alpha_{A'},k_{A'},n_{A'};A')$ 
& $H^-(\alpha_{B,A},k_{B,A},n_{B,A};B)$ 
& $\gamma_A A$ 
& $H^-(\alpha_{A,B},k_{A,B},n_{A,B};A)$ 
& $\gamma_B B$ \\
\hline
f$^{+}$FBL$^{+-}$ 
& $H^+(\alpha_{A'},k_{A'},n_{A'};A')$ 
& $H^+(\alpha_{B,A},k_{B,A},n_{B,A};B)$ 
& $\gamma_A A$ 
& $H^-(\alpha_{A,B},k_{A,B},n_{A,B};A)$ 
& $\gamma_B B$ \\
\hline
f$^{+}$FBL$^{-+}$ 
& $H^+(\alpha_{A'},k_{A'},n_{A'};A')$ 
& $H^-(\alpha_{B,A},k_{B,A},n_{B,A};B)$ 
& $\gamma_A A$ 
& $H^+(\alpha_{A,B},k_{A,B},n_{A,B};A)$ 
& $\gamma_B B$ \\
\hline
f$^{-}$FBL$^{++}$ 
& $H^-(\alpha_{A'},k_{A'},n_{A'};A')$ 
& $H^+(\alpha_{B,A},k_{B,A},n_{B,A};B)$ 
& $\gamma_A A$ 
& $H^+(\alpha_{A,B},k_{A,B},n_{A,B};A)$ 
& $\gamma_B B$ \\
\hline
f$^{-}$FBL$^{--}$ 
& $H^-(\alpha_{A'},k_{A'},n_{A'};A')$ 
& $H^-(\alpha_{B,A},k_{B,A},n_{B,A};B)$ 
& $\gamma_A A$ 
& $H^-(\alpha_{A,B},k_{A,B},n_{A,B};A)$ 
& $\gamma_B B$ \\
\hline
f$^{-}$FBL$^{+-}$ 
& $H^-(\alpha_{A'},k_{A'},n_{A'};A')$ 
& $H^+(\alpha_{B,A},k_{B,A},n_{B,A};B)$ 
& $\gamma_A A$ 
& $H^-(\alpha_{A,B},k_{A,B},n_{A,B};A)$ 
& $\gamma_B B$ \\
\hline
f$^{-}$FBL$^{-+}$ 
& $H^-(\alpha_{A'},k_{A'},n_{A'};A')$ 
& $H^-(\alpha_{B,A},k_{B,A},n_{B,A};B)$ 
& $\gamma_A A$ 
& $H^+(\alpha_{A,B},k_{A,B},n_{A,B};A)$ 
& $\gamma_B B$ \\
\hline
\end{tabular}
\vspace{3pt}
\caption{Flux terms for the forced double feedback loop architecture under Hill-function models. 
We define the increasing (activation) Hill function as 
$H^+(\alpha,k,n;\zeta) = \alpha{\zeta^n}/({\zeta^n+k^n})$ 
and the decreasing (repression) Hill function as 
$H^-(\alpha,k,n;\zeta) = \alpha{k^n}/({\zeta^n+k^n})$. 
Degradation terms assume mass action kinetics.}
\label{tab:dfbl_motifs}
\end{table}

\subsubsection{Dynamics in time: general near-equilibrium properties}
Near equilibrium $(A'_e,A_e,B_e)=\overline{\textbf{e}}$, the dynamics of forced dFBL from Equation \eqref{dfbl_odes} collapses into the following lctiODE form:
\begin{shaded}
\begin{subequations}
   \begin{alignat}{2}
        \dfrac{d\Delta A}{dt} 
        &= 
        \left. \dfrac{\partial f_{A'}}{\partial A'} \right|_{\overline{\textbf{e}}} \Delta A' + &\left. \dfrac{\partial f_{B,A}}{\partial B} \right|_{\overline{\textbf{e}}} \Delta B - \left. \dfrac{\partial g_{_A}}{\partial A} \right|_{\overline{\textbf{e}}} \Delta A
        \\
        \dfrac{d\Delta B}{dt} 
        &= 
        &\left. \dfrac{\partial f_{A,B}}{\partial A} \right|_{\overline{\textbf{e}}} \Delta A - \left. \dfrac{\partial g_{_B}}{\partial B} \right|_{\overline{\textbf{e}}} \Delta B
   \end{alignat}
   \label{dfbl_odes_lin}
\end{subequations}
\end{shaded}
Denoting 
\begin{equation}
    m_{A'} 
    = 
    \dfrac{{A'}_e^{n_{_{A'}}-1} \alpha_{_{A'}} k_{_{A'}}^{n_{_{A'}}} n_{_{A'}}}{\left({A'}_e^{n_{_{A'}}}+k_{_{A'}}^{n_{_{A'}}}\right)^2}
    \quad , \quad
    m_{B, A} 
    = 
    \dfrac{B_e^{n_{_{B,A}}-1} \alpha_{_{B,A}} k_{_{B,A}}^{n_{_{B,A}}} n_{_{B,A}}}{\left(B_e^{n_{_{B,A}}}+k_{_{B,A}}^{n_{_{B,A}}}\right)^2} 
    \quad , \quad 
    m_{A,B} 
    = 
    \dfrac{A_e^{n_{_{A,B}}-1} \alpha_{_{A,B}} k_{_{A,B}}^{n_{_{A,B}}} n_{_{A,B}}}{\left(A_e^{n_{_{A,B}}}+k_{_{A,B}}^{n_{_{A,B}}}\right)^2}
\end{equation}
the near-equilibrium forms of its eight motif variants under Hill-function assumptions are given by:
\begin{equation}
    \textbf{f$^{\textcolor{red}{\pm}}$dFBL$^{\textcolor{teal}{\pm}\textcolor{violet}{\pm}}$:} 
    \quad 
    \begin{cases}
        \dfrac{d\Delta A}{dt} 
        = 
        \textcolor{red}{\pm} m_{A'} \Delta A' \textcolor{violet}{\pm} m_{B, A} \Delta B - \gamma_A \Delta A
        \\\\
        \dfrac{d\Delta B}{dt} 
        = 
        \textcolor{teal}{\pm}m_{A,B} \Delta A - \gamma_B \Delta B
    \end{cases}
\end{equation}

\subsubsection{Transfer function}
The species upon which the feedback loop acts is $A$, because it is both the entry point of the external driver $A'$ and the node that closes a single loop path: $A$ to $B$ to $A$. We therefore wish to find the transfer function from forcing species $A'$ to measured species $A$. Laplace transforming Equation \eqref{dfbl_odes_lin} and solving for the ratio of transformed measurement fluctuations $\Delta A$ to transformed forcing fluctuations $\Delta A'$ gives the transfer function of a general fdFBL architecture:
\begin{shaded}
\begin{equation}
    \text{TF}_\text{fdFBL} 
    = 
    \dfrac
    {
    \left. \dfrac{\partial f_{A'}}{\partial A'} \right|_{\overline{\textbf{e}}} \cdot \left(s+\left. \dfrac{\partial {g_{_B}}}{\partial B} \right|_{\overline{\textbf{e}}}\right)
    }
    {
    \left(s+\left. \dfrac{\partial {g_{_A}}}{\partial A} \right|_{\overline{\textbf{e}}}\right) \left(s+\left. \dfrac{\partial {g_{_B}}}{\partial B} \right|_{\overline{\textbf{e}}}\right) - \left. \dfrac{\partial f_{A,B}}{\partial A} \right|_{\overline{\textbf{e}}} \cdot \left. \dfrac{\partial f_{B,A}}{\partial B} \right|_{\overline{\textbf{e}}}
    }
    \label{dFBL_genTF}
\end{equation}
\begin{equation}
    =
    \dfrac{
    \left( \dfrac{\left. \dfrac{\partial f_{A'}}{\partial A'} \right|_{\overline{\textbf{e}}}}{s+\left. \dfrac{\partial {g_{_A}}}{\partial A} \right|_{\overline{\textbf{e}}}} \right)
    }
    {
    1 - 
    \left(
    \dfrac{\left. \dfrac{\partial f_{A,B}}{\partial A} \right|_{\overline{\textbf{e}}}}{s+\left. \dfrac{\partial {g_{_B}}}{\partial B} \right|_{\overline{\textbf{e}}}}
    \right)
    \left(
    \dfrac{\left. \dfrac{\partial f_{B,A}}{\partial B} \right|_{\overline{\textbf{e}}}}{s+\left. \dfrac{\partial {g_{_A}}}{\partial A} \right|_{\overline{\textbf{e}}}}
    \right)
    }
    =
    \dfrac{
    \left. \dfrac{\partial f_{A'}}{\partial A'} \right|_{\overline{\textbf{e}}}}{
    \left. \dfrac{\partial f_{B,A}}{\partial B} \right|_{\overline{\textbf{e}}}
    } \cdot
    \dfrac{ 
    \left( \dfrac{\left. \dfrac{\partial f_{B,A}}{\partial B} \right|_{\overline{\textbf{e}}}}{s+\left. \dfrac{\partial {g_{_A}}}{\partial A} \right|_{\overline{\textbf{e}}}} \right)
    }{
    1 - 
    \left(
    \dfrac{\left. \dfrac{\partial f_{A,B}}{\partial A} \right|_{\overline{\textbf{e}}}}{s+\left. \dfrac{\partial {g_{_B}}}{\partial B} \right|_{\overline{\textbf{e}}}}
    \right)
    \left(
    \dfrac{\left. \dfrac{\partial f_{B,A}}{\partial B} \right|_{\overline{\textbf{e}}}}{s+\left. \dfrac{\partial {g_{_A}}}{\partial A} \right|_{\overline{\textbf{e}}}}
    \right)
    }
    \label{fdFBL_blockform}
\end{equation}
\end{shaded}
The functional form of Equation \eqref{fdFBL_blockform} is why we can equivalently construct forced double feedback loop transfer functions using the closed-loop rule from block diagram algebra involving forward and backward paths with transfer functions
\begin{equation*}
    \left(
    \dfrac{\left. \dfrac{\partial f_{A,B}}{\partial A} \right|_{\overline{\textbf{e}}}}{s+\left. \dfrac{\partial {g_{_B}}}{\partial B} \right|_{\overline{\textbf{e}}}}
    \right)
    \quad \text{and} \quad 
    \left(
    \dfrac{\left. \dfrac{\partial f_{B,A}}{\partial B} \right|_{\overline{\textbf{e}}}}{s+\left. \dfrac{\partial {g_{_A}}}{\partial A} \right|_{\overline{\textbf{e}}}}
    \right)
\end{equation*}
respectively. From Equation \eqref{dFBL_genTF}, all eight fdFBL motif variants can be compactly described by:
\begin{equation}
    \text{TF}_{\text{f$^{\textcolor{red}{\pm}}$dFBL$^{\textcolor{teal}{\pm}\textcolor{violet}{\pm}}$}}
    = 
    \dfrac{\textcolor{red}{\pm} m_{A'} \cdot (s+\gamma_B)}{(s+\gamma_A)(s+\gamma_B)-(\textcolor{teal}{\pm}m_{_{A,B}})(\textcolor{violet}{\pm}m_{_{B,A}})}
\end{equation}

\subsubsection{Motif stability and asymptotic response}
Recall that marginal stability occurs when each pole in a system a system contains at least one pole lies on the imaginary axis, and that the imaginary part of such poles defines the system’s resonant frequency. For all architectures considered so far, marginally stable poles had zero imaginary part, meaning no true resonance was possible. Forced double feedback is the first architecture that enables nonzero imaginary poles, enabling genuine resonance. Rewriting the denominator of Equation \eqref{dFBL_genTF} as
\begin{equation*}
    \delta = 
    s^2 
    + 
    \left( \left. \dfrac{\partial {g_{_A}}}{\partial A} \right|_{\overline{\textbf{e}}} + \left. \dfrac{\partial {g_{_B}}}{\partial B} \right|_{\overline{\textbf{e}}} \right)s 
    + 
    \left( \left. \dfrac{\partial {g_{_A}}}{\partial A} \right|_{\overline{\textbf{e}}} \cdot \left. \dfrac{\partial {g_{_B}}}{\partial B} \right|_{\overline{\textbf{e}}} -\left. \dfrac{\partial f_{A,B}}{\partial A} \right|_{\overline{\textbf{e}}} \cdot \left. \dfrac{\partial f_{B,A}}{\partial B} \right|_{\overline{\textbf{e}}} \right) 
    = 
    a s^2 + b s + c \text{ .}
\end{equation*}
the quadratic formula shows that purely imaginary, nonzero conjugate poles arise when $b=0$ and $c>0$. Since the magnitude of degradation rates can only be nonnegative, the first condition enforces
\begin{equation*}
    \left. \dfrac{\partial {g_{_A}}}{\partial A} \right|_{\overline{\textbf{e}}} = \left. \dfrac{\partial {g_{_B}}}{\partial B} \right|_{\overline{\textbf{e}}} = 0 \text{ .}
\end{equation*}
implying no degradation of either $A$ or $B$. As such, the second condition becomes
\begin{equation*}
    \left. \dfrac{\partial f_{A,B}}{\partial A} \right|_{\overline{\textbf{e}}} \cdot \left. \dfrac{\partial f_{B,A}}{\partial B} \right|_{\overline{\textbf{e}}} > 0 \text{ .}
\end{equation*}
Among the eight motif variants, this corresponds to both A-to-B and B-to-A branches having identical regulation modes. From our two results, the frequencies at which resonance occurs in a forced double negative feedback loop is
\begin{shaded}
\begin{equation}
    \omega_{\text{res}} = \sqrt{\left. \dfrac{\partial f_{A,B}}{\partial A} \right|_{\overline{\textbf{e}}} \cdot \left. \dfrac{\partial f_{B,A}}{\partial B} \right|_{\overline{\textbf{e}}}} 
    \quad , \quad 
    \text{sign} \left(\left. \dfrac{\partial f_{A,B}}{\partial A} \right|_{\overline{\textbf{e}}} \right) = \text{sign} \left( \left. \dfrac{\partial f_{B,A}}{\partial B} \right|_{\overline{\textbf{e}}} \right)
\end{equation}
\end{shaded}
When the conditions for purely imaginary complex conjugate poles are not met --- namely, if at least one degradation rate is nonzero or the $A$–$B$ regulation modes have opposite signs --- the set of one zero and two poles yields one of three behaviours: local band-pass behaviour, staircase filtering, or decay with inflection; identical to the FFL cases in Table \ref{tab:ffl_breaks}. Regardless of these intermediate filtering behaviours, two asymptotic behaviours are guaranteed. In the low frequency limit ($s\rightarrow 0)$,
\begin{equation*}
    \text{TF}_\text{fdFBL} 
    \rightarrow 
    \dfrac
    {
    \left. \dfrac{\partial f_{A'}}{\partial A'} \right|_{\overline{\textbf{e}}} \cdot \left. \dfrac{\partial {g_{_B}}}{\partial B} \right|_{\overline{\textbf{e}}}
    }
    {
    \left. \dfrac{\partial {g_{_A}}}{\partial A} \right|_{\overline{\textbf{e}}} \cdot \left. \dfrac{\partial {g_{_B}}}{\partial B} \right|_{\overline{\textbf{e}}} - \left. \dfrac{\partial f_{A,B}}{\partial A} \right|_{\overline{\textbf{e}}} \cdot \left. \dfrac{\partial f_{B,A}}{\partial B} \right|_{\overline{\textbf{e}}}
    }
    =c \in \mathbb{R} \text{ .}
\end{equation*}
This indicates that under quasi-static forcing the system exhibits constant amplitude modulation by a factor $c$. Depending on the regulation modes between the three species, $c$ may be positive or negative, corresponding to a phase shift of $0^\circ$ (in phase) or $180^\circ$ (out of phase) between input and output oscillations. In the high frequency limit ($s\rightarrow \infty)$, 
\begin{equation*}
    \text{TF}_\text{fdFBL} 
    \rightarrow 
    \lim_{s\rightarrow \infty}
    \dfrac
    {
    \left. \dfrac{\partial f_{A'}}{\partial A'} \right|_{\overline{\textbf{e}}}
    }
    {
    s 
    + 
    \left( \left. \dfrac{\partial {g_{_A}}}{\partial A} \right|_{\overline{\textbf{e}}} + \left. \dfrac{\partial {g_{_B}}}{\partial B} \right|_{\overline{\textbf{e}}} \right)
    }
    =0 \text{ .}
\end{equation*}
Therefore, the architecture filters out high frequency oscillations. Our transfer function has asymptotic order of $s^0/s^1$ in this high-frequency regime, suggesting the asymptotic phase shift is $-90^\circ$ from Equation \eqref{phi_lim}.

\subsubsection{Frequency response function}
As was with FFLs, given the number of parameters present in our parameter space, plotting phasor and Bode trajectories to characterise the effect of varying parameter values can cause misleading simplifications. In this regard, we leave the reader with the full analytical frequency response functions for the general forced double feedback loop architecture:
\begin{equation*}
    M(\omega) 
    = 
    \dfrac{\Bigg| \left(\left. 
    \left. \dfrac{\partial f_{A'}}{\partial A'} \right|_{\overline{\textbf{e}}} \cdot \dfrac{\partial g_{_B}}{\partial B} \right|_{\overline{\textbf{e}}}\right) \Bigg|}
    {
    \left( \sqrt
    {
    \Big[\Re(p_{_+})\Big]^2 + \Big[\omega - \Im(p_{_+})\Big]^2
    }\right)
    \left(\sqrt
    {
    \Big[\Re(p_{_-})\Big]^2 + \Big[\omega - \Im(p_{_-})\Big]^2
    }\right)
    }
\end{equation*}
\begin{equation*}
    \phi(\omega) = \arg \Bigg( \left.\dfrac{\partial f_{A'}}{\partial A'} \right|_{\overline{\textbf{e}}} \Bigg) 
    +
    \arctantwo\Bigg(\omega,\left. \dfrac{\partial g_{_B}}{\partial B} \right|_{\overline{\textbf{e}}}\Bigg)
    -
    \arctantwo\Bigg(\omega - \Im(p_{_+}),-\Re(p_{_+})\Bigg)
    -
    \arctantwo\Bigg(\omega - \Im(p_{_-}),-\Re(p_{_-})\Bigg)
\end{equation*}
where
\begin{equation*}
    p_\pm = \dfrac{1}{2} \left[ -\left( \left. \dfrac{\partial {g_{_A}}}{\partial A} \right|_{\overline{\textbf{e}}} + \left. \dfrac{\partial {g_{_B}}}{\partial B} \right|_{\overline{\textbf{e}}} \right) \pm \sqrt{\left( \left. \dfrac{\partial {g_{_A}}}{\partial A} \right|_{\overline{\textbf{e}}} + \left. \dfrac{\partial {g_{_B}}}{\partial B} \right|_{\overline{\textbf{e}}} \right)^2 - 4\left( \left. \dfrac{\partial {g_{_A}}}{\partial A} \right|_{\overline{\textbf{e}}} \cdot \left. \dfrac{\partial {g_{_B}}}{\partial B} \right|_{\overline{\textbf{e}}} -\left. \dfrac{\partial f_{A,B}}{\partial A} \right|_{\overline{\textbf{e}}} \cdot \left. \dfrac{\partial f_{B,A}}{\partial B} \right|_{\overline{\textbf{e}}} \right)} \right]
\end{equation*}

%% file: 5_Appendix.tex
\section{Appendix A: Frequency response plots under biological values}

In the following figures, frequency response plots of each motif are generated under representative in vivo conditions observed in E. coli and yeast, where synthesis rates $\alpha$ span $[0.02,0.1]$ nM$\cdot$min$^{-1}$, dissociation constants $k$ range within $[0.01,0.10]$ nM and loss rates $\gamma$ occur between $[0.01.0.24]$ min$^{-1}$. The representative values of each quantity is assigned to its median; that is: $\alpha = 0.05$ nM$\cdot$min$^{-1}$, $k = 0.03$ nM and $\gamma = 0.05$ min$^{-1}$. For simplicity the median concentration of our driving species $A_e$ and cooperativity constant $n$ are assumed to be $1$ nM/min and $2$ (unitless), respectively.

\begin{figure}[ht]
    \centering
    \includegraphics[width=1\linewidth]{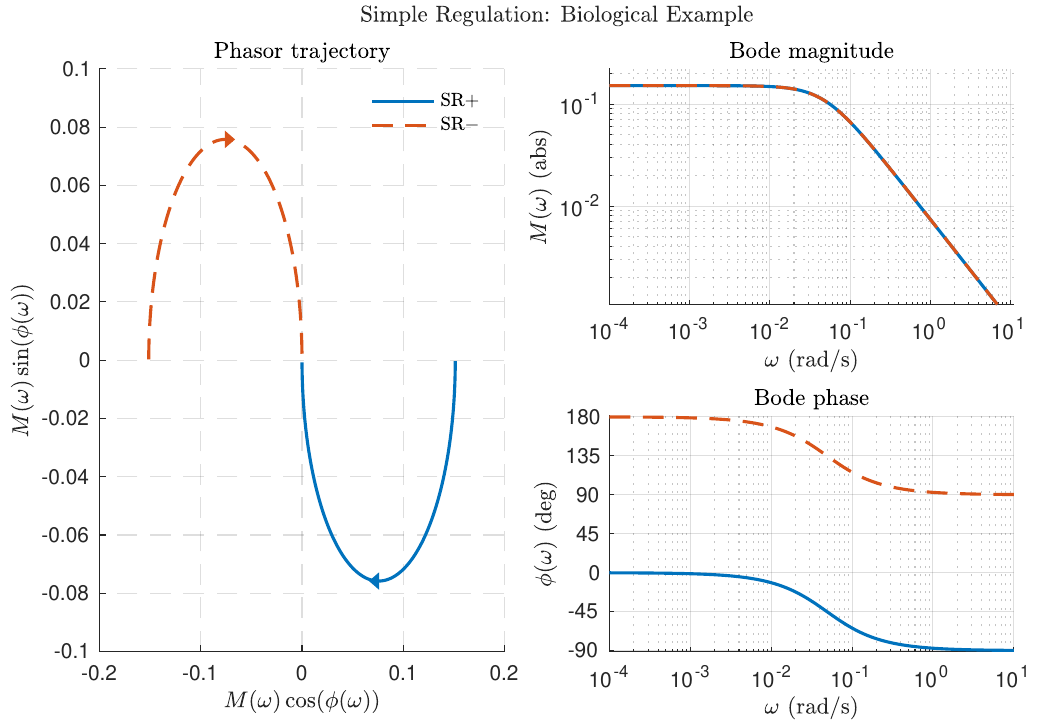}
    \caption{Phasor (left) and Bode (right) plots of positive (blue) and negative (orange) SR, under representative parameter values.}
    \label{fig:Biology_SR}
\end{figure}

\begin{figure}[ht]
    \centering
    \includegraphics[width=1\linewidth]{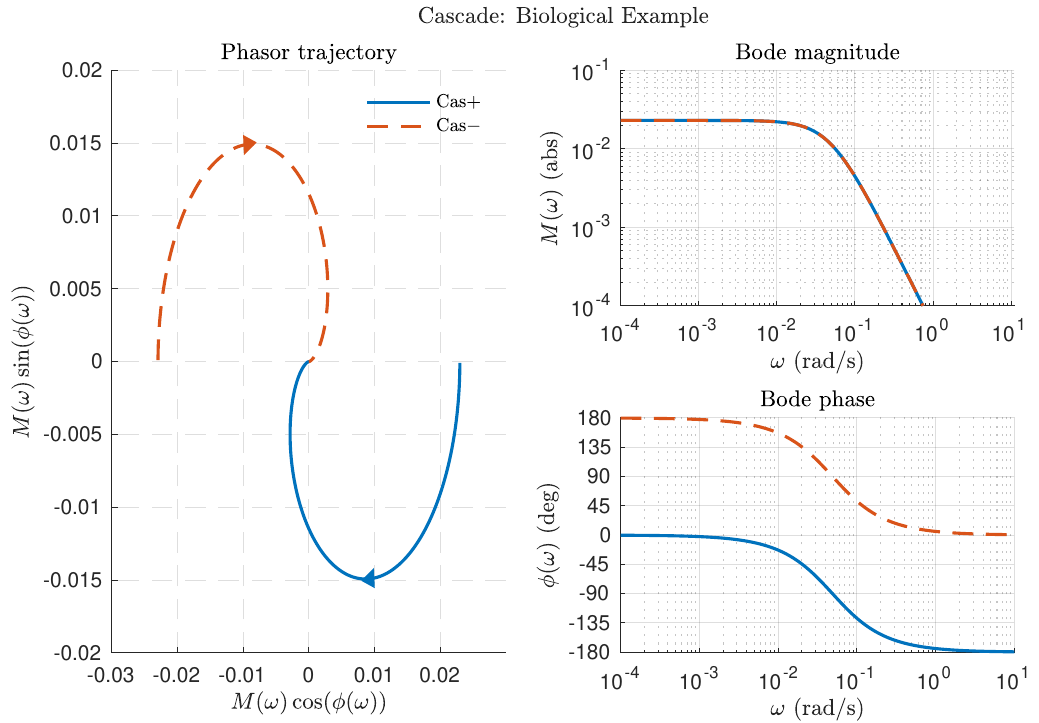}
    \caption{Phasor (left) and Bode (right) plots of positive (blue) and negative (orange) cascades, with each SR motif under the same representative parameter values.}
    \label{fig:Biology_Cas}
\end{figure}

%% file: 6_AppendixB.tex
\section{Appendix B: autonomous autoregulation}
\subsubsection{General architecture}
Unlike forced AR, the autonomous counterpart does not possess an external upstream force. Its dynamics are solely governed by the single self-regulating species via the equation:

\begin{equation} \label{AR_general}
    \frac{dA}{dt} = f_1(A) - f_2(A) \text{ .}
\end{equation}

\begin{figure}[ht]
    \centering
    \includegraphics[scale=0.7]{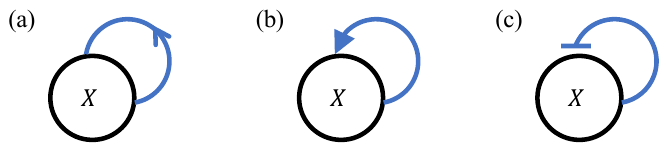}
    \caption{General autoregulation motif architecture (a), and its positive (b) and negative (c) modes.}
    \label{fig:AR_archs}
\end{figure}

Here, $f_1$ and $f_2$ describe the rates of production and loss of population A due to itself, respectively. This interaction is often illustrated with a single node looping onto itself (see Figure \ref{fig:AR_archs}(a)). Like SR, AR also occurs in two modes — promotion and repression — which we call positive AR and negative AR, respectively. In positive AR, each A molecule reinforces the production of more A molecules --- increasing the system's production rate until a maximum is reached. In negative AR, the converse happens. Each A molecule inhibits the production of more A molecules --- continuing in a perpetual cycle until population reaches a minimum. These are illustrated as Figure \ref{fig:AR_archs}(b) and (c).

\subsubsection{Dynamics in time: a common model}
In positive AR, the production rate of species A increases with its own concentration until saturation. This monotonic rise toward saturation in $f_1$ is commonly modelled using an increasing Hill function \cite{alon_introduction_2019}:

\begin{equation}
    f_1(A)=\alpha \frac{A^n}{A^n+k^n} \quad \text{,} \quad \alpha,k,n\in\mathbb{R^+}
\end{equation}

In contrast, in negative autoregulation, the production rate of A decreases as its concentration increases, approaching a basal production rate. This monotonic decline is often modelled using a decreasing Hill function \cite{alon_introduction_2019}:

\begin{equation}
    f_1(A)=\alpha \frac{k^n}{A^n+k^n} \quad \text{,} \quad \alpha,k,n\in\mathbb{R^+} \text{ .}
\end{equation}

In both cases, the loss rate in A is proportional to its concentration and modelled as linear degradation:

\begin{equation}
    f_2(A)=\gamma A \quad \text{,} \quad \gamma\leq0 \text{ .}
\end{equation}
\begin{shaded}
The resulting dynamical models are:
\begin{align}
\textbf{Positive AR:} \quad &\frac{dA}{dt} = \alpha \frac{A^n}{A^n + k^n} - \gamma A \label{eq:ARpos} \\
\textbf{Negative AR:} \quad &\frac{dA}{dt} = \alpha \frac{k^n}{A^n + k^n} - \gamma A \label{eq:ARneg}
\end{align}
\end{shaded}

\subsubsection{Dynamics in time: general near-equilibrium properties}
Autoregulation drives A towards equilibrium $A_e$. Local to $A_e$, the dynamics of Equation \eqref{AR_general} collapse into its linear regime:

\begin{equation}
    \frac{d \Delta A}{d t} \sim \left.\left(\frac{\partial f_1}{\partial A}-\frac{\partial f_2}{\partial A}\right)\right|_{A_e} \cdot \Delta A \text{ .}
    \label{eq:ar_taylor1}
\end{equation}

For each motif case, this becomes:
\begin{shaded}
\begin{align}
\textbf{Positive AR:} \quad &\frac{d\Delta A}{dt} \sim +m - \gamma \label{eq:ARpos_dDeltaA}\\
\textbf{Negative AR:} \quad &\frac{d\Delta A}{dt} \sim -m - \gamma
\label{eq:ARneg_dDeltaA} \end{align}
where 
\begin{equation*}
    \frac{A_e^{n-1} \alpha k^n n}{\left(A_e^n+k^n\right)^2} = m\left(A_e, \alpha, k, n,\right) \in \mathbb{R}^{+} \text{ .}
\end{equation*}
These two equations describe how fluctuations in A influence its own rate of change --- that is, how a deviation from equilibrium $A_e$ determines the evolution of that same deviation over time.
\end{shaded}

\subsubsection{Transfer function}
Recall from Equation \eqref{TF} that the transfer function is defined as the ratio between the Laplace transforms of two unique species. One might therefore question how a transfer function can be defined from Equation \eqref{eq:ARpos_dDeltaA}, which involves only a single species. A pair still exists --- A and itself --- but this pairing is subtly hidden in time. To see this pair, we must first unravel autoregulation into a recursive chain of events --- beginning with $\Delta A$ at some initial time $t_0$ to $\Delta A$ at a later time $t_1$, and so on. In this view, the relevant pair of species consists of any two adjacent instances of $\Delta A$. What the transfer function captures, then, is how the shape of a given fluctuation is transformed by the autoregulatory mechanism into a new version of itself --- altered in shape, but not species. This differs from the previously discussed non-feedback architectures, in which transfer functions described transformations into new species.

To start this derivation, we first move Equation \eqref{eq:ARpos_dDeltaA} from the time domain into the frequency domain via the Laplace transform, as follows:

\begin{align}
    \mathcal{L}\left\{\frac{d(\Delta A)}{d t}\right\}(s)&=\mathcal{L}\bigg\{(m-\gamma) \cdot \Delta A\bigg\}(s) \nonumber\\
    \Rightarrow s \mathcal{L}\{\Delta A\}(s)-\left.\Delta A\right|_{t=0}&=(m-\gamma) \mathcal{L}\{\Delta A\}(s)
    \label{eq:ARpos_sLA}
\end{align}

Here lies the crucial realisation: unlike the motifs discussed previously, the initial condition in this case is not necessarily zero. That is, $\left.\Delta A\right|_{t=0} \neq 0$. Each iteration through the feedback loop alters this initial condition. This nonzero value is a reflection of our earlier conceptual unravelling of the loop in time. This unravelling created a recursive structure, where each recursive unit inherits its initial condition from the end of the previous one and produces a new initial condition for the next. As such, the value of $\left.\Delta A\right|_{t=0}$ is therefore only meaningful within the context of a given recursive unit. We must first decide on a recursive unit to work on before interpreting its initial condition. Thus, we interpret the nonzero initial condition $\left.\Delta A\right|_{t=0}$ as a quasi-input --- not imposed externally but arising intrinsically from the system’s internal recursion --- and interpret $\Delta A$ as the quasi-output. Note that it does not matter which recursive unit we pick, since all units have the relative transformation from start to end. Transposing Equation \eqref{eq:ARpos_sLA} to extract the ratio of output $\mathcal{L}\{\Delta A\}(s)$ to input $\left.\Delta A\right|_{t=0}$ gives:

\begin{align}
(s-(m-\gamma)) \mathcal{L}\{\Delta A\}(s)&=\left.\Delta A\right|_{t=0} \nonumber \\
\Rightarrow \frac{\mathcal{L}\{\Delta A\}(s)}{\left.\Delta A\right|_{t=0}}&=\frac{1}{s+\gamma-m} \text{ .}
\end{align}

Repeating the same procedure on Equation \eqref{eq:ARneg_dDeltaA}, we can obtain the transfer function for negative AR.

\begin{shaded}
The transfer function for positive and negative autonomous autoregulations are:
\begin{equation}
    \text{TF}_{\text{aAR}^{\pm}}(s)=\frac{1}{s\mp m + \gamma} \hspace{1em},\hspace{1em} m,\gamma\in \mathbb{R}^+ \text{ .}
    \label{TF_aAR}
\end{equation}
Notice that these two autonomous transfer functions are just the denominators of their forced counterparts from Equation \eqref{fAR_TFs}. That is: 
\begin{equation*}
    \text{TF}_{\text{f}^{\textcolor{red}{\pm}} \text{AR}^{+}} = (\pm m_1) \cdot \text{TF}_{\text{aAR}^{+}}(s) 
    \quad \text{and} \quad 
    \text{TF}_{\text{f}^{\textcolor{red}{\pm}} \text{AR}^{-}} = (\pm m_1) \cdot \text{TF}_{\text{aAR}^{-}}(s) \text{ .}
\end{equation*}
\textbf{This correspondence shows why it is natural and physically meaningful to define an intrinsic transfer function for autoregulation (and feedback loops in general). It captures the core dynamical contribution of the autoregulatory loop itself, independent of the external driving branch, and makes explicit how forced motifs inherit their behaviour directly from their autonomous backbone. Conceptually (with a slight abuse of mathematical rigour), one can view forced autoregulation as a cascade of two elements: a constant gain factor given by the upstream driving force $(\pm m_1)$, and the intrinsic transfer function }$\text{TF}_{\text{aAR}^{\pm}}(s)$\textbf{. This provides a useful representational tool for reasoning about how the forced autoregulation architecture decomposes into its external forcing and internal dynamics --- connecting the seemingly incompatible definitions of autoregulation in biological and control-oriented literatures}.
\end{shaded}

\subsubsection{Motif stability and asymptotic response}
$T F_{A R^{+}}(s)$ has a single pole at $s=m-\gamma$. Depending on the relationship between the feedback production strength $m$ and the degradation rate $\gamma$, \textbf{the positive AR motif can exhibit all three stability modes}:
\begin{center}
\begin{tabular}{ccc}
\toprule
\textbf{Condition} & \textbf{Pole Location} & \textbf{Stability of AR$^+$ Motif} \\
\midrule
$\gamma > m$ & $s = m - \gamma < 0$ & Asymptotically stable \\
$\gamma = m$ & $s = 0$ & Marginally stable \\
$\gamma < m$ & $s = m - \gamma > 0$ & Unstable \\
\bottomrule
\end{tabular}
\end{center}
The positive autoregulation mechanism dampens oscillations to zero when the rate of degradation $\gamma$ exceeds that of production $m$. The time-domain intuition for this comes from realising that the condition $\gamma>m$ implies there is more expenditure than production and, so, each autoregulatory cycle expends more $A$ molecules than the number created. This deficit carries over to the next cycle, reducing the size of the next input fluctuation $\Delta A$. This repeating deficit continues until $\Delta A$ dies out and the system settles back into equilibrium $A_e$. 

In the intermediate case, when production and degradation are exactly balanced, oscillations do not occur. The system remains locked in a steady state, with no net accumulation or loss of A over time.

Finally, when the rate of production exceeds that of degradation, the mechanism produces unbounded oscillations. Converse to the first case, since there is more production than expenditure, each autoregulatory cycle creates a surplus of A molecules. This surplus compounds into the input of the next cycle, causing $\Delta A$ to grow without bound over time.

$T F_{A R^{-}}(s)$ has a single pole at $s=-m-\gamma$. Unlike its positive counterpart, \textbf{the negative AR motif is always asymptotically stable} --- even in the absence of degradation rate. This stability is intuitive --- the production mechanism, described by rate $f_1(A)$, decreases as A increases, thereby self-limiting the accumulation of A.

\subsubsection{Slow and high limit forcing response}
Taking the limits as $|s|\rightarrow0$ and $|s|\rightarrow\infty$, we can deduce the following table. 

\begin{table}[H]
\renewcommand{\arraystretch}{2}
\centering
\begin{tabular}{|c|c|c|c|c|c|c|}
\hline
\textbf{Motif} & 
\textbf{\makecell{Slow Forcing\\Limit $s \to 0$}} & 
\makecell{Quasistatic\\Gain Limit} & 
\makecell{Low-Frequency\\Phase Shift\\Limit} & 
\textbf{\makecell{Rapid Forcing\\Limit $s \to \infty$}} & 
\makecell{Asymptotic\\Gain Limit} &
\makecell{High-Frequency\\Phase Shift\\Limit}\\
\hline
\textbf{\makecell{Positive\\AR}} 
& $\dfrac{1}{\gamma-m} + 0i$ 
& $\left| \dfrac{1}{\gamma-m} \right|$
& \makecell{$0^\circ$ (asym. stab.)
\\ $-90^\circ$ (marg. stab.)
\\ $-180^\circ$ (unstable)}
& $0 + 0i$ 
& $0$ 
& $-90^\circ$\\
\hline
\textbf{\makecell{Negative\\AR}} 
& $\dfrac{1}{\gamma+m} + 0i$ 
& $\dfrac{1}{\gamma+m}$ 
& $0^\circ$ 
& $0 + 0i$ 
& $0$ 
& $-90^\circ$\\
\hline
\end{tabular}
\vspace{3pt}
\caption{Frequency response characteristics of positive and negative AR motifs in both low- and high-frequency forcing regimes.}
\label{tab:AR_freq_response}
\end{table}

\subsubsection{Frequency response functions}

\begin{figure}[ht]
    \centering
    \includegraphics[width=1\linewidth]{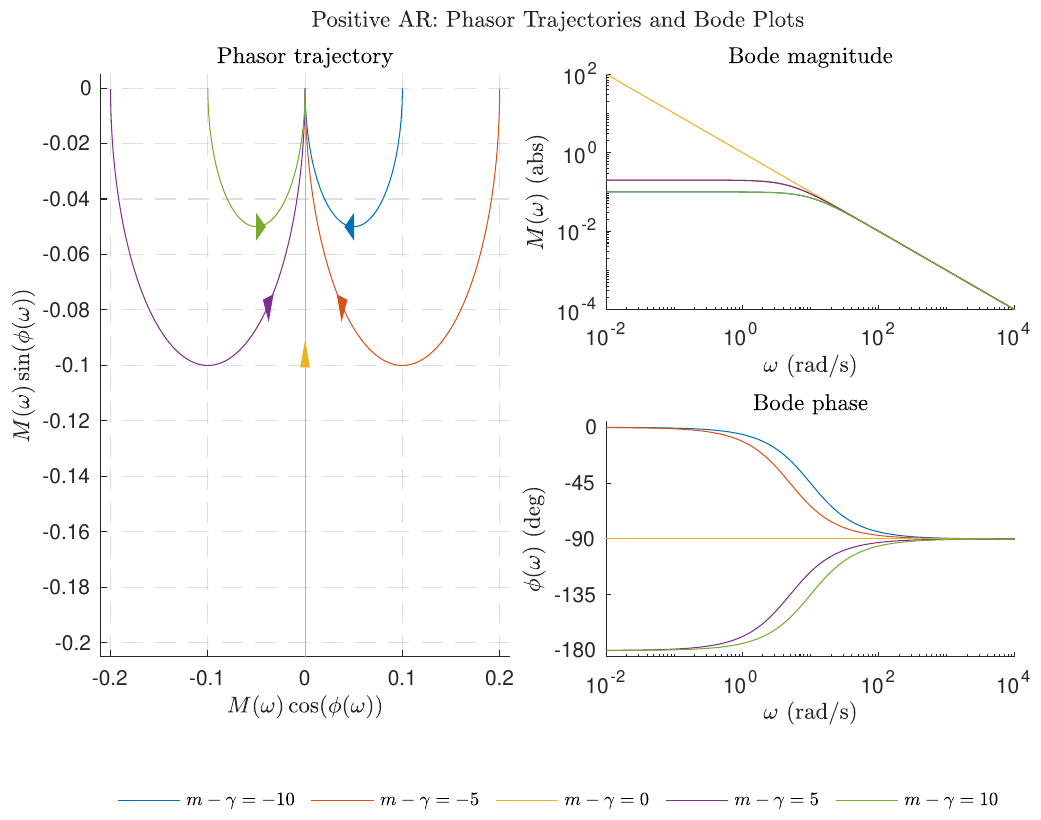}
    \caption{Phasor and Bode plots for positive AR. The blue-green and orange-purple pairs share identical modulus profiles, as seen in their Phasor trajectories, and so overlap each other when plotted as Bode magnitude curves.}
    \label{fig:ARpos}
\end{figure}

The analytical amplitude modulation and phase shift frequency-response functions are

\begin{subequations}
    \begin{align}
        M_{A R^{+}}(\omega) &= 
        \frac{1}
        {\sqrt{\left( (\gamma-m)^2 +\omega^2 \right)}}
        \\
        \phi_{A R^{+}}(\omega) &= 
        - \arctantwo\left(\omega,\gamma-m\right)
    \end{align}
\end{subequations}

for the positive AR motif, and 

\begin{subequations}
    \begin{align}
        M_{A R^{-}}(\omega) &= 
        \frac{1}
        {\sqrt{\left( (m+\gamma)^2 +\omega^2 \right)}}
        \\
        \phi_{A R^{-}}(\omega) &= 
        - \arctantwo\left(\omega,m+\gamma\right)
    \end{align}
\end{subequations}

for the negative AR motif. The corresponding phasor and Bode plots for positive autoregulation (AR) are shown in Figure \ref{fig:ARpos}. In this regime, both the quasi-static low-frequency amplification factor and the break frequency are fully determined by the magnitude of the difference between the degradation rate $\gamma$ and the linear-regime production rate $m$. The outputs of marginally stable and unstable realisations of the positive AR motif are always completely out of phase with one another, at low frequencies, but converge toward the same phase quadrature in the rapid forcing limit. Note that, since the poles the negative AR motif are strictly non-negative, their plots form a subset of those in Figure \ref{fig:ARpos} with $m - \gamma \geq 0$. The quasi-static amplification and break frequency for negative AR is governed by the sum of the degradation rate $\gamma$ and local production rate $m$.

%% file: 7_AppendixC.tex
\subsection{Appendix C: autonomous double feedback loops}
\subsubsection{General architecture}
\begin{figure}[ht]
    \centering
    \includegraphics[width=0.7\linewidth]{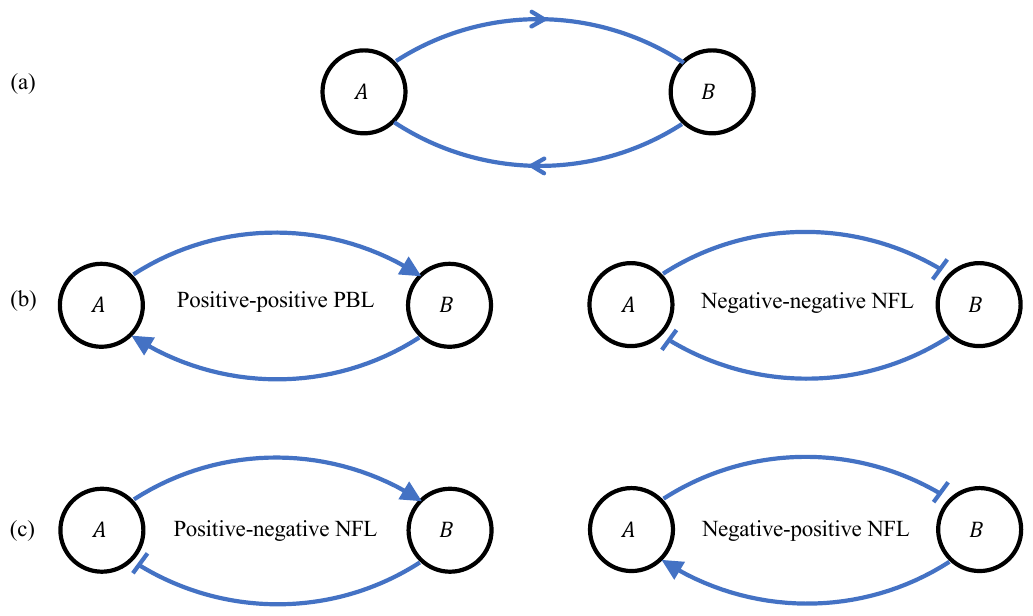}
    \caption{(a) General architecture of two-species feedback loop motifs. (b) Positive feedback loops. (c) Negative feedback loops.}
    \label{fig:FBL_archs}
\end{figure}

In autonomous double FBLs (adFBLs), the production rate of each species is solely dependent on the other, while the degradation of each is solely dependent on itself. This a architecture of interaction is described with:
\begin{subequations}
   \begin{align}
   \dfrac{dA}{dt} &= f_{B,A}(B) - g_A(A) \\
   \dfrac{dB}{dt} &= f_{A,B}(A) - g_B(B)
   \end{align}
   \label{fbl_odes}
\end{subequations}

where subscript $A,B$ has been adopted to indicate the direction of interaction from species $A$ to $B$, and vice versa. FBLs occur in four types: positive-positive and negative-negative, which together form the class of two-species positive feedback loops (PFLs) in the literature; and positive-negative and negative-positive, which together form the class of two-species negative feedback loops (NFLs) in the literature. Figure \ref{fig:FBL_archs} illustrates each architecture. We adopt the convention adFBL$^{\textcolor{black}{\pm}\textcolor{black}{\pm}}$, where the superscripts denote the regulation modes from $A \to B$ and $B \to A$, respectively.

\subsubsection{Dynamics in time: general near-equilibrium properties and common models}
Near equilibrium $(A_e,B_e)$, the local dynamics of each species take on lctiODE form:
\begin{subequations}
   \begin{align}
   \dfrac{d\Delta A}{dt} 
   &= \left. \dfrac{\partial f_{B,A}}{\partial B} \right|_{(A_e,B_e)} \Delta B + \left. \dfrac{\partial g_{A}}{\partial A} \right|_{(A_e,B_e)} \Delta A
   \\
   \dfrac{d\Delta B}{dt} 
   &= \left. \dfrac{\partial f_{A,B}}{\partial A} \right|_{(A_e,B_e)} \Delta A + \left. \dfrac{\partial g_{B}}{\partial B} \right|_{(A_e,B_e)} \Delta B
   \end{align}
   \label{fbl_odes_lin}
\end{subequations}

Under the common autonomous double feedback model in literature, with Hill production and mass-action degradation terms Table \ref{tab:fbl_motifs}, the near equilibrium dynamics for all four adFBL$^{\textcolor{black}{\pm}\textcolor{black}{\pm}}$ motifs are described by:
\begin{equation}
    \textbf{adFBL$^{\textcolor{teal}{\pm}\textcolor{violet}{\pm}}$:} 
    \quad 
    \begin{cases}
        \dfrac{d\Delta A}{dt} 
        = 
        \textcolor{violet}{\pm} m_{B, A} \Delta B - \gamma_A \Delta A
        \\\\
        \dfrac{d\Delta B}{dt} 
        = 
        \textcolor{teal}{\pm}m_{A,B} \Delta A - \gamma_B \Delta B
    \end{cases}
\end{equation}

where
\begin{equation*}
    m_{_{A,B}} = \dfrac{A_e^{n_{_{A,B}}-1} \alpha_{_{A,B}} k_{_{A,B}}^{n_{_{A,B}}} n_{_{A,B}}}{\left(A_e^{n_{_{A,B}}}+k_{_{A,B}}^{n_{_{A,B}}}\right)^2}
    \quad \text{and} \quad 
    m_{_{B,A}} = \dfrac{B_e^{n_{_{B,A}}-1} \alpha_{_{B,A}} k_{_{B,A}}^{n_{_{B,A}}} n_{_{B,A}}}{\left(B_e^{n_{_{B,A}}}+k_{_{B,A}}^{n_{_{B,A}}}\right)^2} 
    \text{ .}
\end{equation*}

\begin{table}[ht]
\renewcommand{\arraystretch}{2}
\centering
\begin{tabular}{|c||c|c|c|c|}
\hline
\textbf{Motif} 
& $f_{B,A}(B)$ 
& $g_1(A)$ 
& $f_{A,B}(A)$ 
& $g_2(B)$ \\
\hline\hline
\makecell{Positive-\\positive} 
& $\alpha_{_{B,A}} \dfrac{B_e^{n_{B,A}}}{B_e^{n_{B,A}}+k_{_{B,A}}^{n_{B,A}}}$ 
& $\gamma_A A$ 
& $\alpha_{_{A,B}} \dfrac{A_e^{n_{A,B}}}{A_e^{n_{A,B}}+k_{_{A,B}}^{n_{A,B}}}$ 
& $\gamma_B B$ \\
\hline
\makecell{Negative-\\negative} 
& $\alpha_{_{B,A}} \dfrac{k_{_{B,A}}^{n_{B,A}}}{B_e^{n_{B,A}}+k_{_{B,A}}^{n_{B,A}}}$ 
& $\gamma_A A$ 
& $\alpha_{_{A,B}} \dfrac{k_{_{A,B}}^{n_{A,B}}}{A_e^{n_{A,B}}+k_{_{A,B}}^{n_{A,B}}}$
& $\gamma_B B$ \\
\hline
\makecell{Positive-\\negative} 
& $\alpha_{_{B,A}} \dfrac{B_e^{n_{B,A}}}{B_e^{n_{B,A}}+k_{_{B,A}}^{n_{B,A}}}$ 
& $\gamma_A A$ 
& $\alpha_{_{A,B}} \dfrac{k_{_{A,B}}^{n_{A,B}}}{A_e^{n_{A,B}}+k_{_{A,B}}^{n_{A,B}}}$
& $\gamma_B B$ \\
\hline
\makecell{Negative-\\positive} 
& $\alpha_{_{B,A}} \dfrac{k_{_{B,A}}^{n_{B,A}}}{B_e^{n_{B,A}}+k_{_{B,A}}^{n_{B,A}}}$ 
& $\gamma_A A$ 
& $\alpha_{_{A,B}} \dfrac{A^{n_{A,B}}}{A_e^{n_{A,B}}+k_{_{A,B}}^{n_{A,B}}}$ 
& $\gamma_B B$ \\
\hline
\end{tabular}
\vspace{3pt}
\caption{Gain and loss rates for each autonomous dual feedback loop motif model, shown in the global nonlinear regime.}
\label{tab:fbl_motifs}
\end{table}

\subsubsection{Transfer function}
As with autonomous autoregulation, we face a similar difficulty in extracting a transfer function from the lctiODEs: it is not straightforward to obtain a clean ratio between the Laplace transforms of our two species. The feedback loop couples perturbations $\Delta A$ and $\Delta B$ in such a way that expressing one solely in terms of the other is impossible. For the individual lines \eqref{fbl_odes_lin}a and \eqref{fbl_odes_lin}b, SR transfer functions can be derived separately, but a single closed-form transfer function for the system as a whole cannot.  

To proceed, we must “disentangle” the implicit input of each feedback cycle $\Delta A|_{t=0}$ from the internal dynamics --- the same approach used in the autonomous autoregulation motif. The key idea is to unravel the loop and isolate the transfer function of a single feedback cycle, recognising that the initial condition of $\Delta A$ need not be zero. An intuition for this comes from the \textit{law of inertia}: an isolated system (our autonomous dynamical loop) at rest will remain at rest unless it is given an initial displacement. Likewise, the biological loop can only undergo feedback if its initial displacement ``force'' $\Delta A|_{t=0}$ is nonzero.

We begin by taking the Laplace transform of the two individual biochemical processes in Equation \eqref{fbl_odes_lin}, giving:

\begin{subequations}
   \begin{align}
   s \mathcal{L}\{\Delta A\}(s) - \Delta A|_{t=0}
   &= \left. \dfrac{\partial f_{B,A}}{\partial B} \right|_{(A_e,B_e)} \mathcal{L}\{\Delta B\}(s) - \left. \dfrac{\partial g_A}{\partial A} \right|_{(A_e,B_e)} \mathcal{L}\{\Delta A\}(s) \label{FBL_AtoB}
   \\
   s \mathcal{L}\{\Delta B\}(s) - \Delta B|_{t=0}
   &= \left. \dfrac{\partial f_{A,B}}{\partial A} \right|_{(A_e,B_e)} \mathcal{L}\{\Delta A\}(s) -\left. \dfrac{\partial g_B}{\partial B} \right|_{(A_e,B_e)} \mathcal{L}\{\Delta B\}(s) \label{FBL_BtoA}
   \end{align}
   \label{fbl_L}
\end{subequations}

Here lies the crucial step that allows us to disentangle the coupling that hides the intrinsic transfer function: we set $\Delta A|_{t=0}\neq0$ and $\Delta B|_{t=0}=0$. This choice is not arbitrary, but rather a statement about the recursive structure of the system. When we ``unravel'' the coupled dynamics --- what we have referred to as the feedback loop architecture --- we see that each cycle corresponds to a process that begins with a perturbation in A, which then influences perturbations in B, and in turn produces the signal that initiates the next perturbation to A. In this picture, setting $\Delta B|_{t=0}=0$ asserts that, at the start of a new cycle, $B$ has not yet responded; while $\Delta A|_{t=0}$ specifies the magnitude of the incoming perturbation to $A$, triggered by the accumulated influence of $B$ from the previous cycle. With this convention, Equation \eqref{fbl_L}(b) yields
\begin{equation}
    \mathcal{L}\{\Delta B\}(s) 
    = 
    \dfrac{\left. \dfrac{\partial f_{A,B}}{\partial A} \right|_{(A_e,B_e)}}
    {s+\left. \dfrac{\partial g_B}{\partial B} \right|_{(A_e,B_e)}}\mathcal{L}\{\Delta A\}(s) \text{ .}
\end{equation}

Substituting back into Equation \eqref{fbl_L}(a) returns a single equation that describes perturbations in $A$ (i.e. $\mathcal{L}\{\Delta A\}(s)$) solely in terms of initial condition $\Delta A|_{t=0}$. This captures a single unit cycle of the autonomous feedback loop, from initial drive $\Delta A|_{t=0}$ to consequent perturbation $\Delta A$. Solving for the ratio of $\mathcal{L}\{\Delta A\}(s)$ to $\Delta A|_{t=0}$ gives the intrinsic transfer function:
\begin{shaded}
\begin{equation}
    \text{TF}_{\text{adFBL}} 
    = \dfrac{\mathcal{L}\{\Delta A\}(s)}{\Delta A|_{t=0}} 
    = \dfrac{s+\left. \dfrac{\partial g_B}{\partial B} \right|_{(A_e,B_e)}}{\left(s+\left. \dfrac{\partial g_A}{\partial A} \right|_{(A_e,B_e)}\right)\left(s+\left. \dfrac{\partial g_B}{\partial B} \right|_{(A_e,B_e)}\right) - \left. \dfrac{\partial f_{A,B}}{\partial A} \right|_{(A_e,B_e)}\left. \dfrac{\partial f_{B,A}}{\partial B} \right|_{(A_e,B_e)}}
    \label{TF_asFBL}
\end{equation}
\textbf{Similar to the relationship between autonomous autoregulations and their forced counterparts, notice that the intrinsic transfer function of the autonomous double feedback loop is simply that of the forced transfer function without its forcing flux coefficient from species $A'$ to $A$ in Equation \eqref{dFBL_genTF}. That is:}
\begin{equation*}
    \text{TF}_{\text{fdFBL}} = \left. \dfrac{\partial f_{A'}}{\partial A'} \right|_{\overline{\textbf{e}}} \cdot \text{TF}_{\text{adFBL}} \text{ .}
\end{equation*}
\end{shaded}
From Equation \eqref{TF_asFBL}, like-sign feedback pairs share one transfer function, while the opposite-sign pairs share another, as follows:
\begin{subequations}
\begin{gather}
    \text{TF}_{\text{adFBL}^{++}} 
    = \text{TF}_{\text{adFBL}^{--}} 
    =
    \dfrac{(s+\gamma_B)}{(s+\gamma_A)(s+\gamma_B)-m_{_{A,B}}m_{_{B,A}}}
    \\
    \text{TF}_{\text{adFBL}^{+-}} 
    = \text{TF}_{\text{adFBL}^{-+}} 
    = \dfrac{(s+\gamma_B)}{(s+\gamma_A)(s+\gamma_B)+m_{_{A,B}}m_{_{B,A}}}
\end{gather}
\end{subequations}